\pdfoutput=1
\documentclass[12pt,letterpaper]{article}
\usepackage[nosort]{cite}
\usepackage{graphicx}
\usepackage{amsfonts}
\usepackage{amssymb}
\usepackage{amsmath}
\usepackage{amsthm}
\usepackage{wgcrev}
\usepackage{afterpage}
\usepackage{verbatim}
\usepackage{color}
\usepackage{slashed}
\usepackage[labelfont={bf},skip=0.8\baselineskip]{caption}
\usepackage{enumerate}
\usepackage[margin=1in]{geometry}
\usepackage{titletoc}%
\usepackage{hyperref}

\numberwithin{equation}{section}

\newtheoremstyle{named}{}{}{\itshape}{}{\bfseries}{.}{.5em}{#3}
\theoremstyle{named}
\newtheorem*{namedconjecture}{Conjecture}

\newcommand{\be}{\begin{equation}}
\newcommand{\ee}{\end{equation}}
\newcommand{\Tr}{{\rm Tr}}

\newcommand{\rmd}{\textrm{d}}

\newcommand{\LQG}{\Lambda_{\rm QG}}

\newcommand{\bb}{\mathbb}
\newcommand{\Mp}{M_{\mathrm{Pl}}}
\newcommand{\wt}{\widetilde}

\usepackage{pifont}
\newcommand{\cmark}{\ding{51}}%
\newcommand{\xmark}{\ding{55}}%

\newcommand{\df}{\mathrel{:=}}

\newcommand{\e}{\mathrm{e}}

\newcommand{\dd}{\mathrm{d}}

\begin{document}

\date{\today}

\title{\bf \LARGE The Weak Gravity Conjecture: A Review}

\institution{MIT}{\centerline{${}^{1}$Center for Theoretical Physics, Massachusetts Institute of Technology, Cambridge, MA 02139 USA}}

\institution{UMASS}{\centerline{${}^{2}$Department of Physics, University of Massachusetts Amherst, MA 01003 USA}}

\institution{HARVARD}{\centerline{${}^{3}$Department of Physics, Harvard University, Cambridge, MA 02138 USA}}

\institution{BERKELEY}{\centerline{${}^{4}$Department of Physics, University of California, Berkeley, CA 94720 USA}}

\authors{Daniel Harlow\worksat{\MIT}\footnote{e-mail: {\tt harlow@mit.edu}}, Ben Heidenreich\worksat{\UMASS}\footnote{e-mail: {\tt bheidenreich@umass.edu}}, Matthew Reece\worksat{\HARVARD}\footnote{e-mail: {\tt mreece@g.harvard.edu}}, and Tom Rudelius\worksat{\BERKELEY}\footnote{e-mail: {\tt rudelius@berkeley.edu}}}

\abstract{The Weak Gravity Conjecture holds that in a theory of quantum gravity, any gauge force must mediate interactions stronger than gravity for some particles. This statement has surprisingly deep and extensive connections to many different areas of physics and mathematics. Several variations on the basic conjecture have been proposed, including statements that are much stronger but are nonetheless satisfied by all known consistent quantum gravity theories. We review these related conjectures and the evidence for their validity in the string theory landscape. We also review a variety of arguments for these conjectures, which tend to fall into two categories: qualitative arguments which claim the conjecture is plausible based on general principles, and quantitative arguments for various special cases or analogues of the conjecture.  We also outline the implications of these conjectures for particle physics, cosmology, general relativity, and mathematics. Finally, we highlight important directions for future research.\\

Submitted to \emph{Reviews of Modern Physics}.}

\preprint{ACFI-T22-01}

\maketitle

\tableofcontents

\enlargethispage{\baselineskip}

\newpage

\section{Introduction}\label{INTRO}

%!TEX root = WGC_Review_arxiv_v2.tex

The Weak Gravity Conjecture is a remarkably simple statement about theories of quantum gravity.  In essence, it says that any gauge force must be stronger than gravity. More precisely, in its mildest form, the Weak Gravity Conjecture holds that any $U(1)$ gauge theory must have at least one object satisfying 
 \begin{equation}
            \frac{|q|}{m} \geq   \left.  \frac{|Q|}{M} \right|_{\mathrm{ext}},
            \label{eq:WGCintro}
 \end{equation}
where $\left. \frac{|Q|}{M} \right|_{\mathrm{ext}}$ is the charge-to-mass ratio of a large extremal black hole.
This simple statement has profound consequences, which touch virtually every aspect of modern fundamental physics, including string theory, cosmology, particle physics, algebraic geometry, black holes, quantum information, holography, scattering amplitudes, and more.

The original paper on the Weak Gravity Conjecture (WGC) from Arkani-Hamed, Motl, Nicolis, and Vafa (AMNV, henceforth) \cite{ArkaniHamed:2006dz} is, by now, more than 15 years old. It sparked a flurry of research shortly after it was released, which slowly tapered off over the course of the next several years. The middle of the 2010's, however, saw a resurgence of interest in the conjecture, which has continued to the present day.

This resurgence of interest was driven in part by the hope that quantum gravity may have something to say about testable low-energy physics, despite the fact that quantum gravitational effects are naively suppressed by powers of energy divided by the Planck mass.  Originally it was hoped that this problem could be circumvented by using string theory to predict low-energy parameters such as Yukawa couplings or the scale of supersymmetry breaking, but the gradual acceptance that string theory has a vast Landscape of four-dimensional vacua has posed a major challenge to this idea: the more possibilities one has, the harder it is to make a unique prediction.

Nonetheless, there may be some simple rules which conclusively exclude particular low-energy actions. The WGC is one such rule, and as we will see below it potentially constrains certain models of particle physics and cosmology and thus offers hope that quantum gravity may yet make decisive predictions for IR physics in the near future.\footnote{The set of low-energy actions which cannot be realized in quantum gravity has been called the ``Swampland''~\cite{Vafa:2005ui}, and many more rules for ruling out such actions have been proposed. Some of these proposals are closely related to the WGC, while others are not.  In this review we focus on the WGC specifically, so our discussion of other parts of the Swampland program will be subjective and incomplete.  Readers interested in a broader discussion might consult, e.g.,~\cite{Brennan:2017rbf,Palti:2019pca,vanBeest:2021lhn,Grana:2021zvf}.}

The WGC has many interesting theoretical implications. In the context of AdS/CFT, it implies nontrivial statements for conformal field theories. In the context of string compactifications, it implies nontrivial statements about Calabi-Yau geometry. In the context of black hole physics, it is intimately related to the preservation of cosmic censorship. These connections, and others that we will review below, suggest that the WGC is pointing us towards deep, fundamental principles of quantum gravity. 

However, despite the recent progress, we are still far from a concrete understanding of such principles, and some of the most basic questions about the WGC remain unanswered. 

First and foremost, we emphasize that the WGC is not really a single, universally-agreed-upon conjecture, but rather a family of distinct but related ``weak gravity conjectures,'' each of which attempts to formalize the idea that ``any gauge force must be stronger than gravity'' in a different way.  These various conjectures have different consequences for particle physics, cosmology, and much more.  Some versions of the WGC have been discarded as counterexamples have been identified, while other versions have seen a growing body of evidence in their favor. Some of the most promising versions of the conjecture are known as the ``tower Weak Gravity Conjecture'' and the ``sublattice Weak Gravity Conjecture,'' and we will elaborate on them shortly.

Moreover, so far no nontrivial version of the conjecture has actually been proven in the sense of being derived from some accepted general principle.  A number of promising routes towards a proof of some version of the WGC have been proposed in recent years, but these routes all suffer from at least one of two drawbacks: either they establish some statement which is qualitatively like the WGC, but without the correct $O(1)$ factors included (i.e., ``no gauge force can be much weaker than gravity''), or they argue for a precise version of the WGC, but rely on additional, unproven assumptions. In particular in the original paper, AMNV motivated the WGC using black hole physics: the requirement that any non-supersymmetric black hole should be able to decay necessitates some version of the WGC. It is not clear however why any non-supersymmetric black hole must be able to decay, and it is also not clear that black hole decay is the fundamental principle underlying the WGC as opposed to an accidental consequence of it. In particular, there is strong evidence for some versions of the conjecture (e.g., the sublattice WGC) with sharp consequences going beyond the minimal requirements of black hole instability. A proof of some form of the WGC---even a mild one---would represent a significant development in our understanding of the conjecture.

Without a proof of the conjecture, or a deeper understanding of why the conjecture must be true, it is difficult to be sure which version(s) of the conjecture are correct, so it is difficult to determine how strong are the constraints imposed by the WGC on particle physics, cosmology, geometry, and more. This means that despite the immense progress in our understanding of the WGC in recent years, the most important discoveries may yet lie ahead. 

The remainder of this review is structured as follows. In Section \ref{NGS}, we review arguments for the absence of global symmetries in quantum gravity, which may be viewed as a sort of precursor to the WGC. In Section \ref{WGC}, we introduce the Weak Gravity Conjecture in its mild and stronger variants. In Section \ref{EVIDENCE}, we outline the evidence for different versions of the WGC, focusing on concrete examples in string theory and Kaluza-Klein theory. In Section \ref{QUALITATIVE}, we present qualitative arguments for approximate versions of the WGC, i.e., without precise $O(1)$ factors included. In Section \ref{DERIVATIONS}, we review the attempted derivations of the WGC, briefly explaining why (in our opinion) each of them falls short of a ``proof'' of the WGC. In Section \ref{IMPLICATIONS}, we discuss broader implications of the WGC for phenomenology, mathematics, and other areas of theoretical physics. In Section \ref{OUTLOOK}, we end with conclusions and outlook. In appendix~\ref{app:extremalitygeneral} we describe a general procedure for determing the black hole extremality bound (needed to correctly normalize the WGC bound) in theories with moduli.

\section{No Global Symmetries}\label{NGS}

%!TEX root = WGC_Review_arxiv_v2.tex

The WGC has its origins in an older conjecture, which says that theories of quantum gravity admit no global symmetries of any kind. One motivation for this conjecture is the following.  An evaporating black hole emits all particles in a theory, without regard to their global charges~\cite{Hawking:1974sw}. This differs from gauge charge, where (at least for continuous gauge group) the electric field outside of a charged black hole provides a chemical potential that favors discharge during evaporation. This insensitivity of black hole evaporation to global charges suggests that black holes can violate global symmetries and destroy global charge~\cite{Zeldovich:1976vq, Zeldovich:1977be}.

\begin{figure}
\centering
\includegraphics[width=100mm]{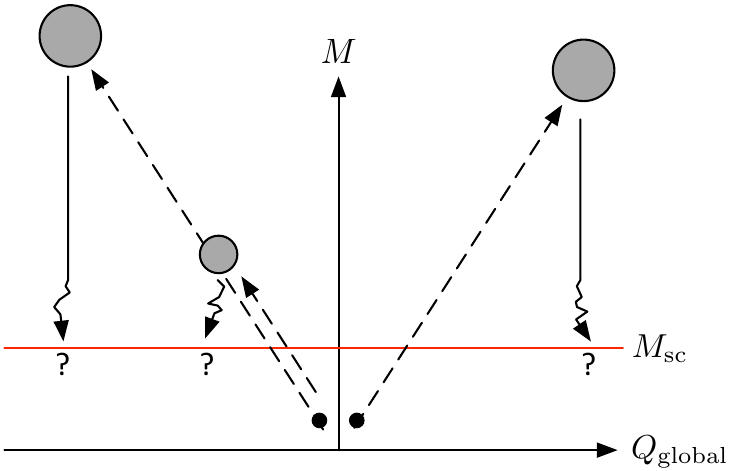}
\caption{Gravitational collapse of global-charged objects creates black holes of arbitrarily large global charge. If subsequently left alone, effective field theory dictates that the resulting black holes decay to objects of size $r\sim r_{\rm sc}$ and corresponding mass $M \sim M_{\rm sc}$ via Hawking radiation without appreciably changing the expected value of their global charge. This implies an infinite number of microstates for black holes of any fixed mass $M \gg M_{\rm sc}$, in violation of the Bekenstein-Hawking entropy formula. (Whether this process eventually results in stable remnants is immaterial.)}
\label{fig:Remnants}
\end{figure}
A more precise argument~\cite{Banks:2010zn} is that a continuous global symmetry would violate the Bekenstein-Hawking formula for black hole entropy. For example, suppose we had a quantum gravity theory with a $U(1)$ global symmetry.  By colliding objects which are charged under this symmetry, one could produce large black holes of arbitrarily large global charge $Q$. The semiclassical calculation of Hawking evaporation implies that these black holes will decay, at least until they reach a radius $r_\mathrm{sc}\gg \ell_{\mathrm{Pl}}$ below which the effective field theory description is invalid. A black hole of initial charge $Q$ will have a final charge $Q' \sim Q$: the Hawking evaporation process may emit charged particles, but it does not preferentially discharge the black hole.  Thus we can prepare black holes of size $r_\mathrm{sc}$ but arbitrarily large charge.  The information which is stored in this charge is arbitrarily large, and in particular exceeds the Bekenstein-Hawking entropy $\frac{\pi r_\mathrm{sc}^2}{G}$.  This argument---illustrated in Figure~\ref{fig:Remnants}---extends directly to any continuous global symmetry, and implies a bound on the size of a finite global symmetry group, albeit one that is exponentially weak in $r_\mathrm{sc}/\ell_\mathrm{Pl}$, which can be a large number in a weakly coupled theory~\cite{Banks:2010zn}.

\begin{figure}
    \centering
    \includegraphics[height=6cm]{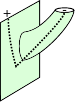}
    \caption{Global symmetry violation by a Euclidean wormhole: a pair of charged particles is created from the vacuum, with the positive charge staying in the asymptotically-flat region but the negative charge ending up in a baby universe.  For someone living in the asymptotic region, this apparently violates the symmetry.  Such a process can't happen for a gauge symmetry, since the baby universe is closed and compact so its gauge charge must be zero.}
    \label{fig:eucwormhole}
\end{figure}
 Another, somewhat more vague, argument for global symmetry violation in quantum gravity is that if certain ``Euclidean wormholes'' are included in the gravitational path integral then apparent global symmetry violation is a consequence \cite{Giddings:1988cx,Abbott:1989jw,Coleman:1989zu,Kallosh:1995hi} (see also \cite{Hawking:1995ag} for an alternative view).  The basic idea is that if there is a finite amplitude for adding a closed connected spatial component to the universe, usually called a ``baby universe,'' then global charge can end up in such a baby universe and therefore charge conservation can appear to be violated in the part of the universe we can actually access (see figure \ref{fig:eucwormhole}).  This statement does not apply to gauge charge, as the gauge charge of a closed universe must be zero.\footnote{This argument for the violation of global symmetries is quite similar to the semiclassical argument that black holes destroy quantum information, so it may seem surprising that the modern consensus is that global symmetries are indeed violated but information is not lost.  The difference is that the global charge of Hawking radiation is a ``simple'' observable, which is the kind the low-energy effective field theory needs to get right, while any extraction of information about the initial state of a black hole requires ``complex'' observables with the capability to invalidate the semiclassical picture.  See \cite{Harlow:2020bee} for more on why global symmetries are not allowed in theories where black hole evaporation is unitary.}

Such general arguments about black hole physics or Euclidean gravity have been supplemented by observations about concrete theories of quantum gravity. In perturbative string theory, given a putative continuous global symmetry, one can create a vertex operator on the worldsheet that creates a gauge field in spacetime coupling to the symmetry current, demonstrating that the would-be global symmetry is, in fact, gauged~\cite{Banks:1988yz}. Similarly, in AdS/CFT, a conserved current for a continuous global symmetry of the CFT implies the existence of a corresponding gauge symmetry in the bulk quantum gravity theory~\cite{Witten:1998qj}. 

\begin{figure}
    \centering
    \includegraphics[height=6cm]{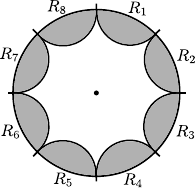}
    \caption{An AdS/CFT contradiction between global symmetry and entanglement wedge reconstruction: the symmetry operators are products of operators supported in the regions $R_1, R_2,\ldots$, but no such operator can implement the symmetry on a charged operator in the center of the space.}
    \label{fig:noGSAdS}
\end{figure}
In the context of AdS/CFT, a holographic argument against global symmetries---both continuous and discrete---was presented in \cite{Harlow:2018jwu, Harlow:2018tng}. Here, a symmetry generator $U_g$ associated with a group element $G$ acting on the boundary $R$ is split into a product,
\begin{equation}
U_g(R) = \prod_i U_g(R_i) U_{\text{edge}}\,,
\label{Usplit}
\end{equation}
where $R = \cup_i R_i$, each $U_g(R_i)$ acts only in the region $R_i$, and $U_{\text{edge}}$ acts at the boundaries of the $R_i$. A charged operator localized in the center of the bulk should transform under $U_g(R)$, but since the entanglement wedge of each $R_i$ will not contain the center of the bulk for $R_i$ sufficiently small, the charged operator cannot transform under the right-hand side of \eqref{Usplit}: a contradiction  (see figure \ref{fig:noGSAdS}). We conclude that such a global symmetry cannot exist under the assumption that entanglement wedge reconstruction holds valid. This argument applies also to the higher-form global symmetries of \cite{Gaiotto:2014kfa}, under which the charged objects are strings or branes instead of particles.

The use of AdS/CFT in \cite{Harlow:2018jwu, Harlow:2018tng} is obviously rather restrictive, but more recently it was observed in \cite{Harlow:2020bee} that essentially the same argument can be used to exclude global symmetries in any theory of quantum gravity where entanglement wedge reconstruction can be applied to an auxiliary reservoir coupled to an evaporating black hole. This assumption is the essential feature of recent calculations of the ``Page curve'' for an evaporating black hole, and thus is closely related to the unitarity of black hole evaporation \cite{Penington:2019npb,Almheiri:2019psf}.  Moreover it was observed, following \cite{Lewkowycz:2013nqa}, that semiclassically this calculation can be interpreted as arising from the appearance of certain Euclidean wormholes in the gravitational path integral \cite{Almheiri:2019qdq,Penington:2019kki}.  Finally in  \cite{Chen:2020ojn,Hsin:2020mfa} it was shown that these Euclidean wormholes can indeed lead to concrete violations of global symmetry, thereby quantifying global symmetry violation in evaporating black hole backgrounds.

Finally, let us remark that the absence of global symmetries in quantum gravity is closely related to another Swampland conjecture, the \emph{Completeness Hypothesis} \cite{polchinski:2003bq}. This hypothesis holds that in any gauge theory coupled to gravity, there must exist charged matter in every representation of the gauge group. The existence of such states is supported by black hole arguments \cite{Banks:2010zn} and holographic arguments in the context of AdS/CFT \cite{Harlow:2015lma,Harlow:2018jwu}. In $G$ gauge theory, if $G$ is compact and connected, or finite and abelian, then the presence of charged matter in every representation is equivalent to the absence of a 1-form symmetry under which Wilson lines are charged. If $G$ is compact but disconnected, or finite and nonabelian, then the presence of charged matter in every representation is equivalent to the absence of ``non-invertible'' global symmetries, which are associated with certain codimension-2 topological operators in the gauge theory \cite{Rudelius:2020orz, Heidenreich:2021tna}. This close connection between the absence of global symmetries and the Completeness Hypothesis means that arguments for one conjecture serve as (indirect) evidence for the other.  An interesting quantitative approach to completeness based on algebraic ideas has been developed in \cite{Casini:2019kex,Casini:2020rgj,Casini:2021zgr}, where relative entropy and conditional expectation are used to diagnose to what extent field theories obey the completeness hypothesis.  

The strongest arguments against the existence of global symmetries in quantum gravity are arguments against {\em exact} global symmetries. For applications, it is important to refine these arguments to ask to what extent {\em approximate} global symmetries are allowed. Recent general arguments along these lines include~\cite{Nomura:2019qps, Fichet:2019ugl, Daus:2020vtf}. As we will see, the Weak Gravity Conjecture is one attempt to address this question: the weak coupling limit of a gauge theory has a global symmetry, and should be forbidden in quantum gravity. As we will discuss in \S\ref{subsec:noapprox} below, the Weak Gravity Conjecture is also related to the breaking of approximate 1-form global symmetries associated with the absence of charged particles. 

\section{Weak Gravity Conjectures}\label{WGC}

%!TEX root = WGC_Review_arxiv_v2.tex

We now consider quantum gravity theories coupled to a $U(1)$ gauge field, in $D>3$ spacetime dimensions, with low-energy actions of the form 
\be\label{eq:action}
S=\int \rmd^D x \sqrt{-g}\left(\frac{R}{2\kappa^2}-\frac{1}{4 e^2(\phi)} F_{\mu\nu}F^{\mu\nu}+\ldots\right).
\ee
Here $e^2(\phi)$ is some function of the scalar fields $\phi^i$ in the theory and
the omitted terms include kinetic terms for these scalars, as well as other possible terms involving additional matter fields and/or higher-derivative terms for the gauge field and the metric.\footnote{In $D=4$ if massless charged particles exist then several aspects of this discussion need to be modified, due to the logarithmic running which eventually drives the renormalized gauge coupling $e$ to vanish in the deep infrared.  The mild WGC still holds in such theories, since after all there are massless charged particles, but to simplify our exposition we will assume that in $D=4$ all charged particles are massive.}  The compactness of the gauge group requires charge to be quantized, and we normalize the gauge field so that the covariant derivative on a field of unit charge is $\partial_\mu-iA_\mu$.  We then define electric charge by
\be
Q=\int_{S^{D-2}_\infty}\frac{1}{e^2(\phi)} \star F,
\ee
where $S^{D-2}_\infty$ is a sphere at spatial infinity, in which case charge is quantized in integer units (i.e., the canonically-normalized electrostatic potential is proportional to $eQ$).

The mildest version of the weak gravity conjecture then says the following:  

           \begin{namedconjecture}[Mild Weak Gravity Conjecture]
 Given any $U(1)$ gauge field coupled to gravity as in \eqref{eq:action}, there must exist an object of charge $q$ and mass $m$ satisfying
            \begin{equation}
            \frac{|q|}{m} \geq   \left.  \frac{|Q|}{M} \right|_{\mathrm{ext}}.
            \label{eq:WGC}
            \end{equation}
            \end{namedconjecture}
    \vspace{.1cm}
\noindent
Here $\frac{|Q|}{M}\big|_{\mathrm{ext}}$ indicates the charge to mass ratio of an extremal black hole of arbitrarily large size (in general there are finite-size corrections to this ratio which are not included in \eqref{eq:WGC}).   We will refer to any object obeying \eqref{eq:WGC} as \textit{superextremal}.  It is convenient to parameterize the extremal charge-to-mass ratio as
\be
\left. \frac{e|Q|}{M}\right|_{\mathrm{ext}}\equiv \gamma^{\frac{1}{2}}\kappa,\label{gammadef}
\ee
where $\kappa>0$ is the gravitational coupling constant appearing in the action \eqref{eq:action}, related to the Planck mass $\Mp$ and the Newton constant $G$ by
\be
\kappa^2=8\pi G=\frac{1}{\Mp^{D-2}},
\ee
and $e^2 = e^2(\langle \phi \rangle)$ denotes the gauge coupling in the vacuum when written without an argument.
$\gamma$ is a dimensionless parameter which in general depends on the function $e^2(\phi)$ and on the metric on moduli space (see Appendix~\ref{app:extremalitygeneral}).  If $e^2(\phi)$ is independent of the moduli, then we simply have
\be\label{gamma0}
\gamma=\frac{D-3}{D-2}.
\ee
Above, as throughout this review, we have of course set $\hbar = c = 1$, but we emphasize that even if we restore them there are no factors of $\hbar$  in \eqref{gammadef} since the extremality bound is a classical notion.

The original motivation for the conjecture is that it provides a kinematic condition that would allow an extremal black hole to shed its charge, which can happen even at zero Hawking temperature via Schwinger pair production~\cite{Gibbons:1975kk,Johnson:2019kda}. However, there is no obvious pathology in a theory that admits infinitely many stable extremal black holes; due to the extremality bound, this would not lead to infinite entropy at finite mass as in the global-charge case in Fig.~\ref{fig:Remnants}. Hence, this motivation falls far short of a proof or even a strong argument

Although the mild Weak Gravity Conjecture has an appealing simplicity, in practice it is too weak to imply anything interesting.  The object which obeys \eqref{eq:WGC} could be very heavy, in which case it would have no substantive consequences for particle physics or cosmology. Moreover it would not even be sufficient to allow ``medium-sized'' near-extremal black holes to decay, and thus would not address the original motivation for the conjecture.  The mild Weak Gravity Conjecture is nonetheless useful to consider, as it is a consequence of all of the various stronger versions of the WGC which have been proposed, which do have other more interesting implications, and so an argument which shows that the mild WGC holds would hopefully also lead to an argument for one or more of the stronger versions.  We now turn to discussing these possible generalizations.  

\subsection{WGC for P-form gauge fields}\label{PFORM}

The mild WGC can be generalized in an obvious way from particles charged under an ordinary 1-form gauge field to $(P-1)$-branes charged under a $P$-form gauge field, with the restrictions $1 \leq P \leq D-3$.  Instead of bounding the charge-to-mass ratio $|q|/m$ of such a particle, the WGC instead bounds the charge-to-tension ratio of the $(P-1)$-brane:

\vspace{.2cm}
           \begin{namedconjecture}[Mild WGC for $P$-form gauge fields]
 Given a $P$-form gauge field coupled to gravity, there must exist a $(P-1)$-brane of charge $Q$ and tension $T_P$, satisfying
            \begin{equation}
            \frac{|Q|}{T_P} \geq   \left.  \frac{|Q|}{T_P} \right|_{\mathrm{ext}}\,. \label{WGCpform}
            \end{equation}
            \end{namedconjecture}
    \vspace{.1cm}
\noindent
Here $\left.  \frac{|Q|}{T_P} \right|_{\mathrm{ext}}$ is the charge to tension of an extremal black brane. It is useful to consider a concrete low-energy theory, with action  
\begin{align}
S = \int \rmd^D x \sqrt{-g}\left(\frac{R}{2\kappa^2}- \frac{1}{4\kappa^2} (\nabla \phi)^2 - \frac{1}{2e_{P}^2} \e^{-\alpha_{P} \phi} F_{P+1}^2\right). \label{eq:generalaction}
\end{align}
Here $F_{P+1} = \rmd A_P$ is the field strength for a $P$-form gauge field $A_{\mu_1 \ldots \mu_P}$, with
\begin{align}
F_q^2 := \frac{1}{q!} F_{\mu_1 \ldots \mu_q} F^{\mu_1 \ldots \mu_q} \,,
\end{align}
and by convention we shift $\phi$ to set $\langle \phi \rangle=0$, so that $e_P$ is indeed the gauge coupling in the vacuum.  In this theory we can write the extremal charge-to-tension ratio as 
\be
\left.  \frac{e_P |Q|}{T_P} \right|_{\mathrm{ext}}=\gamma_P^{\frac{1}{2}} \kappa, \label{eqn:PBraneBound}
\ee
with
\begin{align}
\gamma_{P} =   \frac{\alpha_{P}^2}{2} + \frac{P(D-P-2)}{D-2}\,.
\label{gammaPD}
\end{align}
If we replace $e_P^2 \mathrm{e}^{\alpha_P \phi}$ by some more general function $e_P^2(\phi)$ then $\gamma_P$ is modified as appropriate (see Appendix~\ref{app:extremalitygeneral}).  For future reference we write in one place the superextremality bound:
\be\label{eq:extremalitybound}
e_P^2 Q^2 \geq \gamma_P\kappa^2 T_P^2.
\ee

\subsection{Magnetic WGC}\label{ssec:magnetic}

The magnetic version of the mild WGC is nothing but the ordinary mild WGC, applied to the electromagnetic dual gauge field. For the case of a $P$-form gauge field, this implies the existence of a superextremal magnetically charged $(D-P-3)$-brane, with magnetic charge $|\tilde Q|$ and tension $T_{D-P-2}$, satisfying
\begin{equation}
  \frac{|\tilde Q|}{T_{D-P-2}} \geq   \left.  \frac{|\tilde Q|}{T_{D-P-2}} \right|_{\textrm{ext}}\,.
  \label{magWGC}
\end{equation}
In four dimensions, for $p=1$, this becomes a statement about the charge-to-mass ratio of a magnetic monopole. The monopole mass can be estimated in terms of the energy stored in its magnetic field. This energy is UV-divergent, but if we cut it off at the semiclassical radius $r_{\text{sc}} \sim 1/\Lambda_{\rm NP}$ associated to the ``new physics'' scale $\Lambda_{\rm NP}$ at which the low-energy EFT breaks down, then we obtain
\begin{equation}
m_{\textrm{mon}} \gtrsim \frac{ \Lambda_{\rm NP} }{ e^2 } \,,
\end{equation}
in the absence of a finely-tuned cancellation between the field energy and the bare mass,
where $e$ is the electric gauge coupling.\footnote{This logic is not valid for electrically charged particles, because the self-energy should be cut off at the Compton radius, which is much larger than $\Lambda_\mathrm{NP}^{-1}$. Stated another way, the classical radius of an electric charge is less than its Compton wavelength, whereas the reverse is usually true for a magnetic charge, unless it is exceptionally light due to a finely-tuned cancellation between bare mass and field energy.}
By Dirac quantization, the magnetic gauge coupling is given by $\tilde e = 2\pi/e$, so the magnetic WGC bound~\eqref{magWGC} becomes
\begin{equation}
\Lambda_{\rm NP} \lesssim e M_{\textrm{Pl}}\,.
 \label{magWGCbound}
\end{equation}
In other words, the magnetic WGC places a cutoff on the new physics scale of the abelian gauge theory, which vanishes (in Planck units) in the limit $e \rightarrow 0$. The magnetic WGC thus quantifies the extent to which effective field theory breaks down in the limit of weak gauge coupling. Without imposing the WGC itself, the conclusion~\eqref{magWGCbound} can also be obtained by requiring that the magnetic monopole is not a black hole, i.e., that its Schwarzschild radius is smaller than $r_\mathrm{sc}$~\cite{ArkaniHamed:2007gg, delaFuente:2014aca}.

We emphasize that the new physics scale $\Lambda_{\rm NP} \sim 1/r_{\textrm{sc}}$ is not a cutoff on effective field theory altogether. The abelian gauge theory may be embedded into another effective field theory with a higher cutoff, such as a Kaluza-Klein theory, a nonabelian gauge theory, etc..
 In Section \ref{ssec:Strong}, we will introduce several strong forms of the WGC, and in Section \ref{QUALITATIVE} we will see that some of these strong forms provide a bound not only on $\Lambda_{\rm NP} \sim 1/r_{\textrm{sc}}$ but also on the energy scale $\LQG$ at which gravity becomes strongly coupled. This latter energy scale represents a cutoff on low energy effective field theory in any form, above which quantum gravity effects cannot be neglected.

Finally, let us note that a similar argument can be applied to $(D-P-3)$-branes magnetically charged under a $P$-form gauge field in $D$ dimensions \cite{Hebecker:2017wsu}. The tension of such an object can be approximated as
\begin{equation}
T_{D-P-3} \sim \frac{\Lambda^P}{e_{P}^2}\,,
\end{equation}
where $r_{\text{sc}}=\Lambda^{-1}$ is again the semiclassical radius of the brane, and $e_{P}$ is the electric coupling constant. On the other hand, the tension of a black brane is given by
\begin{equation}
T_{\text{BB}} \sim M_{\mathrm{Pl}}^{D-2} R_S^P\,,
\end{equation}
where $R_S$ is the Schwarzschild radius of the black brane. If we then demand that the magnetic brane is not itself a black hole, so that $\Lambda^{-1} = r_{\text{sc}} \gtrsim R_S$, we then have
\begin{equation}   \label{eq:magWGCboundgeneral}
\Lambda \lesssim (e_{P}^2 M_{\mathrm{Pl}}^{D-2})^{\frac{1}{2 P}}\,.
\end{equation}
This reduces to \eqref{magWGCbound} in the familiar case $D=4$, $P=1$.

\subsection{The convex hull condition}\label{ssec:CHC}

So far, we have focused on theories with a single gauge field. In general, however, a quantum gravity theory will have more than one gauge field, so the statement of the WGC must be generalized to this case. For simplicity, we focus on the case of particles charged under $1$-form gauge fields, though analogous statements hold for branes charged under higher-form gauge fields.

In a theory of $N$ abelian gauge fields, the charge of a given particle may be represented by an $N$-vector $\vec{Q}$, where $Q_i$ is the charge under the $i$th gauge field. The set of all possible charges $\vec{Q}$ consistent with charge quantization forms a lattice $\Gamma \simeq \mathbb{Z}^N \subset \mathbb{R}^N$. We define a ``charge direction'' $\hat{Q}$ as a unit vector in $\mathbb{R}^N$, and we say that such a charge direction is ``rational'' if $\lambda \hat{Q}\in \Gamma$ for some $\lambda \in \mathbb{R}$. 

Finally, we define a ``multiparticle state'' as consisting of one or more actual particles in the theory with ``mass'' $m$ and ``charge'' $\vec{q}$ equal to the sums of the masses and charges of the constituent particles. 
This corresponds to a limit where the particles in question are taken infinitely far from each other, so that they do not interact. A multiparticle state is superextremal if
$\vec{z} \df \vec{q}/m$ has a length which is greater than or equal to the charge-to-mass ratio of an extremal black hole in the $\hat{Q}$ charge direction. The length of this vector is measured with the inverse of the kinetic matrix of the $U(1)$ gauge fields, i.e., given a Lagrangian $-\frac{1}{4} K_{ij} F^i_{\mu \nu}F^{j\mu \nu}$, the length of $\vec{z}$ is $\left(K^{ij}z_i z_j\right)^{1/2}$, where $K^{ij}K_{jk} = \delta^i_{~k}$.

\begin{figure}
\centering
\includegraphics[width=75mm]{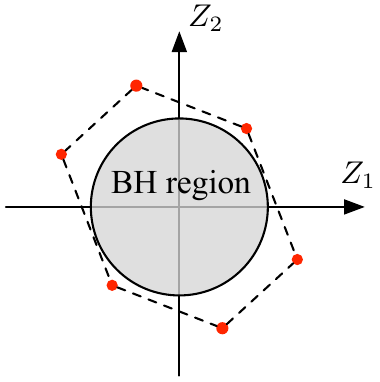}
\caption{The Convex Hull Condition. In theories with multiple $U(1)$s, the WGC is equivalent to the statement that the convex hull of the charge-to-mass vectors of the various particle species must contain the black hole region.}
\label{fig:CHC}
\end{figure}

With this, we may define a mild WGC in such a theory as follows:
\vspace{.2cm}
\begin{namedconjecture}[Mild WGC for multiple gauge fields]
For every rational direction $\hat Q$ in charge space, there is a superextremal multiparticle state with $\vec{z}\propto \hat{Q}$.
\end{namedconjecture}\label{eq:multipleWGC}
    \vspace{.1cm}
\noindent
When there are a finite number of stable particles in the theory, this statement admits an equivalent, geometric formulation known as the \emph{convex hull condition} (CHC) \cite{Cheung:2014vva}. The CHC considers the set of all charge-to-mass vectors $\vec{z}_i := \vec{q}_i/m_i$ for the particles in the theory, and it holds that the convex hull of this set should contain the region in $\vec{Z}$-space where black holes live. This condition is depicted graphically in Figure~\ref{fig:CHC}. Note that in the absence of massless scalar fields, the black hole region is simply the interior of an ellipsoid, $K^{ij}z_i z_j \leq \gamma \kappa^2$.  If massless scalar fields are added to the theory, the black hole region will generically grow in size, and it may change its shape as well. Thus, the CHC gives stronger bounds in theories with massless scalar fields than those without.

\subsection{Strong forms of the WGC}\label{ssec:Strong}

So far all versions of the WGC which we have discussed are still ``mild'' in the sense of not having particularly interesting implications.  From the very first paper on the WGC, however, there has been interest in stronger versions of the WGC. This interest is not just wishful thinking: as we will see in Section \ref{EVIDENCE}, all known examples in string theory seem to satisfy stronger statements than the mild WGC.  Moreover the heuristic arguments we will review in Section \ref{QUALITATIVE} also give support to the idea that something stronger than the mild WGC is true.  

A first strong form to mention, which is at times implicit in AMNV, is the statement that the WGC should be satisfied by superextremal particles \emph{which are not themselves black holes}. Higher-dimension operators in the action can modify the extremality bound of finite-sized black holes, as we will discuss further in Section \ref{DERIVATIONS}. If the charge-to-mass ratio of these finite-sized extremal black holes decreases as their mass is taken to infinity, the mild form of the WGC can be satisfied by stable, finite-sized black hole states. This scenario satisfies the letter of the WGC law, but not the spirit of it, which holds that all black holes should be able to decay by emitting charged particles. This points to a first strong form of the WGC: the particles satisfying the WGC bound should not be black holes.

AMNV suggested two additional possible strong forms of the WGC:
The first held that the lightest charged particle should be superextremal. The second held that the particle of smallest charge should be superextremal.
Neither of these statements hold in general, however: they are violated, for instance, in certain $T^n$ orbifold compactifications of type II and heterotic string theory~\cite{Heidenreich:2016aqi}.

However, a growing body of evidence points to another pair of strong forms~\cite{Heidenreich:2016aqi,Montero:2016tif,Andriolo:2018lvp,Heidenreich:2019zkl}:
\begin{namedconjecture}[Tower Weak Gravity Conjecture]
For every site in the charge lattice, $\vec{q} \in \Gamma$, there exists a positive integer $n$ such that there is a superextremal particle of charge $n \vec{q}$.
\end{namedconjecture}
\noindent
\begin{namedconjecture}[Sublattice Weak Gravity Conjecture]
There exists a positive integer $n$ such that for any site in the charge lattice, $\vec{q} \in \Gamma$, there is a superextremal particle of charge $n \vec{q}$.
\end{namedconjecture}
\noindent
A few remarks about these conjectures are in order. First, note that the tower WGC implies that in any charge direction $\hat{q}$, there must exist an infinite tower of superextremal particles. Indeed, the tower WGC is often defined by this latter statement. In the following section, however, we will see that consistency under dimensional reduction requires the formal definition we have given here.

Second, note that the sublattice WGC is strictly stronger than the tower WGC: the sublattice WGC implies that the integer $n$ appearing in the definition of the tower WGC can be chosen independently of $\vec{q}$. The sublattice WGC is equivalent to the statement that there is a (full-dimensional) sublattice of the charge lattice such that there is a superextremal particle at each site in the sublattice. The integer $n$ is sometimes referred to as the ``coarseness'' of the sublattice. If $n=1$, we say the theory satisfies the lattice WGC. However, the lattice WGC is false in general; we will exhibit a counterexample in Section~\ref{ssec:counterexample}.

\begin{figure}
\centering
\includegraphics[width=0.48\textwidth]{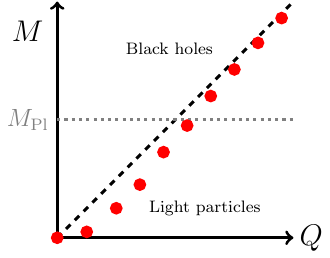}
\caption{Schematic illustration of WGC-satisfying particles (red dots) if the tower/sublattice WGCs hold. The black hole extremality bound is the dashed diagonal line. At small $Q$, the WGC is satisfied by light particles described by EFT. At large $Q$, black holes with small corrections obey the WGC; these asymptotically approach the extremality bound at large $Q$.}
\label{fig:tower}
\end{figure}

Third, note that the tower WGC and the sublattice WGC require an infinite set of superextremal particles in each rational charge direction, whereas the ordinary WGC may be satisfied in a given charge direction by multiparticle states. We will see in the following section that the existence of superextremal particles, rather than merely multiparticle states, is required for consistency under dimensional reduction. For small charges, the necessary particles are ordinary, quantum-mechanical particles, represented by fields in the effective field theory. Very far out on the charge lattice, the ``particles'' are actually black holes. The tower and sublattice WGCs thus interpolate between the effective quantum field theory regime and the gravitational regime of the quantum gravity theory in question. This is schematically illustrated in Figure~\ref{fig:tower}.

Fourth, note that it is possible (and, in fact, quite common in string theory examples) for the particles satisfying the tower/sublattice WGCs to be unstable resonances rather than stable states of the theory. Unstable resonances are not as easy to define as stable, single-particle states, since they do not correspond to states in the Hilbert space of the theory, but rather to localized peaks in the S-matrix of some scattering process. If the theory is weakly coupled, such a peak will be localized at a particular energy scale---the mass of the unstable particle---and the lifetime of this particle will be long. If the theory is strongly coupled, however, such a peak will be spread out across a range of energy scales, and it is not so easy to define the mass of the resonance. Correspondingly, the tower WGC and sublattice WGC are not so easy to define in this case.

Fifth and finally, note that the tower/sublattice WGCs are modified in the presence of a few very light charged particles in 4d due to the logarithmic running of the gauge coupling. Such charged particles appear near special loci in the moduli space where they become massless (e.g., where the Coulomb and Higgs branches of an $\mathcal{N}=2$ theory intersect). In $D\ge 5$, this has a mild effect---generating finite threshold corrections---but in 4d the log running reduces the infrared gauge coupling gradually to zero as the massless locus is approached. A naive reading of the tower/sublattice WGCs would then suggest that an infinite tower of charged particles becomes light near the massless locus, but this does not always occur, in particular when the massless locus lies at finite distance in the moduli space.\footnote{The absence of an infinite tower of light charged particles in such cases agrees with the Emergence Proposal \cite{Heidenreich:2017sim,Grimm:2018ohb,Heidenreich:2018kpg}.} While this seems to be a counterexample to the 4d tower/sublattice WGCs as originally stated, replacing the infrared gauge coupling in the WGC bound with its renormalized value resolves the problem \cite{Heidenreich:2017sim}, suggesting that the conjectures are subtly modified rather than being invalidated in 4d. By contrast, this problem is absent in $D\ge 5$ and no modification seems to be needed there (see, e.g., \cite{Alim:2021vhs}).\footnote{The difference between the 4d and higher-dimensional cases can also be explained by noting that the tower/sublattice WGCs are related to the mild WGC in one lower dimension (see \S\ref{REDUCTION}), whereas the mild WGC requires modification in 3d---if it continues to exist at all---due to the absence of asymptotically flat black holes.}

In closing, let us mention one other proposed ``strong form'' of the WGC: a superextremal state can saturate the WGC bound (i.e., be extremal) only if the theory is supersymmetric and the state in question is a BPS state~\cite{Ooguri:2016pdq}. This conjecture is a very mild extension of the ordinary WGC, since there is no good reason why the mass of a superextremal particle should be tuned precisely to extremality unless the state is a BPS state in a supersymmetric theory. Nonetheless, this extension is interesting, as it suggests that extremal black holes can be (marginally) stable only if they are BPS. When applied to the WGC for $p$-form gauge fields, the analogous statement further implies that any non-supersymmetric anti-de Sitter (AdS) vacuum supported by fluxes must be unstable.

\subsection{WGC for nonabelian gauge fields}\label{ssec:nonabelian}

Thus far, our definition of the WGC has dealt exclusively with particles charged under continuous, abelian gauge groups. We now want to discuss its extension to continuous, nonabelian gauge groups.  For $D=4$ this discussion is complicated by the fact that nonabelian gauge fields are often confined, in which case the notion of a charged particle is not well-defined, so this topic is of most interest for $D>4$. 

The mild form of the WGC extends in a rather trivial way: one simply decomposes the irreducible representations of the gauge group $G$ into charges under the $U(1)^{\textrm{rk}(G)}$ Cartan and demands that the ordinary WGC should be satisfied with respect to this Cartan subgroup. This requirement is automatically satisfied by the massless gluon fields of the theory. The sublattice WGC, on the other hand, is somewhat more subtle to define in the nonabelian context. We will use the following definition~\cite{Heidenreich:2017sim}:
\begin{namedconjecture}[Sublattice WGC for nonabelian gauge fields]
Given $G$ gauge theory (with $G$ a connected Lie group) coupled to quantum gravity, there is a finite-index Weyl-invariant sublattice $\Gamma_0$ of the weight lattice $\Gamma_G$ such that for every dominant weight $\vec{Q}_R \in \Gamma_0$, there is a superextremal resonance transforming in the $G$ irrep $R$ with highest weight $\vec{Q}_R$.
\end{namedconjecture}
\noindent
This statement is stronger than simply requiring that the abelian sublattice WGC should be satisfied with respect to the Cartan of $G$, as the latter can be satisfied by particles transforming under a sparse set of representations provided they are sufficiently light. One argument for this stronger statement is that it is satisfied in perturbative string theory; this follows from the modular invariance argument discussed in Section~\ref{MODULAR} below. This conjecture has also been shown to hold in certain 6d F-theory compactifications~\cite{Cota:2020zse}.

A natural question, now that we have defined the sublattice WGC for continuous abelian and nonabelian gauge groups, is whether there are further extensions for finite groups (or disconnected groups, more generally). Thought experiments involving the evaporation of black holes carrying charge under finite gauge groups suggest bounds on UV cutoffs that are similar in spirit to WGC bounds~\cite{Dvali:2007hz,Dvali:2007wp,Dvali:2008tq,Craig:2018yvw}. WGC bounds can also be applied separately to the $A$ and $B$ fields associated with a massive gauge field in BF-theory~\cite{Reece:2018zvv}, which can lead to conclusions consistent with black hole thought experiments in $\mathbb{Z}_N$ gauge theory~\cite{Craig:2018yvw}. These considerations may hint at the existence of a formulation of the WGC encompassing all gauge groups.

\subsection{WGC in asymptotically AdS spacetimes}\label{ssec:aAdS}

Thus far, we have focused on the WGC in flat (Minkowski) spacetimes. It is also worthwhile to define the conjecture in spacetimes with nontrivial curvature. Here, with an eye towards AdS/CFT, we restrict ourselves to possible definitions of the WGC in AdS spacetimes.

The flat space definition~\eqref{eq:WGC} depends on the mass $m$ of the particle, but in AdS$_{D}$ with AdS radius $R$ a more natural quantity is its rest energy $\frac{\Delta}{R}$ (in AdS/CFT $\Delta$ is the scaling dimension of the CFT operator which is dual to the field which creates the particle).  The relation between $m$ and $\Delta$ depends on the dimensionality of spacetime and the spin of the particle; for a scalar field in AdS$_{D}$ the relationship is
\begin{equation}
\Delta = \frac{D-1}{2} + \sqrt{\frac{(D-1)^2}{4} + R^2 m^2 }\,.
\end{equation}

A minimal requirement of any WGC bound in AdS$_{d+1}$ is that it reduces to the flat space bound in the limit where $R\to \infty$.  One obvious proposal which does this was noted by \cite{Nakayama:2015hga};
\begin{equation}
    e^2 q^2 \geq \gamma \kappa^2\frac{\Delta^2}{R^2}\,.
    \label{eq:WGCAdS}
\end{equation}
As in~\eqref{eq:WGC}, $\gamma =  \frac{D-3}{D-2}$ in the absence of massless scalar fields.  Using the AdS/CFT correspondence, this bound can be recast in terms of data of the CFT$_{D-1}$ as a bound on the charge $q$ and dimension $\Delta$ of the operator $\mathcal{O}$ dual to the charged field. In $D=5$, the CFT bound is \cite{Nakayama:2015hga}:
\begin{equation}
\frac{q^2}{b} \geq \frac{\Delta^2}{12 c}\,,
\end{equation}
where $c \sim \langle T T \rangle$ is the central charge of the CFT and $b \sim \langle J J \rangle$ is the beta function coefficient of the conserved current associated to the gauge field in the bulk.  On the other hand there is no particular reason why \eqref{eq:WGCAdS} is more likely than some other expression which has the same flat space limit, so the proper formulation of the WGC in AdS remains an open problem.

The Weak Gravity Conjecture in AdS/CFT is closely related to the recently formulated ``Abelian Convex Charge Conjecture''~\cite{Aharony:2021mpc}. Given a CFT with a $U(1)$ global symmetry, if we define $\Delta(n)$ to be the dimension of the lowest dimension operator of charge $n$, then this conjecture holds that
\begin{align}
    \Delta(n_1 q_0 +n_2 q_0) \geq \Delta(n_1 q_0) + \Delta(n_2 q_0)\,,
\end{align}
for $q_0 \geq 1$ an order-one integer. A similar statement is conjectured to hold for nonabelian gauge groups. Semiclassical tests of this statement were carried out in~\cite{Antipin:2021rsh}. If true, this conjecture implies that there must exist a particle in the AdS bulk theory with non-negative self-binding energy, which is very similar to the Repulsive Force Conjecture discussed below. Strong forms in which $q_0$ is 1 or is the charge of the lowest dimension charged operator were also briefly considered in~\cite{Aharony:2021mpc}, but such statements (as currently formulated) are in tension with a flat-space example, as we will discuss in~\ref{ssec:counterexample}.

\begin{figure}
\centering
\includegraphics[width=75mm]{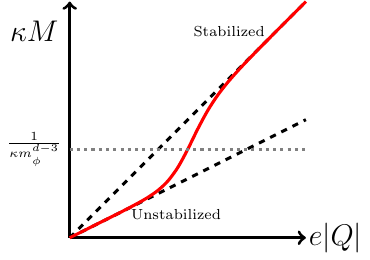}
\caption{Modulus stabilization and a non-convex spectrum of charged black holes. In a theory where a modulus $\phi$ is stabilized with mass $m_\phi$, the extremal black hole spectrum (red curve) should interpolate between small black holes that follow the ``unstabilized'' extremality bound (lower dashed black line) with slope $\gamma_\phi^{-1/2}$ and large black holes that follow the ``stabilized'' extremality bound (upper dashed black line) with larger slope $\gamma_0^{-1/2}$. The red curve indicates the smallest possible mass for a given charge. The detailed shape depends on the potential and couplings of $\phi$, but any spectrum that interpolates between the two linear regimes must fail to be convex for some intermediate values of $|Q|$.}
\label{fig:modulusconvexity}
\end{figure}

In comparing the Convex Charge Conjecture and various strong forms of the WGC, it is important to remember that not every CFT operator corresponds to a single-particle state in AdS. A convex spectrum of charged {\em single-trace} operators would have important implications for moduli stabilization. Consider a theory in which the gauge coupling $e(\phi)$ is a function of a stabilized modulus $\phi$ with mass $m_\phi$, and which has a separation of length scales $L \gg m_\phi^{-1} \gg r_\mathrm{sc}$, where $L$ is the curvature radius of an AdS (or dS) vacuum and $r_\mathrm{sc}$ is the size of the smallest black hole we can treat as semiclassical. In this case, there are black hole solutions that can be approximated as flat-space black holes with a massless modulus $\phi$ when the black hole radius $r$ obeys $m_\phi^{-1} \gg r \gg r_\mathrm{sc}$ and as flat-space black holes with no modulus when $L \gg r \gg m_\phi^{-1}$. Consequently, the black hole spectrum includes a range of extremal black holes that effectively have a modulus-dependent constant $\gamma_\phi$ in the extremality bound~\eqref{gammadef}, and another range with the modulus-independent value $\gamma_0$~\eqref{gamma0}. The modulus-dependent constant $\gamma_\phi$ is larger, as in~\eqref{gammaPD}, so that the WGC becomes weaker in the infrared than in the UV. As a result, the minimum mass as a function of charge for any black hole spectrum that interpolates between these limits must fail to be convex, as illustrated in Fig.~\ref{fig:modulusconvexity}. On the other hand, at large $|Q|$, one could consider states consisting of multiple small black holes instead of a single large black hole, which could then have a lower mass following the ``unstabilized'' line. From the CFT viewpoint, these would correspond to multi-trace, rather than single-trace, operators. A better understanding of the Convex Charge Conjecture in CFTs and its relationship to large-$N$ expansions, then, could potentially have important implications for the existence of vacua with stabilized moduli and scale separation.

\subsection{WGC for axions and axion strings}\label{ssec:axions}

In Section \ref{PFORM}, we extended the WGC to the case of a $P$-form gauge field. An especially interesting case to consider is $P=0$, in which the gauge field $A_0$ is a periodic scalar field ($A_0 \sim A_0 + 2 \pi$), also known as an axion.

This case is somewhat degenerate however, since the objects charged under this gauge field must be $(-1)$-branes, also known as instantons, with tension given by the instanton action $T_0 \equiv S_{\textrm{inst}}$.\footnote{A potential source of confusion here is that in general these instantons have nothing to do with the topologically-nontrivial gauge field configurations introduced in \cite{Belavin:1975fg}, but they happen to coincide for the particular case of the QCD axion in four dimensions.  More broadly however there can be axions without gauge fields and gauge fields without axions, and for $D\neq 4$ these two meanings of ``instanton'' do not even correspond to objects with the same dimensionality.  The instantons we discuss here are \textit{always} zero-dimensional dynamical objects in the Euclidean path integral with the property that their instanton number as defined by equation \eqref{eq:instanton} is nonzero.}  The instanton charge, also called the instanton number, is given (in Euclidean signature) by 
\be\label{eq:instanton}
n=i \int_{S^{d-1}} f^2 \star \rmd A_0,
\ee
where $f\equiv \frac{1}{e_0}$ is sometimes called the axion decay constant and $S^{d-1}$ is a small sphere surrounding the instanton.  In attempting to formulate an axion version of the WGC, however, we run into the problem that there is no immediately obvious notion of extremality. Indeed, naively plugging in $P=0$ to \eqref{gammaPD} (assuming the absence of massless scalar moduli), we see that $\gamma_{0}$ is zero, so the naive WGC bound \eqref{WGCpform} is trivial. Most likely, this does not indicate the absence of any sort of axion WGC bound, but rather that the $O(1)$ coefficient $\gamma_{0}$ must be fixed by some other means. In the absence of a clear notion of extremality, the axion WGC bound is typically written simply as follows: 
\begin{namedconjecture}[Axion WGC]
Given an axion (i.e., a periodic scalar) with axion decay constant $f$ coupled to quantum gravity, there must exist an instanton of instanton number $n$ satisfying 
\begin{equation} \label{eq:axionWGC}
\frac{ n }{ f }  \gtrsim S_{\mathrm{inst}} \kappa\,.
\end{equation}
\end{namedconjecture}
\noindent
Note, in particular, that the sharp bound in the $P$-form WGC \eqref{WGCpform} has been replaced by a $\gtrsim$, to account for the unknown $O(1)$ coefficient $\gamma_{0}$.

There have, however, been proposals for what this $O(1)$ coefficient should be. In the case of a 1-form, the WGC bound is the opposite of the black hole extremality bound, which sets the maximal charge-to-mass ratio of a macroscopic object in the low-energy theory (namely, a black hole). When it comes to instantons charged under an axion gauge field, there is once again a family of macroscopic solutions in the low-energy theory, known as gravitational instantons, which ostensibly can be used to fix $\gamma_{0}$ and define the extremality bound.

How exactly this should be done is not quite clear, however, and there are (at least) two proposals on the table. The confusion deals with the question of which class of gravitational instanton should be used to define the extremality bound, as there are three such classes:
\begin{enumerate}[1)]
\item Solutions with a singular core, also known as ``cored'' solutions.
\item Solutions with a flat metric (which we will refer to as ``extremal'' solutions).
\item Wormhole solutions, with two different asymptotic regions connected by a smooth throat.
\end{enumerate}
The metric for these solutions takes the form
\begin{equation}
\rmd s^2 = \left( 1 + \frac{ C }{ r^{2D-4} } \right)^{-1} \rmd r^2 + r^2 \rmd\Omega_{D-1}^2\,,
\end{equation}
where $\rmd \Omega_{D-1}^2$ is the metric on the unit $(D-1)$-sphere, and $C$ is positive, vanishing, and negative for cored, extremal, and wormhole solutions, respectively.

These solutions can all be obtained when we consider theories with a massless, dilatonic modulus. Starting from the action \eqref{eq:generalaction} for $P=0$, the action of the extremal instanton solution is given by
\begin{equation}
S_{\textrm{ext}} = \frac{\sqrt{2} |n|}{\alpha f \kappa }\,,
\label{eqn:Sext}
\end{equation}
where $n$ is the instanton number. Meanwhile, the lower bound on the action of a cored solution is given by \cite{Bergshoeff:2004fq, Bergshoeff:2004pg}:
\begin{align}
\label{eqn:Sextremal}
S_{\rm min} = \frac{\sqrt{2} |n|}{f \kappa} \times \begin{cases} \frac{1}{\alpha} & \alpha \ge \widetilde{\alpha} \\ \frac{1}{\widetilde{\alpha}} \sqrt{\frac{2 \widetilde{\alpha}}{\alpha} - 1} & \alpha < \widetilde{\alpha} \end{cases} \,,
\end{align}
where 
\begin{equation}
\widetilde{\alpha} := \sqrt{ \frac{2(D-2)}{D-1}  }\,.
\end{equation}
Finally, the instanton action for half of a wormhole solution is given by \cite{Gutperle:2002km}:
\begin{equation}
S_{\rm \frac{1}{2} wh} = \frac{\sqrt{2} |n|}{\alpha f \kappa} \times \sin \left( \frac{\pi}{2} \frac{ \alpha }{ \widetilde{\alpha} } \right) \,,
\end{equation}
where $\widetilde{\alpha}$ is as above.

With this brief review, we are now in a position to ask: what is the $O(1)$ coefficient for the axion WGC bound in this theory? It is very natural to suppose that the extremal instanton should set the axion WGC bound, just as the extremal black hole sets the ordinary WGC bound. From the instanton action \eqref{eqn:Sext}, this gives the bound:
\begin{equation}
 \frac{|n|}{f S} \geq \frac{|n|}{f S_{\textrm{ext}}} =  \frac{\alpha \kappa}{\sqrt{2}}  \,,~~~~ \alpha > \widetilde{\alpha}\,.
  \label{eq:axionWGC1}
\end{equation}
This bound is a very plausible candidate for the axion WGC when $\alpha \geq \widetilde{\alpha}$. By \eqref{eqn:Sext}, cored instantons have a larger action than the extremal instanton of the same instanton number, just as subextremal black holes have a larger mass than an extremal black hole of the same charge.

For $\alpha < \widetilde{\alpha}$, however, things become more complicated. Cored instantons now have a \emph{smaller} action than the extremal solution. Thus, the axion WGC bound should perhaps be given by the cored instanton of smallest action, which means
\begin{equation}
 \frac{|n|}{f S} \geq \frac{|n|}{f S_{\textrm{min}}} =  \frac{\widetilde{\alpha} \kappa}{\sqrt{2}}  \frac{1}{\sqrt{\frac{2 \widetilde{\alpha}}{\alpha} - 1}} \,,~~~~ \alpha < \widetilde{\alpha} \,.
 \label{eq:axionWGC2}
\end{equation}

However, the half-wormhole solution has an even smaller action than the cored and extremal instanton solutions. If the WGC bound is to be set by the macroscopic object of smallest action, then perhaps the axion WGC bound should be set by the half-wormhole solution, so that
\begin{equation}
 \frac{|n|}{f S} \geq \frac{|n|}{f S_{\textrm{min}}} =  \frac{\alpha \kappa}{\sqrt{2}} \frac{1}{\sin \left( \frac{\pi}{2} \frac{ \alpha }{ \widetilde{\alpha} } \right)} \,.
 \label{eq:axionWGC3}
\end{equation}
Note that the right-hand side of this bound remains finite in the $\alpha \rightarrow 0$ limit.

It is not clear which of these bounds should be viewed as the ``correct'' version of the axion WGC. Reference \cite{Heidenreich:2015nta} proposed the bounds \eqref{eq:axionWGC1} and \eqref{eq:axionWGC2}, whereas \cite{Hebecker:2016dsw, Hebecker:2018ofv} suggested the bound \eqref{eq:axionWGC3}. One difference in viewpoint is that the former paper assumed that only true instanton solutions, not wormholes, can contribute to an axion potential, because a wormhole is effectively an instanton/anti-instanton pair with no net charge. The latter argues that, because the instanton and anti-instanton ends of the wormhole can be very distant from each other in Euclidean time, they do in fact generate an axion potential. The latter perspective has a close affinity with the heuristic argument that wormholes violate global symmetries discussed in Section~\ref{NGS}.

Just as the precise statement of the axion WGC is somewhat difficult to define, so too is its magnetic version. Naively, we would like to say that there must exist a $(D-3)$-brane whose charge-to-tension ratio is greater than or equal to that of a large, extremal black $(D-3)$-brane (e.g., a string in $D=4$). Such objects do not exist in asymptotically flat spacetime, as we will discuss further shortly. Hence, rather than assuming an inequality with an exact coefficient determined by an extremality bound, it is natural to suppose that the WGC should imply the existence of some charged $(D-3)$-brane (i.e., a vortex) of charge $\tilde Q$ and tension $T_{D-2}$, satisfying
\begin{equation}
\frac{e_{D-2} | \tilde Q| }{ T_{D-2} }  \gtrsim \kappa\,,
\label{axionmag}
\end{equation}
where $e_{D-2} = 2\pi f$ is the magnetic coupling. Again, the inequality has a $\gtrsim$, and there is an $O(1)$ coefficient that remains to be fixed. Specializing to $D =4$ for convenience, an argument for this has been given in terms of axionic black holes, i.e., those with a nonvanishing integral $\int_\Sigma B$ of the axion's dual $B$-field over the horizon~\cite{Bowick:1988xh}. It has been argued that axionic strings obeying~\eqref{axionmag} are needed to allow this axionic charge to change and avoid a remnant problem in black hole evaporation~\cite{Hebecker:2017uix, Montero:2017yja}.

Because the magnetically charged object in this case has codimension two (e.g., a string in $D = 4$ or a 7-brane in $D = 10$), the classical tension stored in the winding axion field is logarithmically divergent in both the IR and the UV, whereas our discussion of the magnetic WGC in \S\ref{ssec:magnetic} incorporated only a UV divergence. Consequently, \eqref{eq:magWGCboundgeneral} is not valid in the case $P = 0$. Revisiting the logic by estimating the classical self-energy with UV {\em and} IR cutoffs and requiring it to satisfy the magnetic axion WGC bound~\eqref{axionmag}, we have
\begin{equation}
T \sim f^2 \log\frac{\Lambda_\mathrm{UV}}{\Lambda_\mathrm{IR}} \lesssim \frac{f}{\kappa},
\end{equation} 
or in other words
\begin{equation}
\frac{\Lambda_\mathrm{UV}}{\Lambda_\mathrm{IR}} \lesssim \exp\frac{O(1)}{\kappa f}.
\end{equation}
This is compatible with the idea that instantons will generate an IR scale $\Lambda_\mathrm{IR} \sim \mathrm{e}^{-S} \Lambda_\mathrm{UV}$, together with the electric axion WGC~\eqref{eq:axionWGC}, which implies $S \lesssim \frac{1}{\kappa f}$. Indeed, once an axion potential is generated through instantons, the axion vortex becomes the boundary of a domain wall, such that the winding of the axion field is localized inside the wall and there is no significant energy density outside the wall. When an axion vortex is attached to a semi-infinite domain wall, we would view the energy outside the axion vortex core as reflecting the finite domain wall tension rather than an infinite correction to the axion vortex tension. In this way, domain walls naturally provide an IR cutoff to the estimate of the axion vortex tension, and there is a relationship between the magnetic and electric WGC that has the same spirit, although more complicated details, as in the cases $1 \leq P \leq D-3$.

The above estimate neglects gravitational backreaction, which is significant for objects of low codimension. In particular, static vortices in gravitational theories produce a deficit angle. Implications of gravitational backreaction on axion strings (in $D = 4$) for the magnetic axion WGC were considered in~\cite{Dolan:2017vmn,Hebecker:2017wsu}. The static axion string solution in general relativity (in the case with zero axion potential, so that the strings are not confined by domain walls) was first found in~\cite{Cohen:1988sg}. The IR and UV divergences of the string without gravity are reflected in singularities of this solution. When $f < \sqrt{2} M_{\textrm{Pl}}$, the IR singularity lies exponentially far away in Planck units from the core of the string, and the deficit angle is positive. Hence one could consider, for example, large loops of closed string, which would be well-behaved in the IR and potentially completed by UV physics in the string core. When $f > \sqrt{2} M_{\textrm{Pl}}$, the deficit angle becomes negative, the singularity is inside the core of the string, and there is no longer a sensible interpretation of stringlike objects in approximately asymptotically flat spacetime with sensible UV completions in the string core. This suggests $f < \sqrt{2} M_{\textrm{Pl}}$ as a possible consistency condition on 2-form gauge theory in four dimensions.

Although the physics of static axion strings is relatively straightforward, one could consider whether the magnetic axion WGC could be satisfied by time-dependent, rather than static, objects~\cite{Dolan:2017vmn,Hebecker:2017wsu}. Non-singular, time-dependent string solutions were written down for a complex scalar $\Phi$ with a $U(1)$ global symmetry in \cite{Gregory:1996dd}, which features a Lagrangian of the form
\begin{equation}
\mathcal{L} = - \frac{1}{2} | \partial \Phi|^2 - \frac{\lambda}{4} (|\Phi|^2 - f^2)^2\,,
\end{equation}
such that the phase of $\Phi$ is an axion with decay constant $f$ in the low-energy theory. For $f$ smaller than some critical $f_{\text{crit}}$, there exist non-singular axion string spacetimes which inflate along the string direction but which have a static field configuration along slices orthogonal to the string. For $f > f_{\textrm{crit}}$, the field $\Phi$ itself becomes time-dependent, and the theory undergoes ``topological inflation.'' This occurs when the core region of a topological defect, of size $R_\textrm{core}$, has a  potential energy density $V_\textrm{core}$ sufficiently large to sustain a Hubble expansion rate $H \sim \sqrt{V_\textrm{core}}/M_\textrm{Pl}$ with $HR_\textrm{core} \gtrsim 1$~\cite{Linde:1994hy,Vilenkin:1994pv}. The numerical analysis of \cite{Cho:1998xy} found $f_{\textrm{crit}} = 1.63 M_{\textrm{Pl}}$ as the critical value for the onset of topological inflation. It was pointed out in~\cite{Dolan:2017vmn} that the computation with $f \gtrsim M_\textrm{Pl}$ is not necessarily under control, but a scenario with axion strings of winding number $n \gg 1$ such that $f \ll M_\textrm{Pl} \ll n f$ provides a controlled setting with similar conclusions. Numerical studies in~\cite{Dolan:2017vmn} confirmed exponential expansion in this scenario. They also demonstrated power-law expansion in a different model, in which the axion is the holonomy  of a higher-dimensional gauge field. In this case, the radial mode associated with the axion is the radion modulus $R$ of an extra dimension. The string core sees a decompactification limit, $R \to \infty$, which lies at infinite distance in field space where $V(R) \to 0$ (hence no exponential expansion). In both the $|\Phi|^4$ case and the radion case, there is no obvious pathology associated with the time-dependent infinite, straight string configurations. However,~\cite{Dolan:2017vmn} argued that the topological inflation of a closed loop of an axion string with $n f \gtrsim M_\textrm{Pl}$ would violate the ``topological censorship theorem'' of \cite{Friedman:1993ty}, suggesting that it would always collapse into a black hole. It would then be impossible for an observer to traverse a loop linking with a closed axion string to measure the field excursion. The impossibility of such a scenario is one candidate for a magnetic axion WGC.

\subsection{Repulsive Force Conjecture}\label{ssec:RFC}

The WGC was motivated by the idea that gravity should be weaker than any gauge force. The definition we have given above, however, deals not with the relative strength of gravity against other forces, but rather with the notion of superextremal particles. These two notions agree if the only forces are gravity and electromagnetism: a particle is superextremal if and only if the long-range electromagnetic repulsion between a pair of such particles is stronger than their gravitational attraction. In theories with massless scalar fields, however, this correspondence breaks down, and the question of whether a particle is superextremal is distinct from the question of whether or not a pair of such particles will repel each other at long distances. With this in mind, we thus define a particle to be self-repulsive if a pair of such particles repel one another at long distances, and we define the Repulsive Force Conjecture (RFC) as follows:
\vspace{.2cm}
\begin{namedconjecture}[Repulsive Force Conjecture (RFC)]
In any theory of a single abelian gauge field coupled to gravity, there is a self-repulsive charged particle.~\cite{Palti:2017elp}
\end{namedconjecture}
    \vspace{.1cm}
\noindent
After being emphasized by~\cite{Palti:2017elp}, this conjecture was further studied by~\cite{Lust:2017wrl,Lee:2018spm}. This statement can be easily generalized from particles charged under 1-form gauge fields to $(P-1)$-branes charged under $P$-form gauge fields. The generalization to theories with more than one gauge field is somewhat subtle; see \cite{Heidenreich:2019zkl} for further explanation.

While the RFC and the WGC are distinct conjectures in the presence of massless scalar fields, close connections remain, e.g., at the two-derivative level extremal black holes have vanishing long-range self-force \cite{Heidenreich:2020upe}, and the same towers of charged particles typically satisfy the tower/sublattice versions of both conjectures \cite{Heidenreich:2019zkl,BenMatteoProof}.

The idea of gravity as the weakest force has also motivated several variations on a scalar weak gravity conjecture, proposing that light scalars should always mediate forces stronger than gravity for some particles~\cite{Li:2006jja, Palti:2017elp,Lust:2017wrl,Gonzalo:2019gjp}. Such conjectures can lead to interesting consequences, including for phenomenology and cosmology. However, because they do not involve gauge fields and have no connection to black hole extremality, we will not discuss them further. Similarly, we will not discuss weak gravity statements related to higher-spin particles, for which there are sharp bounds from causality~\cite{Kaplan:2020ldi}.

In this review, we will primarily focus on the WGC and the notion of superextremality, but much of our analysis applies equally well to the RFC and the notion of self-repulsiveness. We stress once again that these conjectures are equivalent---and the notions of superextremality and self-repulsiveness are equivalent---in the absence of scalar fields.

\section{Evidence for the WGC}\label{EVIDENCE}

%!TEX root = WGC_Review_arxiv_v2.tex

The WGC was originally motivated by the idea that non-supersymmetric extremal black holes should be able to decay. As we have discussed, this motivation is not very compelling, since there is no obvious reason why stable extremal black holes present a problem for a theory. Nonetheless, this motivation seems to have gotten people to start digging in the right place, since by now there are a number of lines of evidence that support the WGC and its variants. In this section, we will focus on four such lines: an argument from dimensional reduction, examples in string theory, a general argument from modular invariance in perturbative string theory, and the relation between the WGC and the Swampland Distance Conjecture \cite{Ooguri:2006in}.

\subsection{Dimensional reduction}\label{REDUCTION}

One approach to assessing the validity of the WGC is to examine its internal consistency under dimensional reduction~\cite{Heidenreich:2015nta}; similar checks under $T$-duality were carried out in~\cite{Brown:2015iha}.
Our starting point is the Einstein-Maxwell-dilaton action~\eqref{eq:generalaction} for a $P$-form gauge field $A_{\mu_1 \ldots \mu_P}$ in $D =d+1$ dimensions.
We could in principle include additional terms in the low-energy action, such as Chern-Simons terms, but for our purposes the above action will suffice.

\subsubsection{Preservation of the $p$-form WGC bound}

We consider a dimensional reduction ansatz of the form,
\be
\rmd s^2 = \e^{\frac{\lambda(x)}{d-2}} \rmd{\hat s}^2(x) + \e^{-\lambda(x)} \rmd y^2,
   \label{eq:dimredansatz}
\ee
where $y \sim y + 2 \pi R$. For now, we do not include a Kaluza-Klein photon in our dimensional reduction ansatz, but we will do so later in this subsection. 
The coefficients of $\lambda(x)$ in the exponentials have been carefully chosen so that the dimensionally reduced action is in Einstein frame, i.e., there is no kinetic mixing between $\lambda$ and the $d$-dimensional metric:
\begin{align}
 \frac{1}{2\kappa_D^2} \int \rmd^D x \sqrt{-g} {\cal R}_D 
 ~~\rightarrow~~  \frac{1}{2\kappa_d^2} \int \rmd^d x \sqrt{- \hat g}  {\cal R}_d  - \frac{1}{2}  \int \rmd^d x \sqrt{- \hat g} \, G_{\lambda \lambda} (\nabla \lambda)^2 \,,
\end{align}
where
\begin{align}
\frac{1}{\kappa_d^2} &= M_d^{d-2} = (2 \pi R) M_D^{D-2} \,, & %\label{Ddconstants} \\
G^{(d)}_{\lambda \lambda} &= \frac{(d-1)}{4\kappa_d^2 (d-2)} = M_d^{d-2} \frac{d-1}{4(d-2)}\,.
%\label{radionkineticterm}
\end{align}

Upon dimensional reduction, the $P$-form gauge field in $D$ dimensions gives rise to both a $P$-form gauge field and a $p=(P-1)$-form gauge field in $d$ dimensions, obtained respectively by taking all of the legs of the gauge field to lie in noncompact directions, or by taking one leg to wrap the compact $S^1$ direction. The gauge couplings of the two gauge fields are given respectively by 
\begin{align}
    e_{P;d}^2 = \frac{1}{2 \pi R}e_{P;D}^2  \,, ~~~~~~~
      e_{p;d}^2 = (2 \pi R)e_{P;D}^2   \,.
      \label{edimred}
\end{align}
Similarly, a charged $(P-1)$-brane in $D$ dimensions reduces to both a $(P-1)$-brane and a $(p-1)$-brane, obtained respectively by taking the brane to lie exclusively in noncompact dimensions, or by taking the brane to wrap the compact direction. 
The tensions of these branes are given respectively by 
\begin{align}
T_P^{(d)} =   T_P^{(D)}\,, ~~~~~~
T_p^{(d)} =  (2 \pi R)  T_P^{(D)}\,.
\label{eq:TPd}
\end{align}
Recall from \eqref{eq:extremalitybound} and \eqref{gammaPD} that the WGC bound is modified by the exponential coupling of the radion to the Maxwell term. The Maxwell term of the $P$-form gauge field couples to a linear combination of both $\phi$ and the radion $\lambda$, and it is useful to rewrite these scalar fields in terms of two canonically normalized fields $\sigma$ and $\rho$, the former of which decouples from the Maxwell term, the latter of which couples to it as $\mathrm{e}^{- \alpha_{P;d} \rho} F_{P+1}^2$. The coefficient $\alpha_{P;d}$ is then given by
\begin{equation}
    \alpha_{P;d}^2 = \alpha_{P;D}^2 + \frac{2 P^2}{(d-1)(d-2)}\,.
\end{equation}
This can be rewritten as
\begin{equation}
\frac{\alpha_{P;d}^2}{2} + \frac{P (d-P-2) }{d-2} = \frac{\alpha_{P;D}^2}{2} + \frac{P(D-P-2)}{D-2}\,,
\end{equation}
which by \eqref{gammaPD} implies
\begin{equation}
\gamma_{P;d}(\alpha_{P;d}) = \gamma_{P;D}(\alpha_{P;D}) \,,
\end{equation}
from which we conclude that the $(P-1)$-brane satisfies the $P$-form WGC bound \eqref{eq:extremalitybound} in $D$ dimensions if and only if it satisfies the $P$-form WGC bound in $d$ dimensions: in other words, the WGC is exactly preserved under dimensional reduction.

A similar story applies to the case of the wrapped brane: a particular linear combination of $\phi$ and $\lambda$ couples to $F_{p+1}^2$, which ultimately leads to the coefficient
\begin{equation}
    \alpha_{p;d}^2 = \alpha_{P;D}^2 + \frac{2 (d-p-2)^2}{(d-1)(d-2)}\,.
\end{equation}
This can be rewritten as
\begin{equation}
\frac{\alpha_{p;d}^2}{2} + \frac{p (d-p-2) }{d-2} = \frac{\alpha_{P;D}^2}{2} + \frac{P(D-P-2)}{D-2}\,,
\end{equation}
which by \eqref{gammaPD} implies
\begin{equation}
\gamma_{p;d}(\alpha_{p;d}) = \gamma_{P;D}(\alpha_{P;D}) \,,
\end{equation}
so again, the WGC bound is exactly preserved: the $(P-1)$-brane satisfies the $P$-form WGC bound in $D$ dimensions if and only if it satisfies the $p$-form WGC bound in $d$ dimensions after wrapping on $S^1$.

\subsubsection{Kaluza-Klein modes and a violation of the CHC}\label{ssec:KK}

Let us now add a Kaluza-Klein photon to our dimensional reduction ansatz:
\be
\rmd s^2 = \e^{\frac{\lambda(x)}{d-2}} \rmd{\hat s}^2(x) + \e^{-\lambda(x)} (\rmd y+R B_1)^2 \,,
\label{eq:ansatz}
\ee
where $y \cong y + 2 \pi R$ and $B_1$ is normalized so that the KK modes carry integral charges. The dimensionally reduced action is then given by
\be
S = \int \rmd^d x \frac{\sqrt{-{\hat g}}}{2\kappa_d^2} \left[{\cal {\hat R}}_d - \frac{d-1}{4(d-2)} (\nabla \lambda)^2 - \frac{R^2}{2} \e^{- \frac{d-1}{d-2} \lambda} H_2^2\right] \,,
\ee
where $H_2 = \rmd B_1$. From this action, we may read off the KK photon gauge coupling and the radion--KK photon coupling:
\be
\frac{1}{e_{\rm KK}^2} = \frac{1}{2} R^2 M_d^{d-2} \;\;,\;\; \alpha_{\rm KK} = \sqrt{\frac{2(d-1)}{d-2}} \,.
\ee
Here, $\alpha_{\rm KK}$ is defined by the coupling to the normalized radion $\hat{\lambda} = \sqrt{\frac{d-1}{2(d-2)}} \lambda$.

The WGC bound for a particle with $n$ units of KK charge is then given by (\ref{eq:extremalitybound}):
\be
\left[\frac{\alpha^2_{\rm KK}}{2} + \frac{d-3}{d-2}\right] m^2 \leq e_{\rm KK}^2 n^2 M_d^{d-2} \,.
\ee
This means that $\gamma_{\rm KK} = 2$, and the WGC bound is simply
\be
m^2 \leq \frac{n^2}{R^2} \,.
\ee
This may be compared to the spectrum of KK modes for a particle of mass $m_0$ in the parent theory:
\be
m^2 = m_0^2 + \frac{n^2}{R^2} \,,~~~ n \in \mathbb{Z}\,,
\ee
where the KK charge $n$ specifies the momentum $n/R$ of the particle along the compact circle. We see therefore that KK modes of massless particles saturate the WGC bound, whereas KK modes of uncharged massive particles violate the WGC bound. The $D$-dimensional parent theory necessarily has at least one massless particle---namely, the graviton---so the dimensionally reduced theory necessarily has superextremal particles charged solely under the KK photon. Indeed, each of the KK modes of the graviton is superextremal, so there is actually an infinite tower of superextremal KK modes, as required by the tower/sublattice WGC.

What happens, however, if we include a $U(1)$ in the parent theory in $D$ dimensions? The resulting $d$-dimensional theory will now have two $U(1)$ gauge fields, and the WGC is equivalent to the convex hull condition (CHC) introduced in Section \ref{ssec:CHC}. 

In the parent theory, a particle of charge $q$ and mass $m$ is superextremal when the dimensionless charge-to-mass ratio $Z_D \df \frac{q}{m} e_D \gamma_D^{-1/2} M_D^{(D-2)/2}$ has magnitude $|Z_D| \geq 1$. Likewise, in the dimensionally reduced theory a particle of charge $(q,q_{\rm KK})$ and mass $m$ is superextremal when the dimensionless charge-to-mass ratio vector
\begin{equation}
    \vec{Z}_d \df \biggl(\frac{q}{m} e_d \gamma_d^{-1/2} M_d^{(d-2)/2}, \frac{q_{\rm KK} - \frac{q \theta}{2\pi}}{m R}\biggr) \,,
\end{equation}
has length $|\vec{Z}_d| \ge 1$, where $\theta = \oint A$ is the vev of the axion descending from the gauge field and $\gamma_d = \gamma_D$ accounting for the radion coupling as above.

\begin{figure}
\centering
\includegraphics[width=0.45\textwidth]{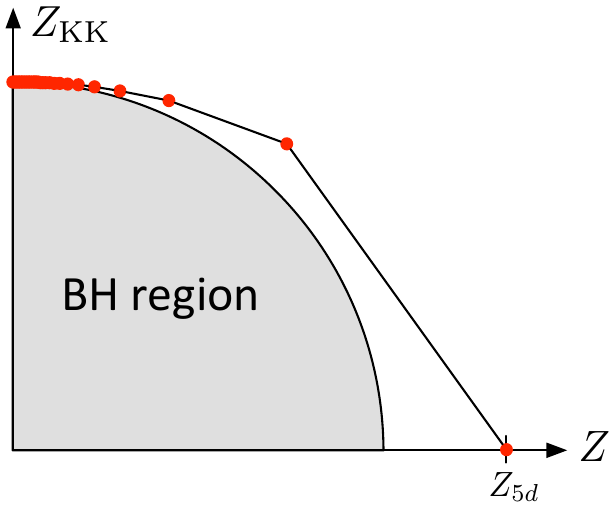} \hspace{0.02\textwidth}
\includegraphics[width=0.45\textwidth]{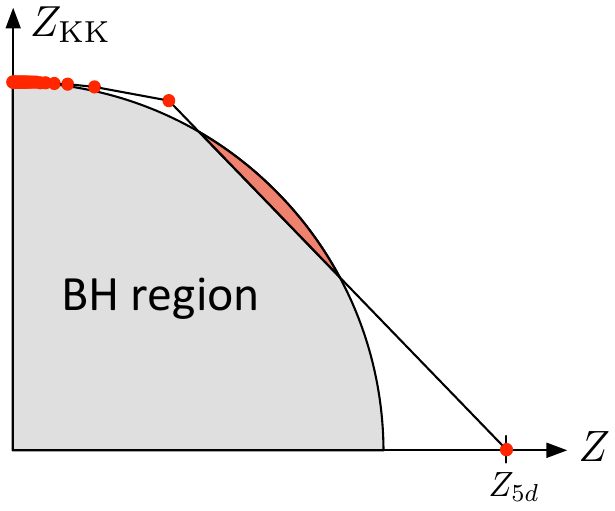}
\caption{(Left) The Kaluza-Klein modes of a superextremal particle with charge $q_F$ in $d+1$ dimensions are superextremal after reduction on $S^1$, as their charge-to-mass vectors $\vec{Z}$ lie outside the elliptical black hole region. (Right) If the $S^1$ is sufficiently small, the convex hull condition is violated, and the Kaluza-Klein modes of the superextremal particle in question do not satisfy the WGC.}
\label{CHCKK}
\end{figure}

The $n$th KK mode of a particle of charge $q$ and mass $m_0$ in the parent theory has mass $m^2 = m_0^2 + \frac{1}{R^2} (n-\frac{q\theta}{2\pi})^2$, and hence the charge-to-mass vector
\begin{equation}
\vec{Z}_d^{(n)} = \frac{ ( \mu Z_D , x_n) }{\sqrt{ \mu^2 + x_n^2 }  }\,,~~~\mu = m_0 R\,,~~~x_n = n - \frac{q \theta}{2 \pi} \,.
\end{equation}
The charge-to-mass vectors of the KK modes, along with the convex hull they generate, are plotted in figure \ref{CHCKK} (left). The vectors lie on the ellipsoid $Z_{d1}^2/Z_D^2 + Z_{d2}^2 = 1$, which lies outside the unit disk provided that $|Z_D|\ge 1$, so each KK mode of a particle that was superextremal in the parent theory is superextremal.

However, the fact that each individual KK mode is superextremal does not ensure that the convex hull condition is satisfied. As shown in figure \ref{CHCKK} (right), as we take the limit $R \rightarrow 0$, the KK modes of the particle are pushed closer and closer towards the point $(0,1)$. Below some critical value of $R$, the convex hull condition is violated. In fact, if $Z_D = 1$, saturating the WGC bound, then the convex hull condition will be violated for any value of $R$. Starting from a theory that satisfies the WGC in $D$ dimensions, we have arrived at a theory that violates the WGC in $d$ dimensions.

It is important to realize that this does not represent a counterexample to the WGC, because there is no good reason to think that the $D$-dimensional theory we started with is in the Landscape as opposed to the Swampland. Rather, we showed that the WGC in $D$ dimensions alone is not sufficient to ensure that the WGC holds in $d$ dimensions. If we want the WGC to hold in $d$ dimensions, we need to impose a stronger constraint than the WGC in $D$ dimensions.

To identify such a constraint, it is worth noting that a violation of the convex hull condition for sufficiently small $R$ will arise whenever the number of superextremal particles in $D$ dimensions is finite. To satisfy the WGC in $d$ dimensions for all $R$, therefore, requires an infinite number of superextremal particles in $D$ dimensions. Indeed, it is not hard to see that the tower WGC, as defined in Section \ref{ssec:Strong}, is a sufficient condition for ensuring that the WGC is satisfied in the dimensionally reduced theory. Indeed, this observation is what originally motivated the tower/sublattice WGC. There are at present no known counterexamples to either of these conjectures in string theory.

One can further check that the tower WGC is satisfied in $d$ dimensions provided that it is satisfied in $D$ dimensions (and likewise for the sublattice WGC), so the tower WGC and sublattice WGC are automatically preserved under dimensional reduction, unlike the mild WGC.

The general idea that a proposed consistency criterion in quantum gravity should apply not just to a single vacuum but to all of its  compactifications, whose application to the WGC discussed here originated in~\cite{Heidenreich:2015nta}, was later dubbed the ``Total Landscaping Principle'' in~\cite{Aalsma:2020duv}, and has been fruitfully applied in several contexts, e.g.,~\cite{Montero:2017yja, Heidenreich:2019zkl, Aalsma:2020duv, Cremonini:2020smy, Rudelius:2021oaz}.

\subsection{Higgsing}\label{ssec:Higgsing}

We have just seen that the mild WGC is not automatically preserved under compactification: starting with a theory that satisfies the WGC, we can produce a theory that violates the WGC by Kaluza-Klein reduction on a circle. This points towards some stronger version of the WGC, such as the tower/sublattice WGCs, which are automatically preserved.

In this subsection, we will see that a similar issue arises from the process of Higgsing. Starting with a theory that satisfies certain forms of the WGC, we can produce a theory that violates these forms of the conjecture by Higgsing. However, other versions of the WGC will be preserved. In particular, we show, following \cite{Saraswat:2016eaz}, that the mild WGC, tower WGC, and sublattice WGC are preserved (barring a very special fine-tuning that we do not expect to occur). In contrast, the statements that the lightest charged particle should be superextremal or that a particle of smallest charge should be superextremal are not preserved under Higgsing.

Consider a theory with two $U(1)$ gauge fields, $A$ and $B$. For simplicity, we assume that their gauge couplings are identical, $g_A = g_B = g$ and that we are working in four dimensions. Suppose that there are two superextremal particles with masses with $m_1 < m_2$ and charges $q_1 = (1, 0)$ and $q_2 = (0, 1)$, respectively, and let these be the lightest charged particles in the theory.

Next, suppose that there is a scalar field of charge $(n, 1)$ which acquires a vev $v$. Under this process, the gauge boson $H = n A + B$ acquires a mass $m_H = g v \sqrt{n^2+1}$, and the gauge boson $L = (A-n B)/(n^2+1)$ remains massless, with gauge coupling $g_{\textrm{eff}} = g/\sqrt{n^2+1}$. After Higgsing, the particle of charge $q_1 = (1, 0)$ has quantized charge $1$ under the massless gauge field $L$. It still has mass $m_1$, so for $n$ sufficiently large, we find that this particle is no longer superextremal after Higgsing, since $g_{\textrm{eff}} \simeq g/n \ll m_1 / M_{\textrm{Pl}}$.

On the other hand, the particle of charge $q_2 = (0, 1)$ has quantized charge $-n$, so it remains superextremal after Higgsing. Thus, the mild version of the WGC remains satisfied in this theory. However, since we assumed $m_1 < m_2$, the lightest charged particle is no longer superextremal, and there is no longer a superextremal particle of charge $1$. We see that the strong forms of the WGC that demand that either the lightest charged particle or the particle of smallest charge should be superextremal are not automatically preserved under Higgsing: they are violated here in the Higgsed theory even though they were satisfied in the unHiggsed theory.

As with the reasoning that led to the tower/sublattice WGCs above, one might be tempted to search for stronger conjectures that ensure the lightest charge particle and/or particle of smallest charge automatically remain superextremal even after Higgsing. However, in the following subsection, will we see an explicit example in string theory in which these latter versions of the WGC are violated, so these conjectures should simply be discarded rather than fixed up with an even stronger consistency condition.

In the Higgsing example we considered here, the mild form of the WGC is preserved by the Higgsing process. Similarly, if we assume that the tower/sublattice WGCs are satisfied before Higgsing, we will find that they are still satisfied after Higgsing by the tower of particles with charge proportional to $(n, 1)$. However, this is no longer automatically true when we generalize our theory. If we assume that there is mixing in the charge lattice between the two $U(1)$s, such that the canonically normalized charge vectors take the form $(g_A, 0)$ and $(g_B^1, g_B^2)$ with $g_B^1/g_A$ irrational, and if we assume that the sublattice WGC is exactly saturated before Higgsing, so there are no particles in the theory charged under $B$ strictly below the WGC bound, then by giving a vev to a scalar field with charge 0 under $B$, we will find that there are no superextremal particles in the Higgsed theory. In this very special case, the tower, sublattice, and mild form of the WGC are all violated after Higgsing.

However, this special scenario is not very likely in practice. It is true that the WGC bound may be exactly saturated in much or all of the charge lattice--this happens, for instance, in theories with extended supersymmetry, where BPS bounds may forbid strictly superextremal particles in certain directions in the charge lattice. However, the very same BPS bound ensures that any Higgs field with the required charges is massive, hence the problematic Higgsing scenario discussed above does not arise.

We conclude that the tower/sublattice WGC and mild form of the WGC are unlikely to be violated by Higgsing in any UV complete theory of quantum gravity. However, even in the example considered previously in this subsection, the sublattice of superextremal particles after Higgsing may be much sparser than the sublattice of superextremal particles before Higgsing, with indices that differ by a factor of $n$. Relatedly, the superextremal particle of charge $q_2 = (0, 1)$ may have a mass $m_2$ which is well above the magnetic WGC scale of the the IR theory, $\Lambda \sim g_{\textrm{eff}} M_{\textrm{Pl}} \sim g M_{\textrm{Pl}}/n$ \cite{Saraswat:2016eaz} (see also~\cite{Furuuchi:2017upe}). To ensure that the lightest superextremal particles do not have parametrically large charge, one must argue for an $O(1)$ upper bound on the charge $n$ of the Higgs field in this theory. Very little work has gone into arguing for such an upper bound (outside of specific string theory contexts~\cite{Ibanez:2017vfl}), though it could be a very worthwhile direction for future research.

\subsection{String theory examples}\label{STRINGEXAMPLES}

\subsubsection{Heterotic string theory}\label{ssec:HETEROTIC}

As a first example, let us consider $SO(32)$ heterotic string theory in ten dimensions. 
The low-energy effective action in Einstein frame is given by \cite{Polchinski:1998rr}
\be
\frac{1}{2 g_s^2 \kappa_{10}^2} \int \rmd^{10}x \sqrt{-g} \left(R - \frac{1}{2} \partial_\mu \Phi \partial^\mu \Phi\right) - \frac{1}{2 g_s^2 g_{10}^2} \int \rmd^{10}x \sqrt{-g}\, \e^{-\phi/2} {\rm Tr}_V\left(|F_2|^2\right),
\ee
where ${\rm Tr}_V$ is the trace in the fundamental representation, normalized so that ${\rm Tr}_V(T^a T^b) = 2 \delta^{ab}$ for the basis of generators $T^a$.

We may then define
\begin{align}
8 \pi G_N &= M_{10}^{-8} =  g_s^2 \kappa_{10}^2 = \frac{1}{2} g_s^2 (2\pi)^7 \alpha'^4, &
e^2 &= \frac{1}{2} g_s^2 g_{10}^2 = g_s^2(2\pi)^7 \alpha'^3,
\end{align}
where $e^2$ is the coupling constant associated with any single $U(1)$ in the maximal torus. Notice that our dilaton coupling parameter is $\alpha = 1/2$, which by \eqref{gammaPD} gives $\gamma = 1$.

The charge lattice of the $SO(32)$ heterotic string consists of all charge vectors of the form:
\begin{align}
{\vec q} = \left(q_1, q_2, \ldots q_{16}\right) \quad \text{or} \quad {\vec q}  = \left(q_1 + \frac{1}{2}, \ldots, q_{16} + \frac{1}{2}\right) \qquad
 \text{with} \quad q_i  \in \mathbb{Z},  \sum_i q_i  \in 2 \mathbb{Z}. 
\end{align}
This lattice is even, i.e., $|q|^2 \in 2 \mathbb{Z}$ for any $\vec{q}$ in the lattice.
States must satisfy the level-matching condition
\be
\frac{\alpha'}{4} m^2 = N_L + \frac{1}{2} \left|{\vec q}\right|^2 - 1 = N_R - \frac{1}{2},
\ee
where $N_{L,R}$ are the occupation number of the left and right-moving oscillators, with $N_L$ a non-negative integer and $N_R$ a positive half-integer. Given any choice of $N_L \ge 0$ and $\vec{q} \ne 0$, we may always choose $N_R$ to satisfy the level-matching condition. Thus, the lightest state with a given $\vec{q} \ne 0$ has
\be
m^2 = \frac{2}{\alpha'} \left(\left|{\vec q}\right|^2 - 2\right).
\ee
The charge-to-mass vector of this state then obeys
\be
\left|{\vec Z}\right|^2 = \frac{2}{\alpha'} \left|\frac{{\vec q}}{m}\right|^2 = \frac{\left|{\vec q}\right|^2}{\left|{\vec q}\right|^2 -2 } > 1\,,
\ee
which shows that the state is superextremal. This means that there is a superextremal particle in every representation of the $SO(32)$ gauge group, so the theory satisfies the nonabelian sublattice WGC (in fact, it even satisfies the lattice WGC). Compactifying this theory on $T^n$ and turning on Wilson lines, the gauge group is generically broken to its Cartan subgroup, and the resulting theory will satisfy the lattice WGC for abelian gauge groups. The same is true for $E_8 \times E_8$ heterotic string theory.

\subsubsection{F-theory}\label{ssec:Ftheory}

Consider F-theory compactified to six dimensions on an elliptically-fibered Calabi-Yau threefold $Y_3$, with base $B_2$.  A gauge symmetry $G$ arises from a stack of 7-branes wrapping a holomorphic curve $C$ in the base, and the gauge coupling and 6d Planck scale are related to the volume of $B_2$ and $C$ via
\begin{equation}
M_6^4 \sim \text{vol}(B_2)\,,~~~~\frac{1}{g_{\text{YM}}^2} \sim \text{vol}(C)\,.
\end{equation}
In \cite{Lee:2018urn}, the authors showed that the limit $g_{\text{YM}} \rightarrow 0$ with $M_6$ finite can be achieved only if the base $B_2$ contains a rational curve $C_0$ whose volume goes to zero as $\text{vol}(C) \rightarrow \infty$. A D3-brane wrapping $C_0$ gives rise to a string with string states charged under the gauge group, and in the tensionless limit $\text{vol}(C_0) \rightarrow 0$, this string is identified with a heterotic string in a dual description.

In some cases, such as when the base $B_2$ is a Hirzebruch surface, the dual heterotic string is weakly coupled. In this case, the sublattice WGC follows from modular invariance, as we will show in Section \ref{MODULAR} below. If the heterotic string is strongly coupled, however, then it is not so simple to compute the spectrum of string states, and the best one can do is to compute an index of charged BPS string states using the elliptic genus. Using properties of the elliptic genus, \cite{Lee:2018urn} argued that the sublattice WGC is necessarily satisfied with respect to the gauge group $G$. This was strengthened by~\cite{Cota:2020zse} to an argument that 6d F-theory compactifications on Calabi-Yau threefolds in fact satisfy the nonabelian sublattice WGC of \S\ref{ssec:nonabelian}.

Subsequent work \cite{Lee:2019tst, Klaewer:2020lfg} analyzed the elliptic genera of tensionless strings coming from wrapped D3-branes in 4d theories coming from F-theory compactified on elliptically fibered Calabi-Yau 4-folds. For generic fluxes, properties of the elliptic genus suffice to prove the sublattice WGC. For non-generic fluxes, there are still superextremal string states, but (unlike in six dimensions) these superextremal particles do not necessarily furnish a sublattice. Indeed, the authors of \cite{Lee:2019tst} identified an example of an F-theory compactification for which the elliptic genus detects no superextremal string states of charge $4 \vec{q}$ for any $\vec{q}$ in the charge lattice. This does not necessarily imply a counterexample to the sublattice WGC, however, as there are other sectors of charged states not visible to the elliptic genus, and it is conceivable that these sectors may contain the requisite superextremal particles to satisfy the sublattice WGC.

\subsubsection{A counterexample to the lattice WGC}\label{ssec:counterexample}

We have seen a number of examples which satisfy the WGC, as well as its stronger variants. We will now present an example which violates a number of proposed strong forms of the WGC. Nonetheless, it still satisfies the WGC, tower WGC, and sublattice WGC.

 The example in question comes from compactifying type II string theory on the $T^6/(\mathbb{Z}_2 \times \mathbb{Z}_2')$ orbifold with orbifold action defined by the two generators:
 \begin{align}
\begin{split}
\theta: \quad & \theta_4 \mapsto \theta_4 + \pi,\, \theta_5 \mapsto \theta_5 + \pi, \\
\omega: \quad &  \theta_6 \mapsto \theta_6 + \pi, \, \theta_i \mapsto -\theta_i, i=1,...,4.
\end{split}
\label{eq:favorite}
\end{align}
Here, the $T^6$ in question is parametrized by the angles $\theta^i \cong \theta^i + 2 \pi$, $i=1,\ldots,6$, and for simplicity we take the metric to be diagonal in the $\theta_i$ basis. Note that the $\omega$ generator acts as a ``roto-translation'': a rotation combined with a translation in a different direction. This roto-translation acts freely, and thus the orbifold geometry is smooth.  As a result, the compactification can be understood within supergravity, as well as on the string worldsheet.

For our purposes, it will suffice to concentrate on the $\theta_4$, $\theta_5$, $\theta_6$ dimensions of the $T^6$.  
Each of these dimensions has a gauge field associated with Kaluza-Klein momentum around the $S^1$; we will denote them respectively by $A^4_\mu$, $A^5_\mu$, and $A^6_\mu$. The action of $\omega$ projects the first of these fields out of the spectrum, leaving $A^5_\mu$ and $A^6_\mu$ as the only Kaluza-Klein gauge bosons in the theory.

Next, consider a field $\phi$ on the $T^3$ parametrized by $\theta_4$, $\theta_5$, $\theta_6$. Its field decomposition is given by
\be
\phi(x^\mu, \theta_4, \theta_5, \theta_6) = \sum \phi_{n_4, n_5, n_6}\!(x^\mu)\, \e^{i n_4 \theta_4 +i  n_5 \theta_5 + i n_6 \theta_6}.
\ee
The orbifold action imposes the identifications
\begin{align}
\begin{split}
\theta: \quad \phi_{n_4,n_5,n_6}(x) &= \left(-1\right)^{n_4 + n_5} \phi_{n_4,n_5,n_6}\!(x), \\
\omega: \quad \phi_{n_4,n_5,n_6}(x) &= \left(-1\right)^{n_6} \sigma(\phi) \phi_{-n_4,n_5,n_6}\!(x).
\end{split}
\end{align}
Here $\sigma(\phi)$ denotes an additional sign that may arise depending on the nature of the field $\phi$--for instance, the graviton, $A^5_\mu$, and $A^6_\mu$ have $\sigma = 1$, whereas $W_\mu$ has $\sigma = -1$ due to the action of $\omega$ (and is therefore projected out of the spectrum).

Now, we look at the sublattice of the charge lattice consisting of the charges $(n_5, n_6)$ under the surviving Kaluza-Klein fields $A^5_\mu$, $A^6_\mu$. 
For $n_5$, $n_6$ both even, Kaluza-Klein modes of the graviton with $n_4 = 0$ are projected in and saturate the extremality bound, so there are indeed superextremal particles of these charges. For $n_6$ odd, $n_5$ even, a mode will be projected out unless it has $\sigma = -1$, but KK modes of the gauge field $A^4_\mu$ satisfy this condition and similarly saturate the extremality bound. For odd $n_5$, however, the action of $\theta$ imposes the constraint that $n_4$ must be odd, which leads to an additional contribution of $(n_4/R)^2$ to the mass squared of such a mode:
\begin{equation}
 m^2 = \left(\frac{n_5}{R_5} \right)^2 + \left(\frac{n_4 }{R_4} \right)^2\,,~~n_5~ \rm{ odd },\, n_4~ \rm{ odd.}
\label{eq:untwistedmass}
\end{equation}
 This additional contribution renders such modes subextremal: there are no superextremal particles of charge $(n_5, n_6)$ for $n_5$ odd. This result is summarized in Table \ref{favtab}.

\begin{table}
\begin{center}
\begin{tabular}{|c|c|c|} \hline
$n_5$ \textbackslash $~n_6$ & even & odd \\ \hline
odd & \xmark & \xmark \\ \hline
even & \cmark &  \cmark \\\hline
\end{tabular}
\caption{\label{favtab} Superextremal particles in type II compactified on the $T^6/(\mathbb{Z}_2 \times \mathbb{Z}_2')$ orbifold exist for $n_5$ even but not for $n_5$ odd. As a result, the lattice WGC is violated, whereas the sublattice WGC is satisfied with coarseness 2.}
\end{center}
\end{table}

This theory represents a counterexample to the lattice WGC: there exist charges in the charge lattice without superextremal particles, namely, any charge $(n_5, n_6)$ with $n_5$ odd. By moving in the moduli space of the theory, the sizes of the $R_i$ of the cycles of the torus can be adjusted freely, and for certain values of the moduli additional proposed ``strong forms'' of the WGC may also be violated. For example, taking $R_4 \gg R_5 > R_6$ and $R_I \gg \sqrt{\alpha'}$, the winding modes become heavy and the lightest charged particle in the spectrum is subextremal with $(n_5,n_6)=(1,0)$. This particle is also the state of smallest charge in its direction in the lattice. Thus, this theory represents a counterexample to both of the strong forms of the WGC considered in AMNV \cite{ArkaniHamed:2006dz}: neither the lightest charged particle nor the particle of smallest charge in the $n_5$ direction in the lattice are superextremal. Furthermore, the masses of the particles of odd $A^5$ charge $n$ violate the convexity condition:
\begin{equation}
   2 m_{n} \geq m_{n+1} + m_{n-1}\,,
   \label{convexitycond}
\end{equation}
where $m_n$ is the mass of the lightest particle of charge $n$. If there exists an AdS analog of this example, it would violate the strong forms of the ``Abelian Convex Charge Conjecture'' of \cite{Aharony:2021mpc} introduced in \S\ref{ssec:aAdS}.

On the other hand, the tower WGC and the sublattice WGC are satisfied in this example: given any charge $\vec{q}$, there exists a superextremal particle of charge $2 \vec{q}$. The sublattice of superextremal particles therefore has coarseness 2.

Finally, let us remark on a puzzling feature of this example: at tree level in string perturbation theory, the lightest particles with odd $n_5$ are in fact stable \cite{Heidenreich:2016aqi}. This suggests that black holes of odd $n_5$ charge cannot decay, in violation of the original motivation of the WGC. It is possible that loop corrections could modify the spectrum so that this conclusion could be avoided; more work is needed to see if this possibility is actually realized.

A number of other counterexamples to the lattice WGC were identified in \cite{Heidenreich:2016aqi}. These counterexamples all involve orbifold compactifications of string theory, and all of them satisfy the sublattice WGC with a superextremal sublattice of coarseness no larger than 3. Furthermore, in all such examples, the majority of sites in the charge lattice have superextremal particles---even sites outside the superextremal sublattice. This means that when it comes to the existence of superextremal particles, quantum gravity seems to impose even \emph{stronger} constraints than the tower/sublattice WGC; such constraints are seldom discussed, simply because it is not so easy to formulate them as precise mathematical statements.

\subsubsection{Axions in string theory}\label{ssec:AxionsST}

Recall that the WGC for axions~\eqref{eq:axionWGC} implies an upper bound on the axion decay decay constant $f$ in terms of the instanton action $S$,
\be
f \lesssim \frac{M_{\rm Pl}}{S}\,.
\ee
Within string theory, the condition $S \gtrsim 1$ is typically required for perturbative control. For instance, the instanton action may represent the size of some compactification cycle in string units, so the $\alpha'$ expansion breaks down when this cycle is smaller than the string scale. The WGC for axions thus amounts to the condition that $f \lesssim M_{\rm Pl}$ within the perturbative regime of string theory.

In fact, this condition was famously pointed out by Banks, Dine, Fox, and Gorbatov \cite{Banks:2003sx} even before the original AMNV paper on the WGC. In particular, \cite{Banks:2003sx} considered axions in heterotic, type I, type IIA, type IIB, and M-theory compactified to four dimensions. In all cases, these axions arise either as the periods of a $p$-form $C_p$ over a $p$-cycle $\Sigma_p$ of the compactification manifold, $\oint_{\Sigma_p} C_p$, or as the dual of a 2-form gauge field $B_{\mu\nu}$ in four dimensions. In all cases, their decay constants were found to be bounded above as $ f \lesssim M_{\rm Pl}$.\footnote{\cite{Banks:2003sx} incorrectly claims the model-independent heterotic axion, i.e., the 4d dual of $B_{\mu \nu}$, has $f = M_{\rm Pl}$. In fact, it has $f \sim  M_s^2/M_\mathrm{Pl}$; see, e.g.,~\cite{Svrcek:2006yi}.}

More recently, a number of works have taken advantage of an improved understanding of Calabi-Yau compactifications to investigate the prospects for super-Planckian decay constants in type IIA/IIB string theory with greater precision \cite{Rudelius:2014wla,Rudelius:2015xta, Montero:2015ofa, Bachlechner:2015qja,  Brown:2015lia, Junghans:2015hba, Palti:2015xra, Conlon:2016aea, long:2016jvd, Hebecker:2017lxm}. In all cases, the axion decay constants appear to be bounded above by $f \lesssim M_{\rm Pl}$. This remains true even in theories with multiple axions (see Section \ref{ssec:axioninflation}). 

The size of decay constants allowed in string theory is important because models of so-called ``natural inflation'' require $f > M_{\rm Pl}$. From the perspective of inflation, however, we are interested not only in the kinematic question of how large an axion decay constant can be, but also in the dynamical question of how an axion rolls in its potential. Obtaining the potential of an axion in a type IIB compactification is not simple, as it requires information about the sheaf cohomology of curves/divisors of the Calabi-Yau compactification manifold, which is not known in general. Some progress in understanding the relevant sheaf cohomology has been made in \cite{Braun:2017nhi}. Finally, note that bounds on axion decacy constants do not directly constrain axion monodromy models~\cite{Silverstein:2008sg, McAllister:2008hb}.

The axion WGC can also be studied outside of the context of specific string constructions. In general, the breakdown of the instanton expansion that arises when $S \lesssim 1$ is always due to the presence of new, light states. This has been argued to follow from the general Lee--Yang theory of phase transitions~\cite{Stout:2020uaf}. If the potential $V(\theta)$ is a smooth function of the axion value $\theta$, then its harmonics asymptotically decay as $\exp(-n S)$ where $S$ is determined by the location of the nearest singularity $\zeta_*$ to the unit circle for the complexified coordinate $\zeta = \exp(i \theta)$. This asymptotic notion of $S$ has been suggested to define the correct formulation of the axion WGC in the limit when instantons are not well-defined semiclassical objects~\cite{Stout:2020uaf}.

\subsection{Modular invariance}\label{MODULAR}

So far, we have seen that the WGC and its tower versions hold true in a large class of examples in string/M-theory.
In this subsection, we present a very general argument for the sublattice WGC in 2d CFTs on the basis of modular invariance. This result can be viewed as either (a) a proof of the sublattice WGC in perturbative string theory, viewing the 2d CFT as the worldsheet theory \cite{Heidenreich:2016aqi}, or (b) a proof of the sublattice WGC in AdS$_3$, viewing the 2d CFT as the boundary dual of an AdS$_3$ theory \cite{Montero:2016tif}.

A 2d CFT has a partition function of the form:\footnote{Here and below, we use $\tilde \mu$ rather than the oft-used $\bar \mu$, as a reminder that $\mu$ and $\tilde{\mu}$ are independent variables.  Indeed, in many cases, such as the worldsheet CFT of heterotic string theory, the number of left-moving and right-moving currents is different, so there is no way to identify the chemical potentials in complex conjugate pairs. }
\be
Z (\mu, \tilde{\mu} ; \tau, \bar{\tau}) \equiv \Tr (q^{\Delta} 
   \bar{q}^{\tilde{\Delta}} y^Q  \tilde{y}^{\tilde{Q}}) \,,
\ee
where $\Delta = L_0 - \frac{c}{24}$, $\tilde{\Delta} = \tilde{L}_0 -
\frac{\tilde{c}}{24}$, $Q$ and $\tilde{Q}$ are the charges carried by
left/right movers under a conserved current, $q = \e^{2 \pi i \tau}$, $y = \e^{2 \pi i \mu}$, and $\tilde y = \e^{2 \pi i \tilde \mu}$. The partition function satisfies
\begin{equation} \label{eqn:rhoperiod}
  Z (\mu + \rho) = Z (\mu) \;, \qquad \forall \rho \in \Gamma_Q^{\ast} \,,
\end{equation}
where $\Gamma_Q^{\ast} = \{ (\rho, \tilde{\rho}) | \rho Q - \tilde{\rho}
\tilde{Q} \in \bb{Z} \}$ is the dual lattice to the charge lattice.

Modular transformations form a group $SL(2, \mathbb{Z})$, generated by the T-transformation $\tau \rightarrow \tau + 1$ and the S-transformation $\tau \rightarrow -1/\tau$.
Following e.g.~\cite{Benjamin:2016fhe}, modular transformations act on the partition function as
\begin{align}
  Z (\mu ; \tau + 1) &= Z (\mu ; \tau) \,,  &
  Z (\mu / \tau ; - 1 / \tau) &=  \e^{\pi i k \frac{\mu^2}{\tau} - \pi i
  \tilde{k}  \frac{\tilde{\mu}^2}{\bar{\tau}}} Z (\mu ; \tau) \,. \label{eqn:SL2Z}
\end{align}
Here $k, \tilde{k}$ are related to the leading term in the current-current
OPE:
\begin{equation}
  J_L (z) J_L (0) \sim \frac{k}{z^2} + \ldots \;, \qquad J_R (\bar{z}) J_R (0)
  \sim \frac{\tilde{k}}{\bar{z}^2} + \ldots \;.
\end{equation}
Unitarity implies that $k, \tilde{k}$ are
non-negative, and positive for non-trivial currents.
For the case of multiple currents, this becomes 
\begin{equation}
  J_L^a (z) J_L^b (0) \sim \frac{k^{a b}}{z^2} + \ldots \;, \qquad J_R^{\tilde{a}} (\bar{z}) J_R^{\tilde{b}} (0)
  \sim \frac{\tilde{k}^{\tilde{a} \tilde{b}}}{\bar{z}^2} + \ldots \;.
\end{equation}
$k^{a b}$ and $\tilde{k}^{a b}$ can be thought of as metrics, which raise and lower indices and define inner products. Thus we may write $\mu \cdot \rho \equiv \mu_a k^{a b}
\rho_b$, $\mu \cdot Q \equiv \mu_a Q^a$, and $\tilde{Q}^2 \equiv \tilde{Q}^{\tilde{a}} \tilde{k}^{- 1}_{\tilde{a} \tilde{b}} \tilde{Q}^{\tilde{b}}$.

Next, we combine the the periodicity condition~(\ref{eqn:rhoperiod}) with the S-duality transformation~(\ref{eqn:SL2Z}), which implies
\begin{align} \label{eqn:Zquasiperiod}
 & Z (\mu + \tau \rho ; \tau) =
 \exp\biggl[- 2 \pi i (\mu \cdot \rho) - \pi i  \tau \rho^2 +
  2 \pi i  (\tilde{\mu} \cdot \tilde{\rho}) + \pi i \tau \tilde{\rho}^2
  \biggr] Z (\mu ; \tau) \,.
\end{align}

The partition function encodes the spectrum of the theory, which means that the
quasi-period $\mu \rightarrow \mu + \tau \rho$ must map the spectrum to
itself. This occurs thanks to a rearrangement of simultaneous changes in charge and conformal weight, a phenomenon known as ``spectral flow'' \cite{Schwimmer:1986mf}. To describe this, we define:
\begin{equation}
  T \equiv \Delta - \frac{1}{2} Q^2 \,, ~~\qquad \tilde{T} \equiv
  \tilde{\Delta} - \frac{1}{2}  \tilde{Q}^2 \,,
\end{equation}
which allows us to rewrite~(\ref{eqn:Zquasiperiod}) as:
\begin{align}
  Z &= \Tr \left( q^{T + \frac{1}{2} Q^2}  \bar{q}^{\tilde{T} + \frac{1}{2} 
  \tilde{Q}^2} y^Q  \tilde{y}^{\tilde{Q}} \right) 
  = \Tr \left( q^{T +
  \frac{1}{2}  (Q + \rho)^2}  \bar{q}^{\tilde{T} + \frac{1}{2}  (\tilde{Q} +
  \tilde{\rho})^2} y^{Q + \rho}  \tilde{y}^{\tilde{Q} + \tilde{\rho}} \right) \; ,
  \label{Zrew}
\end{align}
where we have introduced the shorthand notation,
\begin{align}
y^Q \equiv \exp[ 2 \pi i \mu_a Q^a] \,,~~~~~~  \tilde{y}^Q \equiv \exp[- 2 \pi i \tilde{\mu}_a \tilde{Q}^a] \,.
\end{align}
By expanding the traces in \eqref{Zrew} in powers of $Q$ and matching the first and second lines of the equation, we find that the spectrum must be invariant under
\begin{equation}
  Q \rightarrow Q + \rho \,, \qquad ~~ \tilde{Q} \rightarrow \tilde{Q} +
  \tilde{\rho} \,,
  \label{Qtransform}
\end{equation}
with $T$ and $\tilde{T}$ held fixed. This implies 
\begin{equation}
  \Gamma_Q^{\ast} \subseteq \Gamma_Q \,,
\end{equation}
i.e., the dual lattice $\Gamma_Q^{\ast}$ is a sublattice of the charge lattice $\Gamma_Q$. From here, beginning from the graviton, which has $\Delta = \tilde \Delta = 0$, $Q = \tilde Q = 0$, we use invariance of the spectrum under the transformation \eqref{Qtransform} to deduce the existence of a state with
\begin{equation}
\Delta = \tilde \Delta = \frac{\alpha'}{4} m^2 \leq \text{max} \left( \frac{1}{2} Q^2 , \frac{1}{2} \tilde Q^2 \right)\,,
\end{equation}
for all $Q \in   \Gamma_Q^{\ast}$. One can show that these states are superextremal \cite{BenMatteoProof}, which means that the sublattice $\Gamma_Q^{\ast} $ is entirely populated by superextremal particles, and the sublattice WGC is satisfied.

\subsection{Relation to the Swampland Distance Conjecture}\label{RELATION}

\cite{Ooguri:2006in}~introduced a number of Swampland conjectures regarding the moduli space $\mathcal{M}$ of a consistent theory of quantum gravity. First, they conjectured that such a moduli space is parametrized by vacuum expectation values of massless scalar fields. Second, they conjectured that such a moduli space has an infinite diameter--there exist points at arbitrarily large geodesic distance. Third, they introduced what is now known as the Swampland Distance Conjecture (SDC) (also see~\cite{Klaewer:2016kiy}):
\vspace{.2cm}
           \begin{namedconjecture}[The Swampland Distance Conjecture (SDC)]
Compared to the theory at some point $p_0 \in \mathcal{M}$, the theory at a point $p \in \mathcal{M}$ has an infinite tower of particles, each with mass scaling as
\begin{equation}
m \sim \exp( -\lambda d(p, p_0) )\,,
\end{equation}
where $d(p, p_0)$ is the geodesic distance in $\mathcal{M}$ between $p$ and $p_0$, and $\lambda$ is some order-one number in Planck units.
            \end{namedconjecture}
    \vspace{.1cm}
\noindent
Consequently, in the infinite distance limit $d(p, p_0) \rightarrow \infty$, an infinite tower of states becomes light.

The appearance of an infinite tower of light particles is reminiscent of the tower WGC, as well as the tower RFC, both of which require that an infinite tower of particles become light in the limit of vanishing gauge coupling. Indeed, in many contexts, these conjectures will be satisfied simultaneously by the same tower of particles. In 4d theories with $\mathcal{N}=2$ supersymmetry, for instance, \cite{Gendler:2020dfp} proved that in any infinite distance limit with a vanishing gauge coupling, there exists an infinite tower of charged particles satisfying
\begin{equation}
\left( \frac{q}{m} \right)^2  \geq \left( \frac{q}{m} \right)^2 \Big |_{\text{ext}} = \frac{1}{2} \frac{1}{M_4^2} + g^{ij} \mu_i \mu_j\,,
\end{equation}
where $\mu_i$ is the scalar ``charge'' of the particle with respect to the $i$th massless scalar field, and $g^{ij}$ is the inverse metric on moduli space. The equality on the right-hand side implies an equivalence between the WGC bound and the RFC bound for these particles: they are superextremal precisely when they are self-repulsive. This in turn implies $g^{ij} \mu_i \mu_j = \alpha^2 m^2$ for some $O(1)$ constant $\alpha$. If the moduli space is one-dimensional, this implies that these states must also satisfy the SDC, with $\lambda = \alpha$. If the moduli space has dimension greater than one, the relationship between $\lambda$ and $\alpha$ is more complicated, as it depends on the path taken. However, in a number of examples considered in \cite{Gendler:2020dfp}, the tower satisfying the tower WGC also satisfies the SDC for some $O(1)$ constant $\lambda$.

It is natural to expect that the towers of particles stipulated by the tower WGC and the SDC should agree whenever a point in moduli space of vanishing gauge coupling is also at infinite distance. It has been shown that all infinite distance limits in K\"ahler moduli space in 4d/5d supergravity theories descending from type II/M-theory on a Calabi-Yau threefold are associated with vanishing gauge coupling \cite{Corvilain:2018lgw, Heidenreich:2020ptx}. More generally, it has been conjectured that every infinite distance limit in moduli space should correspond to the vanishing of some $p$-form gauge coupling \cite{Gendler:2020dfp}, and more concretely, it has been conjectured that every infinite distance limit should correspond either to a decompactification limit (at which point a tower of KK modes becomes light) or to a tensionless string limit \cite{Lee:2019wij}. This latter conjecture goes by the name of the ``Emergent String Proposal,'' and if true, it suggests that either a 1-form gauge field or 2-form gauge field (or both) must become weakly coupled at any infinite distance limit in moduli space. Further evidence for this conjecture in the context of 4d $\mathcal{N}=1$ string compactifications was provided in~\cite{Lanza:2020qmt, Lanza:2021udy}. The authors of that work postulated the closely related ``Distant Axionic String Conjecture,'' that any infinite distance limit of a 4d $\mathcal{N}=1$ quantum gravity theory corresponds to the tensionless limit of a string which is charged under some 2-form gauge field. 

Conversely, in compactifications of M-theory to 5d, it has been shown that every point of vanishing gauge coupling is at infinite distance in moduli space \cite{Heidenreich:2020ptx}. An analogous statement holds in the context of AdS/CFT: in any SCFT in $d > 2$ dimensions, every point on the conformal manifold at which some sector of the theory becomes free is at infinite distance in the Zamolodchikov metric \cite{Perlmutter:2020buo}. This translates to the statement that the vanishing of a gauge coupling must occur at infinite distance in the moduli space of the bulk AdS dual theory. Additional progress toward classifying infinite distance limits in string compactifications can be found in e.g.~\cite{Grimm:2018ohb, Grimm:2018cpv, Marchesano:2019ifh, Lee:2019xtm, Grimm:2019wtx, Grimm:2019ixq, Baume:2020dqd, Alvarez-Garcia:2021pxo, Lee:2021usk}.

In the case of infinite distance, weak coupling limits, therefore, the WGC and SDC can be essentially unified. More general, qualitative arguments can be given in support of some sort of unification between these two conjectures: we will elaborate on these arguments in the following section.

\section{Qualitative Arguments for the WGC}\label{QUALITATIVE}

%!TEX root = WGC_Review_arxiv_v2.tex

Having introduced various versions of the WGC and given some ``empirical''  evidence for them, we now turn to the question of why any of them might be true.  In our view the most compelling arguments of this type are those which are more qualitative, not attempting to reproduce some version of the WGC in detail but instead giving some intuition for why a statement of this type should hold.  In this section we will review several such arguments, and then in the following section we will review attempts at a more precise derivation.

\subsection{Emergence}\label{ssec:emergence}
It has long been suspected that spacetime itself must be emergent in any theory of quantum gravity which is nontrivial enough to have some kind of black hole thermodynamics.  One simple argument in this direction is that the Bekenstein-Hawking formula
$$S=\frac{\mathrm{Area}}{4G}$$
tells us that the maximal entropy in a region of spacetime scales only like the surface area of the region, which is different from the volume scaling we have in quantum field theory \cite{tHooft:1993dmi,Susskind:1994vu,Bousso:2002ju}.  This idea is concretely realized in the AdS/CFT correspondence, which formulates quantum gravity in asymptotically-AdS spacetime as the quantum mechanics of a dual CFT living on the asymptotic boundary \cite{Maldacena:1997re}. In \cite{Harlow:2015lma} it was observed that this emergence can be used to motivate a qualitative version of the WGC.  The basic idea is that if spacetime itself is emergent, then surely any bulk gauge fields must also be emergent.  It is impossible however to have an emergent gauge field without the presence of charged particles, and moreover these charged particles cannot be too heavy.  

\begin{figure}
    \centering
    \includegraphics[height=4.5cm]{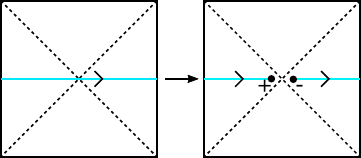}
    \caption{A Wilson line threading an AdS wormhole.  In order for such an operator to respect factorization, we need to be able to split it into a product of ``left'' and ``right'' pieces, each consisting of a partial Wilson line ending on a charged operator.}
    \label{fig:wilson}
\end{figure}

More concretely, we can consider the maximally extended AdS-Schwarzschild geometry, which in AdS/CFT is dual to the thermofield double state
\be
|TFD\rangle\equiv \frac{1}{\sqrt{Z(\beta)}}\sum_i \mathrm{e}^{-\beta E_i/2}|i^*\rangle_L|i\rangle_R
\ee
that lives in the tensor product Hilbert space
\be\label{eq:TP}
\mathcal{H}=\mathcal{H}_L\otimes \mathcal{H}_R
\ee
of two copies of the CFT on a spatial sphere \cite{Maldacena:2001kr}.  In the gravity picture this geometry describes a spatial wormhole connecting two AdS boundaries.  If there is a $U(1)$ gauge field in the gravity description, we therefore can have Wilson line operators 
$$ W_{LR}=\mathrm{e}^{i\int_{c_{LR}}A}$$
which are integrated on a curve $c_{LR}$ which connects the two boundaries through the wormhole (see figure \ref{fig:wilson}).  These Wilson lines however are somewhat puzzling from the point of view of the tensor product Hilbert space \eqref{eq:TP}: in a tensor product Hilbert space every operator can be written as a sum of product operators, but $W_{LR}$ does not seem to have such a decomposition.  Indeed if we try to view it as a product of two ``half-Wilson lines,'' these parts are not gauge-invariant and thus should not act on the physical Hilbert space \eqref{eq:TP}.  The way out of this puzzle is that if charged objects exist, then we can indeed split the Wilson line into two gauge-invariant operators, each of which consists of a Wilson line connecting an asymptotic boundary to a charged operator (see figure \ref{fig:wilson}).   

So far this is only an argument for some version of the completeness hypothesis, but we have not yet really used the emergence of the gauge field.  The idea of \cite{Harlow:2015lma} is as follows: since the gauge field is emergent, there must be some scale $\Lambda_{U(1)}\leq \Lambda_\mathrm{QG}$ at which the coefficient of the Maxwell term in the Wilson action flows to zero.  Here $\Lambda_\mathrm{QG}$ is the scale at which gravity becomes strongly coupled, which in general can be much less than $\Mp$ if there are many degrees of freedom.  If the infrared value of the Maxwell coupling $e$ is small, then it must undergo a substantial RG flow between the scale $\Lambda_{U(1)}$ where it is large and the mass scale $m$ of the lightest charge particles, since below the scale $m$ the gauge coupling can no longer flow.  Therefore there can be a small infrared gauge coupling if and only if there are light charged particles, which is the essence of the stronger versions of the WGC.  Quantifying this argument in general is difficult, but it can be illustrated in concrete models of an emergent $U(1)$ gauge field.  For example the $\mathbb{CP}^{N-1}$ $\sigma$-model is a theory of $N$ complex scalar fields $z_a(x)$ obeying the constraint $\sum_a z_a^*z_a=1$ and interacting with Lagrangian 
\be
\mathcal{L}=-\frac{N}{g^2}\left(D^\mu z\right)^\dagger D_\mu z,
\ee
where the ``gauge field'' $A_\mu$ appearing in the covariant derivative $D_\mu=\partial_\mu-iA_\mu$ is given by
\be
A_\mu \equiv \frac{1}{2i}\left(z^\dagger \partial_\mu z-\partial_\mu z^\dagger z\right).
\ee
This theory is renormalizable for $D=2$, while for general $D$ it can be understood as a lattice model with cutoff energy $\Lambda_{U(1)}$.  Either way, there is a critical value of $g$ near which (for large $N$) its infrared description is as $N$ charged scalars of mass $m$ interacting via Maxwell interactions of strength \cite{Harlow:2015lma}
\be\label{CPe}
\frac{1}{e^2}=\begin{cases}\frac{N}{6\pi m^2} & D=2\\ \frac{N}{12\pi m} & D=3\\ \frac{N}{12\pi^2}\log \left(\frac{\Lambda_{U(1)}}{m}\right) & D=4\\
N \Lambda_{U(1)}^{D-4} & D>4\end{cases} 
\ee
If we now couple the model to gravity, in order for it to work we need $\Lambda_{U(1)}\leq \Lambda_\mathrm{QG}$, where $\Lambda_\mathrm{QG}$ is the scale where gravity becomes strongly coupled.  In a large $N$ theory this is related to the infrared Newton contant $G$ by (see e.g.~\cite{Veneziano:2001ah,ArkaniHamed:2005yv,Distler:2005hi,Dimopoulos:2005ac,Dvali:2007hz,Dvali:2007wp,Kaplan:2019soo})
\be\label{CPG}
\frac{1}{G}\sim \frac{1}{G_{\mathrm{bare}}}+N \Lambda_\mathrm{QG}^{D-2}.
\ee
In order for the massive scalars to be qualitatively superextremal in the sense of \eqref{eq:WGC} we need to have
\be
\frac{1}{e^2}\lesssim \frac{1}{m^2G},
\ee
which indeed follows from \eqref{CPe}, \eqref{CPG} together with $\Lambda_{U(1)}\leq \Lambda_\mathrm{QG}$ and $G_\mathrm{bare}>0$. 

There are a few things to note about this argument. First of all, it is \textit{not} sufficient just to have one very heavy super-extremal particle.  In order to break the Wilson line in the unit charge representation as in figure \ref{fig:wilson}, we need to have objects of unit charge.   This thus gives some qualitative support to the stronger versions of the WGC such as the tower and sublattice WGCs.  Secondly one might worry that this verification of the WGC is accidental, since only a few parameters were involved.  In fact it is quite robust. In particular one can consider $M$ copies of the $\mathbb{CP}^{N-1}$ model, each with different values of $N$ and $m$, and a similar argument shows that the convex hull version of the multiple-$U(1)$ WGC we discussed in Section \ref{ssec:CHC} holds throughout a large parameter space of theories \cite{Harlow:2018tng}.  Moreover one can also show that the ``Grassmannian'' generalization of the model, which flows to an $SU(N)$ gauge theory in the infrared, has objects in the fundamental representation that obey the non-Abelian WGC discussed in Section \ref{ssec:nonabelian}.

We can develop this idea further to make it less dependent on the details of the $\mathbb{CP}^{N-1}$ model.  Specializing for convenience to $D=4$, at one loop order the gauge coupling $e_{\rm UV}$ at an energy scale $\Lambda_{\rm UV}$ is related to the low-energy gauge coupling $e$ according to:
\begin{equation}
\frac{1}{e_{\rm UV}^2} = \frac{1}{e^2} - \sum_i \frac{b_i}{8\pi^2} q_i^2 \log \frac{\Lambda_{\rm UV}}{m_i}. \label{eq:Landaupole}
\end{equation}
Here, $m_i$ and $q_i$ are the mass and charge of the particles in the tower and $b_i$ a beta function coefficient. For $\Lambda_{\rm UV}$ sufficiently large, the right-hand side of this equation vanishes, and correspondingly $e_{\rm UV}$ diverges: this is the well-known Landau pole of $U(1)$ gauge theory coupled to charged matter. The energy scale $\Lambda_{U(1)}$ of the Landau pole thus represents a UV cutoff on the $U(1)$ gauge theory.  Gravity has a similar UV cutoff $\Lambda_{\rm QG}$, which can be thought of as a result of divergent 1-loop corrections to the Einstein-Hilbert term.  This is described by \eqref{CPG}, which for $D=4$ implies that
\begin{equation}
M_{\rm Pl}^2 \gtrsim N_{\rm d.o.f.} \Lambda_{\rm QG}^2.
   \label{eq:4dspeciesbound}
\end{equation}
The energy scale $\Lambda_{\rm QG}$ is the energy scale at which quantum effects significantly modify gravity. It is also sometimes referred to as the ``species bound'' scale, as it scales inversely with the number of light particle species $N_{\rm d.o.f.}$ in the theory.

Now suppose our theory has a superextremal particle of each integer charge, so that it satisfies the sublattice WGC (and in fact, the lattice WGC). Then, the number of particles below a mass scale $\Lambda$ is given by $N(\Lambda) \geq \Lambda/(e M_{\rm Pl})$, so the species bound satisfies
\begin{equation}
M_{\rm Pl}^2 \gtrsim N(\Lambda_{\rm QG}) \Lambda_{\rm QG}^2 \geq \frac{\Lambda_{\rm QG}}{e M_{\rm Pl}} \Lambda_{\rm QG}^2 \,,
\end{equation}
or equivalently,
\begin{equation}
\Lambda_{\rm QG} \lesssim e^{1/3} M_{\rm Pl}.
\end{equation}
We see that in the weak coupling limit $e \rightarrow 0$, the species bound scale tends to zero and effective field theory breaks down due to the tower of superextremal particles.

This tower of charged particles also affects the Landau pole of the gauge theory, $\Lambda_{U(1)}$. Treating the logarithms and numerical prefactors as parametrically order one, the gauge coupling $e_{\rm UV}$ in \eqref{eq:Landaupole} diverges when
\begin{equation}
\frac{1}{e^2} \sim \sum_{q = 1}^{Q} q^2 \sim Q^3,
\end{equation}
where $Q \sim \Lambda_{U(1)}/(e M_{\rm Pl})$ is the largest charge in the tower. Again, this leads to the conclusion
\begin{equation}
\Lambda_{U(1)} \sim e^{1/3} M_{\rm Pl}.  \label{eq:4dLandauBound}
\end{equation}
Thus we see that the tower of superextremal particles leads to UV cutoffs on both gauge theory and gravity. Moreover, for the simple spectrum of charged particles we have considered here, the UV cutoffs for gauge theory and gravity are at parametrically the same energy scale, namely $\Lambda_{\rm QG} \sim e^{1/3} M_{\rm Pl}$.\footnote{Intriguingly, the same parametric cutoff has appeared in a quite different EFT context involving photons with a Stueckelberg mass~\cite{Craig:2019zkf}. It would be very interesting to understand whether this is a coincidence or something deeper.} In a sense, gauge theory and gravity are ``unified'' at this energy scale, as both of them emerge in the infrared from a strongly coupled theory at the energy scale $\Lambda_{\rm QG}$ by integrating out a tower of charged states.

Conversely, let us now assume that the gauge theory becomes strongly coupled at or below the energy scale $\Lambda_{\rm QG}$:
\begin{equation}
\frac{1}{e^2} \sim  \sum_{i | m_i < \Lambda_{\rm gauge}} q_i^2\,, \qquad \mbox{for $\Lambda_{\rm gauge} \lesssim \Lambda_{\rm QG}$,}  \label{eq:strongatLambda}
\end{equation}
where again we are ignoring $O(1)$ factors. We may rewrite this in terms of the average charge-squared $\langle q^2 \rangle_\Lambda$ of the particles with mass below $\Lambda$ as
\begin{align}
\frac{1}{e^2} \sim N(\Lambda_{\rm gauge})\, \langle q^2 \rangle_{\Lambda_{\rm gauge}} 
\lesssim \frac{1}{\Lambda_{\rm gauge}^2} M_{\rm Pl}^{2} \langle q^2 \rangle_{\Lambda_{\rm gauge}} \,,
\end{align}
where we have used the definition of the species bound \eqref{eq:4dspeciesbound}. Finally, we may rearrange this result in a form reminiscent of the WGC bound:
\begin{align}
\Lambda_{\rm gauge}^2 &\lesssim e^2 \langle q^2\rangle_{\Lambda_{\rm gauge}} M_{\rm Pl}^{2}. \label{eqn:averagechargebound}
\end{align}
Since all of the particles contributing to $\langle q^2\rangle_{\Lambda_{\rm gauge}}$ have a mass below $\Lambda_{\rm gauge}$, we see that, in a sense, the ``average'' charged particle in the theory is superextremal. This is not quite the same as the condition that the theory satisfies the Tower WGC, but it points in that direction.

We have considered here only one very simple case: $U(1)$ gauge theory in four dimensions with a single superextremal particle of each integer charge. However, as shown in \cite{Heidenreich:2017sim}, this phenomenon of gauge-gravity unification generalizes to theories in $d \geq 4$ spacetime dimensions, multiple $U(1)$'s, nonabelian gauge groups satisfying the sublattice WGC for nonabelian WGC (see Section \ref{ssec:nonabelian}), theories that satisfy the tower and sublattice WGCs but not the lattice WGC (provided the tower of superextremal states is not too sparse), and theories with a more general density of states (provided the density of states is sufficiently well-behaved). In this wide array of theories, gauge theory and gravity become strongly coupled at parametrically the same energy scale $\Lambda_{\rm QG}$. Conversely, demanding that gauge theory and gravity become strongly coupled at parametrically the same energy scale implies that, in the same sense as \eqref{eqn:averagechargebound}, the ``average'' particle should satisfy the WGC bound.

Finally, let us remark that a similar emergence picture applies to scalar field theories that satisfy the Swampland Distance Conjecture (SDC): just as loop effects from a tower of superextremal particles lead to a strongly coupled gauge theory at the scale $\Lambda_{\rm QG}$, loop effects from a tower of particles satisfying the SDC lead to a strongly coupled scalar field theory at the scale $\Lambda_{\rm QG}$ \cite{Grimm:2018ohb, Heidenreich:2018kpg}. This concept of emergence thereby unifies the SDC and the sublattice WGC: indeed, in many cases, the tower of particles that satisfies the sublattice WGC also satisfies the SDC, and integrating out this tower of particles produces both a weakly coupled gauge theory and a weakly coupled scalar field theory in the IR.

\subsection{No approximate global symmetries}
\label{subsec:noapprox}

As is familiar from a first-year course on electromagnetism, Gauss's law holds that the total electric flux through a closed two-dimensional surface $S$ is equal to the charge enclosed. In particular, the size and shape of the surface is irrelevant: the surface may be continuously deformed, and the total electric flux will not change provided the charge enclosed remains constant.

The modern notion of a higher-form global symmetry offers another perspective on this scenario. The fact that such deformations of the surface $S$ do not affect the total electric flux through it signals the existence of a family of topological surface operators in the theory, which are labeled by an angle $\alpha \in [0, 2\pi)$ and given by the exponentiated electric flux integral,
\begin{equation}
U_\alpha(S) = \exp \left(  i \frac{\alpha}{e^2} \oint_S \star F  \right)\,,~~~\alpha \in [ 0, 2 \pi )
\end{equation}
Such a surface operator signals the existence of a 1-form global symmetry, which is associated with a conserved charge: namely, the electric flux through $S$ counts the charge of any probe particles contained in such a surface. The conserved Noether current associated with this symmetry is given by the electric flux density $J=\frac{1}{e^2} \star F$.

This symmetry is broken in the presence of dynamical charged particles, which screen the charge of a probe particle. At long distances, however, the charge is approximately conserved: the divergence of the Noether current $\partial_\mu F^{\mu\nu}$ is small, and the flux through a closed surface enclosing a probe particle has only weak dependence on the size of the surface. This is encoded by the effective electromagnetic potential in QED at distances large compared to the mass of the electron, also known as the Uehling potential: \cite{Uehling:1935uj}
\begin{equation}
V(r) = \frac{-e^2}{4 \pi r} \left( 1 + \frac{e^2}{16 \pi^{3/2}} \frac{{\rm e}^{-2 m r}}{ (m r)^{3/2}} + ... \right)\,,~~~r m \gg 1.
\label{eq:largem}
\end{equation}
Here, $e$ is the renormalized coupling constant in the IR. We see that corrections to the leading order Coulomb potential are exponentially suppressed at long distances, and the charge $\oint_{S^2(r)} \star F \propto \oint_{S^2(r)} V'(r)$ is approximately conserved.

At distances $r \sim 1/m$, on the other hand, one starts to penetrate the polarization cloud and see the bare charge. The gauge coupling runs logarithmically, and the corrections to the effective Coulomb potential from the electron are $O(e^2)$.

More generally, corrections to the Coulomb potential at a distance scale $r = 1/\Lambda$ from a tower of charged particles are given roughly by
\begin{equation}
\Pi(\Lambda^2)  = \sum_{i | m_i < \Lambda} e^2 q_i^2 \,.
\end{equation}
This is the same expression we saw above in our discussion of emergence, so the corrections become $O(1)$ precisely when the gauge theory becomes strongly coupled. We showed there that a $U(1)$ gauge theory satisfying the tower/sublattice WGCs will become strongly coupled at the scale $\Lambda_{\rm QG}$ at which gravity becomes strongly coupled, so by the same token the approximate 1-form symmetry of such a gauge theory will be badly broken at $\Lambda_{\rm QG}$. This gives us a new intuitive understanding of the tower/sublattice WGCs: these conjectures are intimately tied to the absence of global symmetries in quantum gravity---including higher-form and approximate global symmetries.

Other versions of the WGC, including the magnetic version and the 0-form version, can be similarly related to the absence of approximate global symmetries in quantum gravity \cite{CordovaOhmoriRudelius}.

\subsection{Axion strings}

As a final qualitative check, we review an argument for the WGC in the presence of Chern-Simons terms \cite{Heidenreich:2021yda}. This argument is somewhat circular from the point of view of establishing the WGC, as it assumes the WGC for axions and charged strings in order to prove the ordinary WGC for charged particles. Nonetheless, it demonstrates an important phenomenon: in the presence of Chern-Simons terms involving multiple gauge fields, the WGC bounds for these different gauge fields are ``mixed up'' with one another. This offers a bottom-up criterion for determining when the tower of superextremal particles demanded by the tower/sublattice WGCs are modes of some fundamental string, which aligns with recent work (reviewed previously in in \S\ref{RELATION}) examining emergent strings in infinite distance limits \cite{Lee:2019wij, Lanza:2021udy}.

The argument relies on five simple assumptions. First, we assume a 4d theory of axion electrodynamics, in which an axion couples to the gauge field via a $\theta F \wedge F$ Chern-Simons coupling:
\begin{equation}
S = \int \Big[ -\frac{1}{2 g^2} F \wedge \star F - \frac{1}{2} f_\theta^2 \rmd \theta \wedge \star \rmd \theta + \frac{1}{8\pi^2} \theta F \wedge F \Big].
\end{equation}
Next, we assume the axion WGC:
\begin{equation}
    f_\theta S \lesssim M_{\textrm{Pl}},
    \label{eq:fAS}
\end{equation}
where $S$ is the instanton action. Third, we assume the WGC for a string of tension $T$ charged magnetically under the axion, also known as an axion string:
\begin{equation}
    T \lesssim 2\pi f_\theta M_{\textrm{Pl}}.
    \label{eq:TAS}
\end{equation}
Fourth, we assume that the instanton action takes the form
\begin{equation}
    S = \frac{8 \pi^2}{g^2}.
    \label{eq:SAS}
\end{equation}
This form of the instanton action is most familiar from Yang-Mills theory, but abelian gauge theories also feature instantons with actions of this type, in the form of monopole loops with dyonic winding~\cite{Fan:2021ntg}, as a consequence of the Witten effect~\cite{Witten:1979ey}. Finally, we assume that the axion $\theta$ is a fundamental axion, meaning that the core of the axion string probes physics in the deep ultraviolet. (For more on the distinction of fundamental vs.~non-fundamental strings, see  \cite{Reece:2018zvv} and also~\cite{Dolan:2017vmn}.)

From here, we may combine \eqref{eq:fAS}-\eqref{eq:SAS} to get a bound on the string scale of the axion string:
\begin{equation}
    M_\mathrm{str} := \sqrt{2 \pi T} \lesssim g M_\mathrm{Pl},
    \label{eq:wgcstring}
\end{equation}
which is precisely the WGC scale associated with the gauge field $A$.
Next, our assumption of the Chern-Simons coupling $\theta F \wedge F$ ensures that the higher-spin string excitations of the axion string carry charge under the gauge field $A$, which follows from anomaly inflow on the string worldsheet \cite{Callan:1984sa}. From \eqref{eq:wgcstring}, we learn that the excitations of the axion string satisfy the WGC (up to $O(1)$ factors).

Finally, invoking our assumption that the axion is a fundamental axion, we further conclude that there is a whole tower of string excitations. This establishes (up to $O(1)$ factors) not only the WGC, but also the Tower WGC for the gauge field $A$. Additionally, local quantum field theory breaks down at the axion string scale $M_\mathrm{str} \sim g M_{\textrm{Pl}}$, which for $g$ small is parametrically below the emergence energy scale $g^{1/3} M_{\textrm{Pl}}$ discussed in \S\ref{ssec:emergence}. This can have important consequences for phenomenology, in that it leads to a tension between effective field theories that require a very high energy scale and those which require a very small gauge coupling.

As noted above, this argument represents more of a consistency check on the WGC than an argument for it, as it assumes the WGC for axions and axion strings. Furthermore, note that it relies heavily on the presence of the Chern-Simons term: without this term, there is no reason for the form of the instanton action in \eqref{eq:SAS} to hold, and there is no guarantee that the excitations of the axion string will carry electric charge. Indeed, the circle compactification of a 5d gravity theory yields a Kaluza-Klein photon which does \emph{not} couple to the axion via a $\theta F \wedge F$ coupling, and consequently the Kaluza-Klein modes are not excitations of the axion string, and effective field theory breaks down at the larger scale $e_{\textrm{KK}}^{1/3} M_{4} = M_{5}$. For more details, see \cite{Heidenreich:2021yda}.

The bottom-up argument of this section coheres nicely with studies of infinite distance limits in string theory, discussed above in \S\ref{RELATION}. In particular, the Emergent String Conjecture \cite{Lee:2019wij} implies that every weak coupling limit should correspond to either an emergent string limit or a decompactification limit: we see here that these two cases are distinguished at low energies by the presence/absence of a Chern-Simons coupling. Relatedly, \cite{Lanza:2021udy} found that in a large class of 4d $\mathcal{N}=1$ string compactifications, any infinite distance limit yields a fundamental axion string whose tension scales with the mass of a tower of light particles as $T^w \sim m^2$, where $w=1$, $2$, or $3$. Here, we see that the case $w=1$ corresponds to the case where the light particles are charged and a $\theta F \wedge F$ coupling is present. The large-radius limit of a Kaluza-Klein compactification of minimal 5d supergravity, where there is no such Chern-Simons coupling involving the KK photon but there is one involving the 4d descendant of the 5d graviphoton, corresponds to the $w=3$ case.

So far, the argument we have sketched in this subsection is unique to four dimensions, as the relation \eqref{eq:SAS} does not have a well-known higher-dimensional parallel. However, supergravity constraints in higher dimensions impose similar relations, so that the argument of this subsection does admit higher-dimensional analogs within the supergravity context. For further details, see \cite{Heidenreich:2021yda} for the 5d case and \cite{mixing} for even higher-dimensional cases. 

\section{Attempted Derivations of the WGC}\label{DERIVATIONS}

%!TEX root = WGC_Review_arxiv_v2.tex

\subsection{The WGC from holography}
The weak gravity conjecture is a proposed restriction on non-perturbative quantum gravity, and thus it is natural to ask if we can show that it holds in the theories of non-perturbative quantum gravity we currently possess.  In particular we can ask if the WGC holds within AdS/CFT, which is our best-understood set of quantum gravity theories.  So far this has not been established, but a holographic argument for something closely related to the WGC was given by Montero in \cite{Montero:2018fns}.  In this subsection we will sketch this argument, as well as make a few related observations.

One of the main motivations for the WGC is the idea that near-extremal black holes in flat space should be unstable.  In AdS/CFT that would be a statement about ``small'' black holes, whose size is small compared to the AdS radius, but unfortunately small black holes are not understood so well in AdS/CFT.  Montero instead argues that \textit{large} near-extremal black holes in AdS must be unstable, as otherwise the thermofield double state of the dual CFT at large chemical potential and small temperature would have rather surprising (and likely impossible) entropic properties.  Unfortunately this argument does not amount to a proof of the WGC, as we will see that there are other ways these black holes could decay besides emitting charged particles obeying \eqref{eq:WGC}, but the argument is still quite suggestive, and we are optimistic that more could be learned from it.  

Charged AdS black holes which are very large compared to the AdS scale asymptotically become charged black branes, which in $D$ bulk Euclidean dimensions have gauge field
\be
A_\tau=\frac{i\rho}{(D-3)}\left(\frac{1}{r^{D-3}}-\frac{1}{r_+^{D-3}}\right)
\ee
and metric
\be
\rmd s^2=f(r)\rmd \tau^2+\frac{\rmd r^2}{f(r)}+r^2 \rmd \vec{x}_{D-2}^2,
\ee
with
\be
f(r)\equiv r^2-\frac{2\kappa^2\epsilon}{(D-2)r^{D-3}}+\frac{\kappa^2\rho^2}{(D-2)(D-3)r^{2D-6}}.
\ee
Here $\epsilon$ and $\rho$ are the boundary energy and charge densities respectively, $r_+$ is the largest positive zero of $f(r)$, and we have set the AdS radius to one.  If we work at fixed inverse temperature $\beta$ and chemical potential $\mu$, then we have
\begin{align}\nonumber
r_+&=\frac{2\pi}{(D-1)\beta}\left(1+\sqrt{1+\frac{(D-1)(D-3)^2\kappa^2\beta^2\mu^2}{4\pi^2(D-2)}}\right)\\\nonumber
\rho&=(D-3)r_+^{D-3}\mu\\
\epsilon&=\frac{(D-2)r_+^{D-3}}{2\kappa^2}\left(r_+^2+\frac{(D-3)\kappa^2\mu^2}{D-2}\right).
\end{align}
This black brane approaches extremality when $\kappa \beta \mu\gg 1$, with the extremal radius being given by
\be
r_+|_{\beta=\infty}\equiv r_e=\frac{(D-3)\kappa |\mu|}{\sqrt{(D-1)(D-2)}}.
\ee
At extremality the function $f(r)$ has a double zero at $r=r_e$, so the radial geodesic distance from $r_e$ to any large but finite radius $r_c$ is logarithmically divergent.

\begin{figure}
    \centering
    \includegraphics[height=5.5cm]{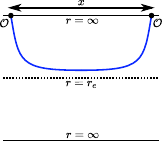}
    \vspace{0.2cm}
    \caption{Exponential decay of spatial correlators for the extremal black hole back brane.}
    \label{fig:geodesic}
\end{figure}
The main point of \cite{Montero:2018fns} is that if we study the Hartle-Hawking state of two such extremal black branes obtained by slicing the Euclidean path integral, there is a tension between two facts which apparently follow from the bulk picture together with the holographic correspondence:\footnote{It has gradually been understood that the semiclassical picture of the bulk needs to be used with some care at very low temperatures, as quantum effects eventually become important \cite{Preskill:1991tb,Maldacena:1998uz,Page:2000dk,Almheiri:2016fws,Iliesiu:2020qvm,Heydeman:2020hhw}.  It would be worthwhile to revisit this argument from the point of view of the modern understanding of these quantum effects via a dimensional reduction to Jackiw-Teitelboim gravity, as they could potentially change the conclusions, but we won't attempt it here.}
\begin{itemize}
    \item \textbf{Exponential correlators:} the fixed-time correlators of boundary operators decay exponentially with distance. 
    \item \textbf{Volume-law entanglement:} the union of a boundary subregion $A_R$ in the right CFT and the same subregion $A_L$ in the left CFT has a von Neumann entropy which grows like the volume of the subregion.
\end{itemize}
There are various ways to understand why the bulk picture implies these results. One nice way is based on the idea that black hole horizons are \textit{extremal surface barriers} \cite{Engelhardt:2013tra}.  What this means is that an extremal surface of any co-dimension greater than or equal to two cannot be smoothly deformed from a surface which does not cross a horizon to a surface which does.  The reason is simple: if such a deformation were possible, then at some point the surface would have to be tangent to the horizon, but then the extremality equations would imply that the surface would be entirely contained in the horizon.  In particular for Euclidean states such as the Hartle-Hawking state, we can approximate the two-point function of as massive field as
\be
\langle\mathcal{O}(x)\mathcal{O}(y)\rangle\sim \mathrm{e}^{-m |x-y|},
\ee
where $|x-y|$ is the geodesic distance between $x$ and $y$.  As shown in figure \ref{fig:geodesic}, the geodesic which is relevant for computing the correlator of two boundary fields necessarily has a length which grows like the boundary distance between the fields: the horizon at $r=r_e$ is an extremal surface barrier, so the geodesic has no choice but to involve a large extensive component which lies just outside the horizon.   Moreover any geodesics which cross the horizon must have infinite length, and therefore there is no correlation between boundary operators on opposite sides. 

\begin{figure}
    \centering
    \includegraphics[height=4cm]{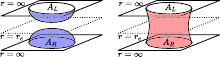}
    \caption{Competing minimal surfaces to compute the von Neumann entropy of $A_L\cup A_R$ in the thermofield double state.  At nonzero temperature the red ``area-law'' surface on the right always wins for large enough volume, but at zero temperature its area is infinite so the blue ``volume-law'' surface on the left wins for any region size.}
    \label{fig:extremal}
\end{figure}
The volume law entanglement can be understood along similar lines: we can compute the von Neumann entropy of the dual CFT on $A_L\cup A_R$ using the Ryu-Takayanagi formula, which tells us that it is given by the area of the minimal-area surface which is homologous to $A_L\cup A_R$.  There are two possible candidates for the RT surface (see figure \ref{fig:extremal}), only one of which gives a volume law, and at finite temperature the ``area-law surface'' eventually wins for big enough regions.  At zero temperature however the ``area-law surface'' always has infinite area due to the infinite distance to the horizon, and so the ``volume-law surface'' always wins.

The reason that exponential decay of correlators and volume-law entanglement are in tension is that the former suggests only short-range entanglement is present, while the latter requires long-range entanglement.  If the entropy of $A_R\cup A_L$ is growing like the volume, then since the total state is pure all this entanglement must be purified by something in the complementary region.  Such a purification is unlikely, given the exponential decay of correlation with distance.  This intuition has been formalized in $1+1$ dimensions into a precise theorem \cite{Hastings:2007iok}, and it quite plausibly holds in general.  

If the tension just described indeed constitutes a contradiction, then the only way out is for the thermodynamic description of the extremal black brane to break down.  There are two ways this can happen.  The first way is that there could be an exactly BPS particle, which turns out to lead to power-law correlators and thus removes the tension.  This is the situation which is realized for BPS branes in supersymmetric theories. The second possibility, which has so far been manifested for all non-supersymmetric extremal branes, is that there is some kind of matter present which causes the brane to be unstable.  Had we been discussing small black holes, such an instability would have immediately required the existence of  superextremal charged particle and thus given a derivation of some version of the WGC.  For large black holes however there are more possibilities for the instability.  Indeed this topic has a long history in the literature on applications of AdS/CFT to condensed matter physics, where the various possibilities go under the names ``holographic superconductor'' \cite{Hartnoll:2008vx, Gubser:2008px, Hartnoll:2008kx} or ``holographic Fermi surface'' \cite{Liu:2009dm, Faulkner:2009wj, Hartnoll:2009ns} depending on whether the particle causing the instability is a boson or a fermion.  The rough idea for the bosonic case is as follows: in the near-horizon region the gauge kinetic term 
\be
-(\nabla^\mu\phi-inA^\mu \phi)^\dagger(\nabla_\mu \phi-inA_\mu\phi)-m^2\phi^\dagger \phi
\ee
for a boson of charge $n$ leads to an effective mass
\be
m^2_\mathrm{eff}=m^2-\frac{n^2e^2}{\kappa^2}
\ee
in the near-horizon $AdS_2$ region.  This leads to an instability if $m^2_\mathrm{eff}$ violates the $AdS_2$ Breitenlohner-Freedman bound $m^2>-\frac{(D-1)(D-2)}{4}$, so in other words there is an instability for masses in the range
\be\label{eq:instability}
-\frac{(D-1)^2}{4}<m^2\leq \frac{n^2 e^2}{\kappa^2}-\frac{(D-1)(D-2)}{4}.
\ee
The first inequality here is the $D$-dimensional Breitenlohner-Freedman bound, which is necessary for the vacuum to be stable.  When $n\neq 0$ the first term on the right-hand side of \eqref{eq:instability} gives something like the WGC inequality, as first noticed in \cite{Denef:2009tp}, but it is missing the factor of $\gamma$.  And moreover due to the second term it is possible to have an instability even if $n=0$, so the brane can be unstable even if there are no charged particles at all!\footnote{In the absence of charged particles it is not so clear what the endpoint of this instability might be.  It would need to be inhomogeneous, since the homogeneous brane solution is unstable.  Likely the final endpoint would be theory-dependent.}  Thus any argument which requires an instability only of large extremal black holes in AdS is not sufficient to imply the validity of the WGC, although it is certainly suggestive.  Various other types of instability for this system have been discussed in the AdS/CMT literature, and the connection to the WGC was also discussed in \cite{Henriksson:2019ifu}.

\subsection{The WGC from thermodynamics}

\subsubsection{The WGC and quasinormal mode frequencies}

An interesting argument linking the WGC to a bound on the imaginary part of the frequencies of black hole quasinormal modes has been given in~\cite{Hod:2017uqc}. The argument relies on a ``Universal Relaxation Bound,'' previously proposed in \cite{Hod:2006jw}. To derive this bound, Hod begins with Bekenstein's bound on information transfer \cite{Bekenstein:1981zz} (which itself derives from Bekenstein's entropy bound \cite{Bekenstein1981}) and places an upper bound on the rate at which an observer can receive information:
\begin{equation}
    \dot{I} \leq \frac{\pi E }{ \log 2 }\,.
    \label{itb}
\end{equation}
Here $I$ is the information and $E$ is the energy of the package containing the information. Using the relations
\begin{equation}
    S = I \log 2\,,~~~\dot{S} = \frac{\Delta S}{\Delta \tau}\,,~~~T = \frac{\Delta E}{\Delta S}\,
\end{equation}
one may rewrite the bound as
\begin{equation}
    \Delta \tau \geq \frac{1}{\pi T}\,,
    \label{urb}
\end{equation}
which Hod interprets as a bound on the time $\Delta \tau$ over which a system can relax to equilibrium, deemed the ``Universal Relaxation Bound.'' Finally, he applies this bound to the quasinormal modes of a black hole by setting $T$ to be the temperature of the black hole and $\Delta \tau$ to be the inverse of the smallest imaginary part of a quasinormal mode frequency, $\Delta \tau = [\mathrm{Min}(\mathrm{Im}(\omega))]^{-1}$, arriving at the bound
\begin{equation}
    \mathrm{Min}[\mathrm{Im}(\omega)] \leq \pi T\,.
    \label{qnmurb}
\end{equation}

In response, it should be noted that while the derivation of Hod's bound \eqref{urb} follows rather straightforwardly from Bekenstein's bound \eqref{itb}, its interpretation as a universal bound on relaxation times is more suspect. Bekenstein derived his bound by imagining a scenario in which one observer sends a package full of information to another. It is not entirely clear how this scenario can be translated into the case of interest at hand, in which a black hole is perturbed and relaxes to equilibrium. A sharper derivation of the proposed Universal Relaxation Bound, particularly in the context of quasinormal mode frequencies, is clearly desirable. However, it should be noted that Hod and others have given both numerical and analytical evidence in favor of the bound \eqref{qnmurb} \cite{Hod:2006jw,Hod:2007tb, Gruzinov:2007ai}, and bounds similar to \eqref{urb} but without the precise $O(1)$ factors have been argued for in other contexts (see e.g. \cite{Lucas:2018wsc} and references therein).

Assuming~\eqref{qnmurb}, Hod argued for the WGC as follows. None of the quasinormal modes arising from gravitational and electromagnetic perturbations of a nearly-extremal Reissner-Nordstr\"om black hole obey the bound. However, if we assume that the bound merely requires that {\em some} mode in the black hole background obeys~\eqref{qnmurb}, then the bound could be satisfied by a quasinormal mode of a matter field. In particular, Hod showed analytically that, to leading order in $T$ in the extremal limit, a charged scalar field has a quasinormal mode which obeys the bound~\eqref{qnmurb} precisely when the scalar satisfies the WGC~\cite{Hod:2017uqc}. A similar conclusion was also obtained analytically in the asymptotically AdS$_2 \times S^2$ near-extremal, near-horizon limit~\cite{Urbano:2018kax}. These are intriguing results, which call for further study. Black holes that are far from extremality have modes that comfortably satisfy the bound. Near extremality, black hole quasinormal modes split into two families:  damped modes, which have $\mathrm{Im}(\omega)$ of order the inverse black hole radius, and zero-damped modes (ZDMs), which have $\mathrm{Im}(\omega) \to 0$ as $T \to 0$~\cite{Yang:2012pj}. Clearly, the bound~\eqref{qnmurb} can only be obeyed by a ZDM. Much of the literature on numerical computation of quasinormal modes focuses on the damped modes, whereas the ZDMs are less well-studied~\cite{Detweiler:1980gk, Yang:2012pj,Konoplya:2013rxa, Richartz:2014jla}. ZDMs for a Reissner-Nordstr\"om black hole have been found in the pure Einstein-Maxwell theory~\cite{Zimmerman:2015trm}, so the precise numerical coefficient in~\eqref{qnmurb} is important for the link to the WGC. Future work could check whether~\eqref{qnmurb} is equivalent to the WGC away from the $T \to 0$ limit. It would also be of interest to explore this correspondence for more general black holes (e.g., dilatonic black holes). Strong numerical evidence for a WGC/quasinormal mode connection would provide a motivation for further study of the quasinormal mode relaxation bound itself.

The inequalities~\eqref{urb} and~\eqref{qnmurb} bear a superficial resemblance to the well-studied chaos bound~\cite{Maldacena:2015waa}, which requires a Lyapunov exponent $\lambda \leq 2\pi T$. However, the relaxation rate and the rate of growth of chaos are not, in general, the same. For example, the Sachdev-Ye-Kitaev model saturates the chaos bound, but its thermal 2-point function falls exponentially with timescale $\tau = q/(2 \pi T)$~\cite{Maldacena:2016hyu}, where $q$ is a positive even integer. This is consistent with~\eqref{urb}, but does not saturate the bound, except when $q=2$.\footnote{BH thanks Zachary Fisher and Ziqi Yan for discussions on this point.}

\subsubsection{The WGC and entropy}

As discussed in Section~\ref{NGS}, one argument against continuous global symmetries is based on the existence of finite-mass black hole states of arbitrarily large global charge, leading to infinite entropy in a finite-size region, in violation of entropy bounds in quantum gravity. This has motivated studies relating the WGC to entropy bounds. For small but nonzero gauge coupling, a WGC-violating theory can have a very large (but finite) number of stable extremal black holes in a finite mass range. For example, it was suggested in~\cite{Banks:2006mm} that the uncertainty in a measurement of the charge of a black hole is of order $1/e$, leading to an entropy scaling as $\log(1/e)$ and eventually violating entropy bounds for sufficiently small $e$. However, it is unclear why one would not be able to measure charge more precisely than $1/e$ (e.g., by measuring the motion of charged particles in the long-range electric field outside the black hole), or why the Bekenstein-Hawking entropy should be the relevant bound for an ensemble with such a large range of possible charges. Furthermore, examples in which exactly stable BPS charged black holes exist (with moduli spaces such that $e$ can be made arbitrarily small) illustrate that the existence of many (marginally) stable species is not, in itself, in contradiction with quantum gravity.

Subsequent studies have examined logarithmic corrections to black hole entropy in the presence of WGC-violating matter~\cite{Shiu:2016weq, Shiu:2017toy,Fisher:2017dbc}. These corrections have interesting properties, but their computation has not led to an undisputed proof of the WGC. In particular,~\cite{Andriolo:2018lvp} claims that the argument of~\cite{Fisher:2017dbc} relies on applying a formula outside its regime of validity.

\subsection{The WGC from corrections to large black holes} \label{subsec:BHcorrection}

The extremality bound for black holes is derived from the two-derivative effective action. From the beginning, it was understood that the WGC could potentially be satisfied by large black holes when higher-derivative corrections to the effective action are taken into account~\cite{ArkaniHamed:2006dz}. Schematically, these modify the extremality bound to take the form $|Q|/M \geq (|Q|/M)|_{\mathrm{ext}} (1 + c / Q^2)$, where $c$ is a linear combination of Wilson coefficients of four-derivative operators and $(|Q|/M)|_{\mathrm{ext}}$ is the charge-to-mass ratio of {\em asymptotically large} extremal black holes, which is computed with the two-derivative action. We continue to define a superextremal state as one for which $|Q|/M \geq (|Q|/M)|_{\mathrm{ext}}$, so that finite-size black holes are superextremal when $c \geq 0$. The corrections to the extremality bound from general four-derivative operators added to Einstein-Maxwell theory were calculated in detail shortly afterward~\cite{Kats:2006xp}, and it was found that certain black holes in heterotic string compactifications do, in fact, become superextremal.\footnote{However, it should be noted that the results of~\cite{Giddings:1993wn,Natsuume:1994hd}, upon which~\cite{Kats:2006xp} relies, are obtained at \emph{string tree level}. Because the string coupling diverges at the horizon of the black holes in question, the string loop expansion may not be under control; see also~\cite{Cvetic:1995bj}.}

Before discussing the technical details, it is useful to describe two quite different interpretations that one might attach to the observation that small corrections to large black holes can allow them to become superextremal. The first is that this trivializes the Weak Gravity Conjecture. The WGC in its most mild form merely requires that {\em some state} in the theory be superextremal. If this state is a large black hole, then the WGC is simply a statement about the signs of some higher-dimension operators in the effective action, and does not imply the existence of any light charged particles below the Planck scale. If a general positivity proof can be constructed for the linear combination of operator coefficients appearing in the corrected extremality bound, the WGC will follow, and as such will be reduced to a statement about gravitational effective field theory. The second viewpoint is that the evidence that we have for  the WGC, as discussed in earlier sections of this review, favors the much stronger tower/sublattice WGCs, involving an infinite tower of charged particles of increasing charge and mass, all of which are superextremal. For very large values of $|Q|$, the charged ``single-particle states'' simply {\em are} black holes, and so the tower/sublattice WGCs require that they be superextremal (as depicted in Figure~\ref{fig:tower}). From this perspective, an EFT argument could explain superextremality far out in the charge lattice, but the tower/sublattice WGCs will also imply the existence of superextremal states at smaller $Q$,  where  the  states are  no longer well-described as black holes in EFT. Arguments in favor of superextremality from higher-derivative corrections cannot decisively favor the former perspective (that EFT is everything) over the latter (that Swampland constraints go beyond EFT). However, if we find consistent theories of quantum gravity (not just EFTs) in which large black holes are {\em subextremal}, this would immediately falsify the tower/sublattice WGCs.

\subsubsection{The corrected  extremality bound}
\label{subsubsec:correctedextremality}

There are several possible four-derivative operators  that may be  added to the Lagrangian of Einstein-Maxwell theory,  built out of $R_{\mu \nu \rho \sigma}$ and $F_{\mu  \nu}$. For this discussion, we work in the normalization $\frac{1}{2\kappa^2} R - \frac{1}{4}F_{\mu \nu}F^{\mu \nu}$ for the two-derivative Lagrangian. If we limit our  attention to CP-conserving terms, there are four independent physical four-derivative terms. Their  contribution to the effective action can be parametrized as
            \begin{align}  \label{eq:S4deriv}
            S_{4\partial} =  \int  \mathrm{d}^Dx \sqrt{-g}  \Big(c_{GB} O_{GB}+ c_{RF} R_{\mu \nu \rho \sigma} F^{\mu \nu} F^{\rho \sigma} 
              + c_{T}  T_{\mu \nu}T^{\mu \nu}  + c_F (F_{\mu \nu}F^{\mu \nu})^2\Big), 
            \end{align}
where $T_{\mu \nu}$ is the two-derivative Maxwell  stress  tensor $F_{\mu  \rho}F_{\nu}^{~\rho} -  \frac{1}{4}g_{\mu \nu} F_{\rho \sigma}F^{\rho \sigma}$,  and $O_{GB} = R^2 - 4R_{\mu \nu}R^{\mu \nu} + R^{\mu\nu\rho\sigma}R_{\mu\nu\rho\sigma}$ is the Gauss-Bonnet term (in $D=4$, this is  a topological term that does not  affect the extremality bound). This is the basis favored by the discussion in~\cite{Arkani-Hamed:2021ajd}. All other four-derivative  operators can be related to these  four terms (up to terms that are of higher order in the derivative expansion) via equations of motion (or, equivalently, field  redefinitions). For example, terms involving the Maxwell stress tensor $T_{\mu \nu}$ can be traded for terms involving $R_{\mu \nu}$ using the Einstein equations, while terms involving $\nabla_\mu F_{\rho \sigma}$ can be transposed via integration by parts into terms that vanish in pure Einstein-Maxwell theory as well as terms involving a commutator of two covariant derivatives, which can be eliminated in favor of the Riemann tensor.

The condition  for extremal Reissner-Nordstr{\"o}m black  holes to become (strictly) superextremal due to four-derivative terms is~\cite{Kats:2006xp}
\begin{align} \label{eq:4derivWGC}
(D-3)\left[(D-2)(D c_T + 16  c_F) +8(D-3)c_{RF} \kappa^2 \right]  
 -  4 (D-4)(3D-7)c_{GB} \kappa^4 > 0.
\end{align}
For $D=4$ this simplifies to $c_T +  4 c_F + c_{RF} \kappa^2 > 0$.\footnote{In the 4d case, stronger constraints can be obtained by considering dyonic black holes; see, e.g.,~\cite{Etheredge:2022rfl}.}

As discussed in~\cite{Charles:2019qqt}, another  convenient basis related to familiar anomalies is
            \begin{align}  \label{eq:S4derivAlt}
            {\tilde  S}_{4\partial} =  \int  \mathrm{d}^Dx \sqrt{-g} & \Big({\tilde c}_{W}  W_{\mu \nu \rho  \sigma}W^{\mu \nu \rho \sigma} + {\tilde c}_{RF} R_{\mu \nu \rho \sigma} F^{\mu \nu} F^{\rho \sigma} 
             +  {\tilde c}_{GB} O_{GB}  + {\tilde c}_F (F_{\mu \nu}F^{\mu \nu})^2\Big). 
            \end{align}
Here $W_{\mu\nu\rho\sigma}$ is the Weyl tensor. The  relationship between the bases~\eqref{eq:S4derivAlt} and~\eqref{eq:S4deriv} is
\begin{align}
{\tilde c}_F  = c_F +  c_T \frac{(D-4)^2}{16(D-1)},  \quad & {\tilde c}_W = \frac{c_T}{\kappa^4} \frac{D-2}{4(D-3)}, \nonumber \\
{\tilde c}_{GB} =  c_{GB} - \frac{c_T}{\kappa^4} \frac{D-2}{4(D-3)}, \quad & {\tilde c}_{RF} = c_{RF}.
\label{eq:S4derivBasisChange}
\end{align}
A more detailed discussion of the field redefinitions  that can convert between operator bases may be found in Appendix B  of~\cite{Cheung:2018cwt}.

Early work on this subject derived the condition~\eqref{eq:4derivWGC} by directly solving the modified equations of motion in the presence of higher-derivative operators and extracting the corrected extremality bound from the perturbed black hole solution. Recently, calculations have been greatly streamlined by the discovery of  elegant formulas relating the change in the extremality bound to integrals evaluated on the {\em uncorrected} black hole solution.
Specifically, the shift in the charge-to-mass ratio $\zeta = |Q|/\sqrt{\gamma} \kappa M$ of an extremal black hole away from $1$ is given by
\begin{equation}
\Delta \zeta = \frac{1}{M} \lim_{\zeta \to 1}\left. \int \rmd^{D-1} x\,N \sqrt{h}\Delta{\cal L}\right|_\text{two-deriv}, \label{eq:MassCorrFormula}
\end{equation}
where $\Delta{\cal L}$ consists of the higher-derivative corrections to the leading order Lagrangian, $N$ and $h$ are the lapse function and spatial metric associated to a fixed-time slice (extending from the horizon to infinity),
 and $|_\text{two-deriv}$ signals that the expression is to be evaluated on the two-derivative solution. 
 %Note that the integral is taken on a fixed-time slice extending from the horizon to infinity, where $g$ is the determinant of the \emph{spacetime} metric (not the determinant of 

Expressions of this form have been derived in multiple ways. One approach (used mainly for 4d Reissner-Nordstr\"om black holes) begins with the Wald entropy of the black hole~\cite{Wald:1993nt}, which is related, through standard thermodynamic arguments, to the Euclidean action evaluated on the solution. This receives corrections only from the corrections to the action evaluated on the uncorrected solution, since the evaluation of the uncorrected action on corrections to the solution vanishes at first order due to the extremality of the uncorrected action at an uncorrected solution~\cite{Cheung:2018cwt, Reall:2019sah}.

The correction to the extremal charge-to-mass ratio is then shown to be related to the change in the black hole entropy~\cite{Cheung:2018cwt}. This has been generalized to rotating and dyonic black holes~\cite{Cheung:2019cwi}, dilatonic black holes~\cite{Loges:2019jzs}, AdS black holes~\cite{Cremonini:2019wdk}, and dyonic Kaluza-Klein black holes~\cite{Cremonini:2020smy}; see further discussion in~\cite{Arkani-Hamed:2021ajd}. In fact, the extremality/entropy relationship has been proven by~\cite{Goon:2019faz} using very general thermodynamic considerations, which imply that when there is a minimal mass for a given charge, $M > M_\text{ext}({\vec Q})$, sensitive to a parameter $\epsilon$ (like the coefficient of a 4-derivative operator),
\begin{equation} \label{eq:GoonPenco}
    \frac{\partial M_\text{ext}({\vec Q},\epsilon)}{\partial \epsilon} = \lim_{M \to M_\text{ext}} \left[-T \left.\left(\frac{\partial S(M,{\vec Q},\epsilon)}{\partial \epsilon}\right)\right|_{M,{\vec Q}}\right],
\end{equation}
even outside the black hole context. Recently, similar results have been derived using the Iyer-Wald covariant phase space formalism \cite{Aalsma:2021qga} (also see~\cite{Aalsma:2020duv}).

Note that it is crucial that the partial derivative on the right-hand side of~\eqref{eq:GoonPenco} is evaluated at fixed \emph{mass}, rather than at fixed \emph{temperature}~\cite{McPeak:2021tvu,Etheredge:2022rfl}. Thus, the mass correction at fixed (zero) temperature, i.e., at extremality, is related to the entropy correction at fixed mass, which takes the black hole away from extremality (since the extremal mass is corrected). A more natural quantity is the extremal entropy correction, evaluated at fixed (zero) temperature. However, this is not related to the extremal mass correction, as has been noted for various explicit stringy black holes (both asymptotically flat and asymptotically AdS)~\cite{Charles:2016wjs, Cano:2019oma, Cano:2019ycn, Bobev:2021oku}. For example, in four dimensions the Gauss-Bonnet term is topological, and contributes to the black hole entropy but does not affect the extremality bound. This is consistent with~\eqref{eq:GoonPenco}, since nonzero contributions to $\partial S/ \partial \epsilon$ that are independent of temperature in the extremal limit make no contribution to the right-hand-size of~\eqref{eq:GoonPenco} due to the explicit $T$ prefactor.

Recently, \eqref{eq:MassCorrFormula} has been obtained without reference to the Wald entropy, via a direct attack on the equations of motion combined with some general reasoning about the Lorentz invariance of the Lagrangian. In this context, the formula was shown to hold for extremal black holes coupled to arbitrary moduli in any dimension~\cite{Etheredge:2022rfl}.

Similar techniques have been adapted to study not only extremality but long-range forces, to assess whether the Repulsive Force Conjecture is satisfied by corrected black holes~\cite{Cremonini:2021upd} (see also \cite{Etheredge:2022rfl}). The results suggest that the RFC may not be automatically satisfied by four-derivative corrections. However, they are obtained in EFT examples, rather than explicit string theory compactifications, so further work should investigate whether these examples can be realized in a full quantum gravity setting (and hence provide a counterexample to the RFC for corrected black holes). A study of the effects of higher derivative corrections on the force between dyonic strings appeared in~\cite{Ma:2021opb}.
            
\subsubsection{Overview of arguments}

The thermodynamic arguments sketched above have provided an efficient tool for computing the correction to the extremal charge-to-mass ratio in a given EFT, as a function of the Wilson coefficients of higher-dimension operators. Such calculations lead to superextremality conditions that take the form of positivity bounds like~\eqref{eq:4derivWGC}. A variety of attempts have been made to prove such bounds from general principles.

In many cases, the general structure of a theory implies that the $c_T$ and $c_F$ terms in~\eqref{eq:S4deriv}, which involve only photons and not gravitons, give the dominant corrections to the extremality bound. When these coefficients are explicitly calculable, they are often positive. Indeed, in quantum field theory (without gravity), one can prove rigorous positivity bounds on the coefficients of four-derivative operators involving $F_{\mu \nu}$~\cite{Adams:2006sv, Cheung:2014ega}. In a regime where the dominant contributions to $c_T$ and $c_F$ arise from low-energy QFT effects which would persist in the $M_\mathrm{Pl} \to \infty$ limit, this is sufficient to prove~\eqref{eq:4derivWGC}, as already noted in~\cite{ArkaniHamed:2006dz}.

Of course, we are interested in gravitational theories, where completely general, rigorous positivity arguments are more elusive. In four dimensions, loop effects involving gravitons can provide dominant contributions to four-derivative operators in the IR, so that taking the $M_\mathrm{Pl} \to \infty$ limit obscures important physics. In general, when the WGC is not satisfied by light charged particles, but is parametrically saturated (or even violated) by all light particles, EFT proofs of~\eqref{eq:4derivWGC} are more difficult to obtain. Below, we will summarize three broad categories of arguments for positivity: those that explicitly compute the coefficients in~\eqref{eq:S4deriv} within a given EFT; those that rely on analyticity, unitarity, and/or causality; and those based on entropy.

\subsubsection{Explicit computations within low-energy EFTs}

\begin{figure}
\centering
\includegraphics[width=0.45\textwidth]{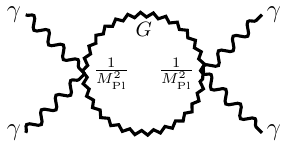}
\caption{Example of a loop diagram leading to logarithmic running of a four-derivative operator in 4d. Photons scatter via a loop of gravitons; due to two couplings each scaling as $1/M_\mathrm{Pl}^2$, this gives rise to a contribution schematically behaving as $\frac{1}{M_\mathrm{Pl}^4} F_{\mu \nu}^4 \log(E)$.}
\label{fig:loopexample}
\end{figure}

In four dimensions, the  Wilson coefficients in~\eqref{eq:S4deriv} exhibit logarithmic renormalization group evolution. This follows from dimensional analysis; for example, $[c_T] = [c_F] = -D$ and $[\kappa^4] = 4-2D$, which agree precisely when $D = 4$. An explicit example of a loop diagram contributing to such running is shown in Figure~\ref{fig:loopexample}. For exponentially large black holes, we expect the Wilson coefficients (evaluated at a renormalization scale  corresponding to the black hole's size) to be {\em dominated} by RG running. As a result, the sign of the  correction should  be determined by such  RG  effects, independent of  details of the UV completion and the  operator coefficients  at the cutoff scale. The consequences were first explored in~\cite{Charles:2019qqt}, and more recently in~\cite{Arkani-Hamed:2021ajd}. The case of multiple $U(1)$s has also been considered~\cite{Jones:2019nev}.

In the basis~\eqref{eq:S4derivAlt} there are logarithmic corrections to ${\tilde c}_W$ and ${\tilde c}_{GB}$, determined by the well-known Weyl anomaly coefficients $c$ and $a$, respectively~\cite{Charles:2019qqt}. The coefficients ${\tilde c}_{RF}$ and ${\tilde c}_F$ do not run. In the basis~\eqref{eq:S4derivAlt}, $c_T$ and $c_{GB}$ run~\cite{Arkani-Hamed:2021ajd}; in either basis, the running of $O_{GB}$ is irrelevant for the extremality bound. In Einstein-Maxwell theory plus any minimally coupled matter of spin $< 3/2$, $c > 0$, ensuring the validity of~\eqref{eq:4derivWGC}. Spin-$3/2$ fields contribute {\em negatively} to $c$, but a single spin-$3/2$ field is insufficient to drive the running negative. However, nonminimal couplings, such as dipole moments for fermions, also contribute negatively to $c$ for a small range of Planck-suppressed couplings. In cases with ${\cal N} \geq 2$ supersymmetry where extremal black holes are BPS, these negative contributions precisely cancel positive ones so that the black hole extremality bound remains uncorrected. However, there are low-energy (non-supersymmetric) effective Lagrangians with no obvious pathologies in which multiple fields with finely-tuned nonminimal couplings could lead to a negative running for $c$, and hence to large black holes that cannot satisfy the WGC.

If the tower or sublattice WGC is true, then the corrections to large black  holes  {\em must} allow them to become superextremal. Thus, there must be a bound on the number of fermionic  fields with dipole couplings in the limited range where the running of $c$ is negative; such theories would lie in the Swampland. In pure QFT, negative coefficients of the $F^4$ operators would violate causality; however, precisely because the negative contributions arise from gravitational-strength interactions, there is no violation of causality in the gravitational context~\cite{Arkani-Hamed:2021ajd}.

Moving beyond log running in the deep IR, we can also consider the threshold corrections induced by integrating out specific massive particles. Neutral bosons coupling to $F_{\mu \nu}F^{\mu \nu}$ or $F_{\mu \nu} {\tilde F}^{\mu \nu}$, exchanged at tree level, generate positive four-derivative operator coefficients consistent with~\eqref{eq:4derivWGC}~\cite{Hamada:2018dde}. Loops of charged particles with sufficiently large charge-to-mass ratio (obeying the WGC themselves, by a safe enough margin) also satisfy~\eqref{eq:4derivWGC}~\cite{Cheung:2014ega}. The challenging case, then, is when there are no light neutral bosons and all of the charged particles have $m \gtrsim e M_\mathrm{Pl}$; then, gravitational-strength ultraviolet contributions can be competitive, and the sign is not obviously determined.

\subsubsection{Arguments from analyticity, unitarity, and/or causality}

\begin{figure}
\centering
\includegraphics[width=0.5\textwidth]{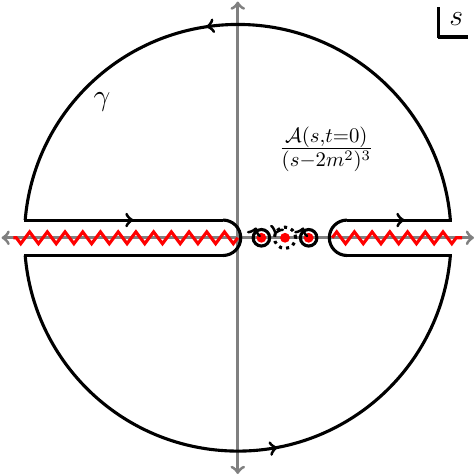}
\caption{The contour integral for a dispersive proof of positivity of four-derivative operators. Here, we illustrate the case of $2 \to 2$ scattering of a particle of mass $m$. The amplitude ${\cal A}(s,t=0)$ has poles at $s = m^2, 3m^2$ and branch cuts at $s \leq 0, s \geq 4m^2$. The dashed contour around the singularity inserted at $s_0 = 2m^2$ can be deformed to the solid contour $\gamma$ surrounding subtractable pole contributions, positive branch cut contributions, and a negligible contour at infinity.}
\label{fig:contour}
\end{figure}

In EFTs embedded within UV-complete quantum field theories, positivity bounds on certain combinations of operator coefficients (or, more invariantly, on derivatives of low-energy scattering amplitudes) may be proven using analyticity, unitarity, and causality~\cite{Pham:1985cr, Adams:2006sv, deRham:2017avq, deRham:2017zjm, Arkani-Hamed:2020blm, Zhang:2020jyn}. A prototypical example is the positivity of the $(\partial \phi)^4$ operator coefficient in the theory of a massive scalar field, derived from a forward dispersion relation. This term contributes an $s^2 + t^2 + u^2$ term in the low-energy amplitude ${\cal A}(s,t)$ for $\phi \phi \to \phi \phi$ scattering. The coefficient of this term can be read off from a second derivative, and in turn related to a contour integral in the complexified $s$ plane by Cauchy's theorem:
\begin{align}
\frac{1}{2}{\cal A}''(s_0,t=0)  = \frac{1}{2\pi i} \oint_\gamma \frac{{\cal A}(s,0)}{(s - s_0)^3}
= \frac{1}{\pi} \int_\mathrm{cuts} \mathrm{d}s\,\frac{s \sigma_\mathrm{tot}(s)}{(s - s_0)^3} > 0.
\end{align}
In the last step, the contour $\gamma$ around $s_0$ has been deformed to enclose the $s$- and $u$-channel branch cuts and two large arcs at large $s$, as illustrated in Figure~\ref{fig:contour}. The integrals along the branch cuts, in the $t \to 0$ limit, are related to positive total cross sections by the optical theorem. The contour at infinity does not contribute, because the Froissart bound (in conjunction with a Phragm{\'e}n-Lindel{\"o}f theorem) constrains the large-$s$ amplitude to obey ${\cal A}(s, t =0) < s^2 \log s$. This argument, given in~\cite{Adams:2006sv}, can be extended to positive $t$ (below the branch cut)~\cite{deRham:2017avq, deRham:2017zjm}. A version of the argument can also be derived in AdS using CFT crossing relations~\cite{Hartman:2015lfa}. Causality constraints, arising from superluminal propagation in nontrivial field backgrounds, lead to similar conclusions~\cite{Adams:2006sv}. Notice that the positivity bound on the $(\partial \phi)^4$ coefficient is a {\em strict} inequality, provided that $\phi$ is not free.

Corrected black holes satisfy the WGC if the inequality~\eqref{eq:4derivWGC} holds. This inequality involves four-derivative operators that contribute to scattering amplitudes of photons and gravitons, so it is natural to seek a general argument, similar to that for $(\partial \phi)^4$, that implies positivity independent of the details of the UV completion. For example, in a theory of only photons, the bounds derived from unitarity of forward scattering of linearly polarized photons (of all possible polarizations) imply
\begin{equation}
c_T \geq 0 \quad \text{and} \quad D c_T + 16 c_F \geq 0,
\end{equation}
in the notation of~\eqref{eq:S4deriv}.
However, the arguments immediately become more difficult in a gravitational context. The relationship between superluminality and causality is more subtle because lightcones are not rigid; cf.~discussions in~\cite{Cheung:2014ega,Goon:2016une, deRham:2020zyh,Bellazzini:2021shn}. Unitarity arguments based on forward dispersion relations face the difficulty that graviton exchange contributes a term $\propto -G_N s^2/t$ to scattering amplitudes, rendering the $t \to 0$ limit ill-defined. Furthermore, high-energy scattering in gravitational theories can produce large black holes, so QFT bounds on the asymptotic UV behavior of amplitudes do not necessarily hold. The fact that the $1/t$ graviton-exchange pole scales as $s^2$ poses a particular problem for bounding four-derivative operators. For example, as shown in~\cite{Bellazzini:2015cra}, if one carries out a contour integral to read off $O(s^4)$ coefficients and then sends $t \to 0$, one obtains candidate positivity constraints on operators involving four Riemann tensors that are compatible with known string theory examples. On the other hand, one cannot isolate the $O(s^2)$ contributions from local operators from those of graviton exchange in this way. Furthermore, if one simply discards the $-G_Ns^2/t$ term and follows the logic of the unitarity bound, one would conclude that  (in $D> 4$, where it affects $2 \to 2$ graviton scattering) the coefficient of the Gauss-Bonnet term must be both $\geq 0$ {\em and} $\leq 0$~\cite{Bellazzini:2015cra}. Theories are known in which this coefficient is nonzero, so it is clear that discarding the $t$-channel pole is not a strictly correct procedure. A plausible interpretation of this result is that the coefficient of the Gauss-Bonnet term cannot be too large with either sign, as further argued in~\cite{Camanho:2014apa} on causality grounds.

In QFT, we can deform a theory by adding relevant operators, without changing the UV behavior. This provides a method for addressing problematic IR divergences. In quantum gravity, we do not have this luxury. Quantum gravity theories are rigid: we cannot simply add terms to the Lagrangian without modifying the entire theory. On the other hand, we can study a consistent theory on different backgrounds. This motivated a novel argument aiming to eliminate the problematic $s^2/t$ pole by compactifying to three spacetime dimensions, where there is no propagating graviton mode~\cite{Bellazzini:2019xts}. A subtlety is that, in resolving the IR problem of gravity, a new UV problem arises: 3d flat-space gravity does not admit localized states of arbitrarily high mass, because a massive particle has a deficit angle that eventually eats up the entire space. In other words, the physics of the 3d theory resembles that of the 4d theory over a range of high energies, but strongly deviates at truly asymptotic energies. Thus, the meaning of ${\cal A}(s,t)$ becomes obscure in high-energy regions, where it seems to be not even well-defined, much less analytic. This was suggested by~\cite{Alberte:2020jsk, Alberte:2020bdz} as a possible culprit behind their observation that the $t$-channel subtracted positivity bounds derived from compactification appear to be overly strong. They require new physics to appear at prematurely small energies, in contradiction with known consistent theories. These works, reinforcing similar arguments made earlier in~\cite{Hamada:2018dde} (see also~\cite{Tokuda:2020mlf}), suggest that positivity arguments can forbid terms of the form $-c^2 \frac{s^2}{M^4}$ with $c \sim O(1)$ and $M$ held fixed in the limit $M_\mathrm{Pl} \to \infty$, but not terms of the form $-c^2 \frac{s^2}{M^2 M_\mathrm{Pl}^2}$, which tend to zero when gravity is decoupled.

Given the subtleties associated with making completely general and rigorous arguments in gravitational theories, much of the work on this subject has focused attention on identifying a sufficient set of conditions to prove~\eqref{eq:4derivWGC}. As discussed above, explicit computations show that tree-level exchange of light bosons interacting with photons and loops of light charged particles both produce corrections to $c_T$ and $c_F$ that satisfy~\eqref{eq:4derivWGC}. In these cases, effects from $c_{RF}$ and $c_{GB}$ are subdominant. This is often the case, as large contributions to $c_{RF}$ or $c_{GB}$ induce causality violation, in the absence of a tower of high-spin states~\cite{Camanho:2014apa} (see also~\cite{Li:2017lmh, Afkhami-Jeddi:2018own} for holographic, CFT-based arguments). The most difficult case to assess is when all contributions to the four-derivative operators arise from an ultraviolet scale, like the string scale. In this case, additional assumptions have been invoked. If Regge states associated with the photon have effects dominating over those associated with the graviton,~\eqref{eq:4derivWGC} can again be derived~\cite{Hamada:2018dde}. Similar arguments have been explored for dilatonic black holes in~\cite{Loges:2019jzs}. Constraints from duality have also been shown to imply positivity conditions~\cite{Andriolo:2020lul, Loges:2020trf}.
            
Recently, new positivity bounds have been derived~\cite{Caron-Huot:2021rmr} that, following~\cite{Camanho:2014apa}, avoid the $t$-channel pole problem by studying scattering at fixed impact parameter, rather than fixed $t$. It remains to be seen whether such an approach can offer a new perspective on the WGC. A crucial test of any completely general future proof of a positivity bound is that it must be compatible with exactly zero correction in the case of BPS black holes.
            
\subsubsection{Arguments from entropy}

As discussed in Sec.~\ref{subsubsec:correctedextremality}, recent work has shown that the higher-derivative correction to the black hole extremality bound is related, in a very general way, to the shift in the Wald entropy of the black hole due to higher-derivative terms. This raises the intriguing prospect of proving~\eqref{eq:4derivWGC} by proving that such corrections to the entropy must be positive~\cite{Cheung:2018cwt}. In particular, an argument based on the Euclidean path integral for black holes with positive specific heat (including Reissner-Nordstr\"om black holes of sufficiently large charge) establishes that the correction to the Wald entropy $\Delta S_{4\partial}$ from four-derivative operators is positive whenever the correction $\Delta F_{4\partial}$ to the free energy of the black hole at fixed temperature is negative. The restriction to positive specific heat allows one to conclude that the classical solution {\em minimizes} (not just extremizes) the Euclidean action. Under these conditions, one can show that any four-derivative operators generated at tree level lead to $\Delta S_{4\partial} > 0$, which in turn implies that corrected black holes satisfy the WGC. Extending these considerations to rotating dyonic black holes leads to a range of inequalities generalizing~\eqref{eq:4derivWGC}~\cite{Cheung:2019cwi}.

The assumptions in this argument are not universally valid, even for tree-level exchange~\cite{Hamada:2018dde}. For example, a massive spin-2 field $h_{\mu \nu}$ with a coupling $h^\lambda_{~\lambda} F_{\mu \nu}F^{\mu \nu}$ generates a negative shift in the entropy. It evades the assumptions because the Euclidean action is not a local {\em minimum} with respect to $h^\lambda_{~\lambda}$. Although this example evades the entropy argument, it violates unitarity, and so cannot be embedded in a consistent quantum gravity theory to provide a counterexample to~\eqref{eq:4derivWGC}. This connection between unitarity and positive contributions to the Wald entropy may hold more generally, and hint at an argument that could extend beyond tree level~\cite{Cheung:2018cwt}. Modular invariance is another supplementary assumption that has been invoked to extend the range of validity of entropy arguments for positivity~\cite{Aalsma:2019ryi}.

\section{Implications and Connections}\label{IMPLICATIONS}

%!TEX root = WGC_Review_arxiv_v2.tex

\subsection{Implications for phenomenology and cosmology}

\subsubsection{Direct application of the WGC}

Neither the WGC nor the tower/sublattice WGCs have immediate novel implications for the Standard Model of particle physics. The electromagnetic coupling constant at low energies is $e = \sqrt{4\pi \alpha} \approx 0.30$, so the electron satisifies the WGC by more than 20 orders of magnitude. Furthermore, because $e$ is an order-one number, the tower of charged particles predicted by the tower/sublattice WGCs could all have mass near or above the Planck scale. If one applies the WGC to the nonabelian gauge groups of the Standard Model (above the QCD scale or the electroweak scale, so that the gauge bosons appear massive), then the gauge bosons themselves obey the WGC, and again a WGC tower could consistently lie around the Planck scale because the coupling constants are order one. Perhaps a more interesting statement is that the WGC implies that a magnetic monopole should exist with a mass near the Planck scale or below (assuming that the bound is not obeyed only by monopoles of very large magnetic charge), but this is not a statement that is readily falsifiable by any conceivable experiment at this time.

Interesting direct applications of the WGC, then, should be sought in new gauge interactions beyond the Standard Model. These could be previously undetected forces through which known particles interact, or hidden sector interactions among particles that are so far unknown (or perhaps detected only indirectly through their gravitational effects, in the form of dark matter).

Given the minimal Standard Model matter content (without right-handed neutrinos), the theory can be extended with a single additional $U(1)$ gauge interaction, coupling to the one of the differences of lepton numbers for different generations: $L_e - L_\mu$, $L_\mu - L_\tau$, or $L_e - L_\tau$. At most one of these symmetries can be consistently gauged, due to mixed 't Hooft anomalies~\cite{Foot:1990mn, He:1990pn}. The case of $L_\mu - L_\tau$ is of particular interest, as it can explain the nearly maximal mixing of muon and tau neutrinos~\cite{Ma:2001md}. Any of these gauge symmetries must be spontaneously broken. The regime of greatest phenomenological interest involves relatively large gauge couplings, where the WGC has little power even assuming it applies in the higgs phase.

A more compelling example is the Standard Model with Dirac neutrino masses, which admits a different $U(1)$ extension, gauging the difference $B - L$ of baryon and lepton number~\cite{Chanowitz:1977ye, Deshpande:1979df}. In this case, the associated gauge field could be exactly massless without contradicting experimental results, provided that it is extraordinarily weakly coupled: $e_{B - L} \lesssim 10^{-24}$~\cite{Wagner:2012ui, Heeck:2014zfa}. The combination of the Planck constraint on the sum of neutrino masses, $\sum m_\nu < 0.12\,\mathrm{eV}$~\cite{Planck:2018vyg} with the values of the neutrino mass-squared differences inferred from neutrino oscillations, $\Delta m_{21}^2 \ll |\Delta m_{31}^2| \approx 2.4 \times 10^{-3}\,\mathrm{eV}^2$~\cite{ParticleDataGroup:2020ssz}, implies that the lightest neutrino has mass $\lesssim 0.03\,\mathrm{eV}$. Thus, the lightest neutrino will obey the WGC for $B-L$ provided that $e_{B-L} \gtrsim 9 \times 10^{-30}$. This provides about five orders of magnitude in allowed $B-L$ coupling in which the mild form of the WGC would be satisfied. The tower/sublattice WGCs, however, provide a significant constraint: an infinite tower of $(B-L)$-charged particles should exist, beginning at masses of order $e_{B-L} M_\mathrm{Pl} \lesssim \mathrm{keV}$ and extending up indefinitely. This implies that, if $B-L$ is an unbroken gauge symmetry in our universe, then billions of undetected particles that interact (albeit very weakly) with ordinary matter exist below the TeV scale. Although this would be surprising, it is not obviously ruled out by data; it would have phenomenology akin to the large extra dimensions scenario~\cite{Arkani-Hamed:1998jmv}. A minimal WGC tower of $B-L$ charged particles would suggest a breakdown of local quantum field theory at energies $\lesssim e_{B-L}^{1/3} M_\mathrm{Pl} \lesssim 10^{10}\,\mathrm{GeV}$, a scaling analogous to that of Kaluza-Klein theory. (However, because Standard Model fermions carry $B-L$ charge and are not accompanied by low-mass excitations of higher $B-L$ charge, we would not expect the $B-L$ gauge group to literally arise as a Kaluza-Klein gauge field from a circle compactification.) The tower/sublattice WGCs put the existence of a massless $B-L$ gauge field in tension with conventional models of GUTs or of high-scale inflation, which postulate local new physics at energy scales above $10^{10}\,\mathrm{GeV}$, but is not ruled out by experimental data.

The WGC might also be applied to possible gauge forces in hidden sectors, possibly related to the dark matter in our universe. One might expect that forces weak enough to have significant WGC constraints would also be too weak to have observable consequences. Somewhat surprisingly, it turns out that very weak forces between dark matter particles can sometimes have observable consequences in astrophysics or cosmology. Dark matter charged under a massless abelian gauge field (or ``dark photon'') has been considered as a simple QFT with rich phenomenology~\cite{Feng:2008mu, Ackerman:2008kmp, Feng:2009mn}. Constraints on the strength of such a coupling arise from evidence that dark matter is approximately collisionless. However, even for very small couplings, there can be collective dark plasma effects~\cite{Ackerman:2008kmp, Heikinheimo:2015kra}. These lead to density fluctuations in the plasma on a time scale of order the inverse plasma frequency, $\omega_p^{-1} \sim \frac{m_d}{e_d} \rho^{-1/2}$, where $m_d$ is the mass of an individual dark matter particle, $e_d$ is the dark photon coupling,  and $\rho$ is the mass density of dark matter (which is directly inferred from observations). If we suppose that the dark matter particles themselves obey the WGC for the dark $U(1)$, this can lead to interesting consequences, as discussed in~\cite{Craig:2019fdy}. In this case, the dark WGC implies that  $\frac{m_d}{e_d} \lesssim M_\mathrm{Pl}$. Quantitative estimates show that dark plasma fluctuations can then lead to shock waves developing on the timescale of a merger of colliding galaxy clusters. Thus, it is conceivable that observations of cluster mergers could reveal dynamical evidence of very weak gauge forces between dark matter particles that approximately saturate the WGC. If the dark matter particles are sufficiently light, then the tower/sublattice WGCs could, in turn, imply important constraints on the UV cutoff of physics in our universe. Dedicated work, including numerical simulations, would be necessary to make more precise statements about observable dark plasma effects.

The tower/sublattice WGCs can also have interesting implications for  nonabelian gauge groups in the dark sector. For example, dark matter charged under such a gauge group can have distinctive cosmological signatures even for quite weak couplings, because the dark gluons constitute a form of interacting dark radiation~\cite{Buen-Abad:2015ova}. The tower/sublattice WGC cutoff on such theories is at most $g^{1/2} M_\mathrm{Pl}$~\cite{Heidenreich:2017sim}. Thus, there can potentially be a tension between cosmological observables associated with interacting dark radiation, and theories of high-scale inflation.

Another topic of substantial recent phenomenological interest has been {\em kinetic mixing} between a dark $U(1)$ and ordinary electromagnetism~\cite{Holdom:1985ag}. Such a mixing can be generated by loops of particles that carry both kinds of $U(1)$ charge. The tower/sublattice WGCs imply the existence of such particles, and hence suggests a minimum kinetic mixing, at least in the absence of gauged charge conjugation symmetries that enforce an exact cancellation. The size of kinetic mixing motivated by such an argument has recently been explored, and compared to concrete string theory examples, in~\cite{Benakli:2020vng, Obied:2021zjc}.

Finally, a direct application of the (magnetic) WGC that is only indirectly relevant for phenomenology was pointed out in \cite{Cribiori:2020use, DallAgata:2021nnr}, which argued that de Sitter critical points in certain gauged supergravity models are incompatible with the magnetic WGC, since by \eqref{magWGCbound} their associated Hubble scale is larger than the scale of new physics, $\Lambda_{\rm NP} \lesssim e M_{\textrm{Pl}} \lesssim H$.

\subsubsection{Bounding the electroweak hierarchy}

A longstanding problem in particle physics is the electroweak hierarchy problem: why is the electroweak energy scale ($v \approx 246\,\mathrm{GeV}$) so many orders of magnitude below the Planck scale ($M_\mathrm{Pl} \approx 2.4 \times 10^{18}\,\mathrm{GeV}$)? An enormous hierarchy between the Planck scale and the masses of the electron, proton, and neutron is necessary in order to have stable, large objects like stars and planets. However, in the Standard Model, the electroweak hierarchy is large only in a tiny subset of the UV parameter space, unlike the hierarchy between the Planck and QCD scales, which is naturally exponentially large due to asymptotic freedom. This has motivated a number of suggested extensions of the Standard Model in which the electroweak hierarchy can naturally become large, ranging from scenarios where electroweak breaking is triggered by dynamical supersymmetry breaking to those where the Higgs boson is a composite particle of a strongly-interacting sector. Traditionally, these models all share the feature that they relate the electroweak hierarchy to a scale generated by dimensional transmutation, and they predict new particles with masses near the electroweak scale. LHC measurements have informed us that the Higgs boson appears to be approximately elementary (i.e., it has Standard Model-like interactions with other fields), and additional electroweak-scale particles have not yet been discovered. This has motivated theorists to pursue novel explanations of the electroweak hierarchy problem.

Because the WGC gives rise to an upper bound on particle masses, it is tantalizing to wonder if it could produce an upper bound on the Higgs scale $v$, thereby explaining why $v$ is so dramatically small compared to $M_\mathrm{Pl}$. This idea was first discussed by~\cite{Cheung:2014vva}, and further explored by~\cite{Lust:2017wrl, Craig:2019fdy}. The simplest, original version of the idea is to suppose that $B-L$ is gauged and that neutrinos, which acquire mass only from electroweak symmetry breaking, are the particles responsible for satisfying the WGC. The (Dirac) neutrino mass must then obey $m_\nu = y_\nu v / \sqrt{2} < \sqrt{2} e_{B-L} M_\mathrm{Pl}$. If we fix $y_\nu$ to its Standard Model value ($\sim 10^{-12}$), and if we postulate a $B-L$ gauge coupling $e_{B-L} \sim 10^{-28}$ (consistent with the experimental limits), this inequality tells us that $v \lesssim 10^{-16} M_\mathrm{Pl}$, and thus requires an electroweak hierarchy of the order that we observe in nature.

While this offers an interesting perspective on how quantum gravity might affect low-energy particle physics in surprising ways, several elements of this argument are unsatisfactory. One is that it seeks to explain the origin of a mysterious factor of order $10^{-16}$ in terms of another small number of order $10^{-28}$, which is unexplained. This is viewed as progress because the electroweak hierarchy is not robust against quantum corrections (the Higgs mass acquires additive corrections of order $\frac{h^2}{16\pi^2} M^2$ when coupled to heavy fields of mass $M$ via interactions of size $h$, which must be ``tuned away'' through cancellations against other contributions), whereas the smallness of the gauge coupling $e_{B-L}$ is ``technically natural'' (its corrections are all proportional to $e_{B-L}$ itself). Nonetheless, if our goal is to understand the origin of small numbers in our theory of nature, this at best shifts the problem to explaining the origin of the small number $e_{B-L}$. One might hope that such a problem has a solution, for instance, in terms of a dynamical mechanism of moduli stabilization. Nonetheless, this shift of the hierarchy problem toward a problem of explaining an exponentially tiny $e_{B-L}$ is in some tension with the spirit of the WGC itself. A very small value of $e_{B-L}$ restores a global symmetry of the theory, and so quantum gravity should resist attempts to generate exponentially tiny gauge couplings. This suggests that perspectives rooted in a very literal interpretation of technical naturalness may encounter obstacles in a quantum gravity setting. A sharper version of this concern is that the magnetic WGC tells us that $e_{B-L} M_\mathrm{Pl}$ serves as an ultraviolet cutoff on our EFT. This is particularly problematic from the viewpoint of the tower/sublattice WGCs, which postulate a tower of $(B-L)$-charged particles appearing at this mass scale. If infinitely many particles in such a tower obey the WGC, then it was unnecessary to require that the neutrinos obey the WGC, destroying the  link between a small $e_{B-L}$ and the Higgs scale $v$.

One refinement of the argument~\cite{Lust:2017wrl, Craig:2019fdy} draws on the Repulsive Force Conjecture in the presence of scalar fields~\cite{Palti:2017elp}, arguing that the bound assumes the schematic form $m \leq \sqrt{g^2  - \mu^2} M_\mathrm{Pl}$ where $g$ is a gauge coupling and $\mu$ is a coupling to scalars. In cases where $g^2 \approx \mu^2$, this can be a much stronger bound than simply $m \leq gM_\mathrm{Pl}$. This offers the opportunity to push the magnetic WGC scale $g M_\mathrm{Pl}$ up to higher energies, where it has less effect on the argument. On the other hand, it introduces yet another small number, the ratio $\sqrt{g^2-\mu^2}/g$, which requires explanation. One must postulate a specific form of the scalar couplings of the light, WGC-obeying matter fields in order to make this argument. It should be different from the scalar couplings of black holes; otherwise, the tower/sublattice WGC tower would begin at the same scale, $\sqrt{g^2  - \mu^2} M_\mathrm{Pl}$, rather than $g M_\mathrm{Pl}$. This application of the RFC requires that the scalar providing the additional force remain light, which itself requires explanation and can lead to additional naturalness constraints on the EFT~\cite{Craig:2019fdy}. 

The most recent variations on the argument~\cite{Craig:2019fdy} explore new forces under which no Standard Model particle is charged. One could, for example, consider a scalar field $\Phi$ charged under a new $U(1)_X$ gauge interaction, which we suppose should satisfy the WGC for $U(1)_X$. If we further posit that $\Phi$ couples to the Higgs boson through a quartic coupling $\kappa |\Phi|^2 |h|^2$, then the additive shift of the $\Phi$ mass-squared by $\frac{1}{2} \kappa v^2$ could cause $\Phi$ to fail to obey the WGC if $v$ is too large. Similar models can be constructed with fermionic fields. These models make distinctive phenomenological predictions, relative to the original $B-L$ model, and could have implications for dark matter dynamics.

These attempts to bound the electroweak scale $v$ using Weak Gravity arguments all invoke a similar set of assumptions. We must assume the existence of very small (but technically natural) couplings. We must also assume that specific particles in the theory, which happen to interact with the Higgs boson, are the ones that satisfy the WGC. If the WGC were satisfied by an independent set of particles, not interacting with the Higgs, then the link to the electroweak hierarchy would be severed. Finally, we must suppose that this restricted set of theories is relevant for the world that we live in. If the landscape of quantum gravity contains many universes resembling our own that do {\em not} contain the postulated $U(1)_{B-L}$ or $U(1)_X$ force and the specific connections assumed between these forces and the electroweak scale, then there is no reason why we would expect our universe to obey the assumptions. The argument that the WGC constrains the electroweak scale would only be plausible if such vacua are overwhelmingly more common than others, or overwhelmingly more likely to be populated by cosmology. In recent years, there has been a proliferation of models that link cosmology to particle physics by postulating the existence of a landscape that takes a very specific form, where for instance certain couplings are assumed to exist and take on fixed values in all vacua, and only a limited set of parameters ``scan'' from one vacuum to another. These have been referred to as ``artificial landscapes''~\cite{Strassler:2014,Strassler:2016}, and in the absence of evidence that they resemble the true landscape of quantum gravity, it is unclear what lessons one can draw from them.

Finally, it may be worth emphasizing that the {\em examples} in which we have checks of the WGC are cases where we compute the mass at leading order in a perturbative expansion, or where the mass is protected by supersymmetry. As a result, we have no explicit examples in which the WGC is satisfied by a state whose mass is fine-tuned to be light due to a cancellation. Indeed, such examples would be extremely difficult to generate. If one could find such examples in the string theory landscape, they would at least serve as an interesting proof of principle that the WGC could require a fine-tuning that would appear accidental from the viewpoint of low-energy effective field theory.

\subsubsection{Other applications to the hierarchy problem}

In \cite{Graham:2015cka}, a dynamical mechanism known as ``cosmological relaxation'' was proposed as a solution to the hierarchy problem. In this scenario, the Higgs field $h$ is coupled to a real scalar field $\phi$ through a potential of the form
\begin{equation}
V = (-M^2 + g \phi) |h|^2 + \left( g M^2 \phi + g^2 \phi^2 + \cdots \right) + \Lambda^4 \cos(\phi/f) \, ,
\label{relaxion}
\end{equation}
where $M$ is the cutoff of the effective field theory and $\Lambda$ depends on the vev of $h$. Initially, the dynamics of $\phi$ are dominated by the polynomial terms, and the cosine term is negligible. When $\phi \sim M^2/g$, however, the Higgs field acquires a vev, and the scale $\Lambda$ for the cosine terms grows, creating a barrier which stabilizes the axion and leaves the Higgs with a mass well below the EFT cutoff $M$.  In order for this mechanism to work, however, the cosine terms must eventually be able to compete with the $g M^2 \phi$ term. In typical relaxation scenarios, this requires $g$ to be roughly of order $10^{-34}$. Furthermore, inflation must last long enough for $\phi$ to scan the entire range of the Higgs mass. This places an additional bound on the cutoff given by $M \lesssim (\Lambda M_{\textrm{Pl}})^{1/2}$, which yields $M \lesssim 10^9$ GeV for $\Lambda = \Lambda_{\text{QCD}}$.

The tiny coupling $g \sim 10^{-34}$ is ``technically natural,'' but this does not necessarily mean that the model can be UV-completed. In particular as we take $g\to 0$ the theory \eqref{relaxion} has an exact global symmetry $\phi'=\phi+2\pi f$, and if the arguments against exact global symmetries have any robustness then they should also rule out sufficiently small values of $g$.  One possible approach to avoiding this problem is to view $\phi'=\phi+2\pi f$ as a gauge symmetry, or in other words to turn $\phi$ into an axion.  This however forbids most of the terms in \eqref{relaxion} (a small explicit violation of a gauge symmetry is just as bad as a large one), and thus kills the feasibility of the model.

So far the most promising proposal for obtaining a large scalar field excursion that is consistent with all versions of the WGC is the ``axion monodromy'' proposal of \cite{Silverstein:2008sg, McAllister:2008hb}.\footnote{This was originally proposed as a model of inflation, as we will discuss later in this section, but it can also be used as a mechanism to implement cosmological relaxation as we discuss here.}  The most basic version of this proposal~\cite{Kaloper:2008fb} uses an axion coupled to a 3-form gauge field $A_3$ via the Lagrangian (which we here write as a $4$-form)
\begin{equation}\label{eq:monodromy}
L = - \frac{1}{2} \rmd\phi\wedge \star \rmd\phi - \frac{1}{2e_3^2} F_4\wedge \star F_4 + \frac{g}{e_3} \phi F_4\,.
\end{equation}
with $F_4 = \rmd A_3$.  Naively one might think that the coupling $g$ should be zero to respect the axion periodicity $\phi\sim \phi+2\pi f$, but as is usual for Chern-Simons type interactions the fact that the integral of $F_4$ obeys the quantization 
\be
\int F_4=2\pi m, \qquad m\in \mathbb{Z}
\ee
means that it is enough that we have
\be\label{gdef}
g=\frac{ke_3}{2\pi f}, \qquad k\in \mathbb{Z}.
\ee
We have normalized $g$ here so that it matches the $g$ in \eqref{relaxion}, so we now have two ways to get a small $g$: either we can take $f$ large in Planck units \textit{or} we can take $e_3$ small.  To avoid trouble with the axion WGC \eqref{eq:axionWGC} we do not want to take $f$ large in Planck units (we'll discuss this more in Section \ref{ssec:axioninflation}), so our task is to understand how constrained we are by the WGC for the three-form gauge field $A_3$. Before discussing that however it is perhaps worth explaining in more detail how the theory \eqref{eq:monodromy} allows for a super-Planckian field excursion.  The equations of motion following from \eqref{eq:monodromy} are
\begin{align}\label{monodromy}
    \star \rmd \star \rmd \phi+\frac{g}{e_3}\star F_4=0 \,,~~~~~
    \rmd(ge_3\phi-\star F_4)=0,
\end{align}
so the quantity
\be
\wt{F}_0\equiv \star F_4-ge_3\phi
\ee
is constant.  In fact it is quantized: the integral of $A_3$ over space is a periodic variable, and $\wt{F}_0$ is proportional to its canonical conjugate.   Working this out gives the quantization
\be
\frac{1}{e_3^2}\wt{F}_0= n,
\ee
with $n\in \mathbb{Z}$.  Substituting this back into the first equation of motion we find
\be\label{monodromyeom}
\star \rmd\star \rmd\phi+g(e_3 n+g\phi)=0,
\ee
which is the equation of motion for a scalar field with a potential that is a second order polynomial just as in \eqref{relaxion}.  This equation may appear not to be gauge-invariant, but it actually is since $n$ is a dynamical variable and the relevant gauge transformation is
\begin{align}
\phi'=\phi+2\pi f \,,~~~~~
n'=n-k.
\end{align}
Here $k$ is the integer defined by \eqref{gdef}.  One way to think about the apparent non-periodicity of the potential is to observe that although $\phi$ is periodic, $\star F_4$, which is not periodic, is also rolling in order to ensure that $\wt{F}_0$ is constant.  Indeed one can say that $\star F_4$ is really the gauge-invariant scalar which is rolling in axion monodromy.

There is an interesting subtlety in this model which is worth mentioning explicitly: although the gauge invariance of the action prevents us from adding arbitrary powers of $\phi$ to the action, $\star F_4$ is perfectly gauge-invariant and thus there is nothing which prevents us from introducing a potential $V(\star F_4)$.  Such a potential presumably is generated by quantum gravity effects, so why does it not ruin the model?  To the extent that axion monodromy can be realized in string theory (which seems unlikely for the relaxion scenario but plausible for inflation), such a potential does exist but is typically of the form 
\be
V(\star F_4)=\frac{1}{e_3^2\ell_s^8} v(\ell_s^4 \star F_4).
\ee
Here $\ell_s$ is the string scale and $v(\cdot)$ is a dimensionless function of a dimensionless variable, expected to be $O(1)$. (In models, the form of this function is known, e.g., from the DBI action~\cite{McAllister:2008hb}.)  The correction to the equations of motion \eqref{monodromy} arising from this potential does not become important until $\star F_4\sim \ell_s^{-4}$.  An axion excursion $\Delta \phi$ gives rise to a change in $\star F_4$ which is of order
\be
\Delta \hspace{-.1cm}\star \hspace{-.1cm}F_4\sim g e_3 \Delta \phi\sim ke_3^2\frac{\Delta \phi}{f},
\ee
so we can have an axion excursion which is large compared to $f$ without feeling the potential $V$ provided that
\be
k e_3^2\ell_s^4\ll 1.
\ee
In string theory, the dimensionless number $e_3^2\ell_s^4$ is often small: it can be  proportional to positive powers of $g_s$, inverse powers of volumes, or warp factors; whatever the reason, as long as it is small we can achieve $\Delta \phi \gg f$ without being sensitive to $V(\star F_4)$.  The robustness of axion monodromy thus relies on high-energy information from string theory: it cannot be established purely using low-energy power counting and symmetries.

We now turn to applying the WGC for 3-form gauge fields to axion monodromy~\cite{Ibanez:2015fcv}.  The objects to which it applies are domain walls of tension $T_3$, across which $\frac{1}{e_3^2}\wt{F}_0$ changes by an integer $Q$, and the WGC says there should be such domain walls with
\begin{equation}
T_3 \leq  \frac{4\pi f g\Mp Q }{k}\,.
\label{WGCdomain}
\end{equation}
The danger here is that an upper bound on the domain wall tension also likely gives some sort of lower bound on the rate for bubbles bounded by the domain wall to nucleate, and if this happens too often it destroys the relaxation mechanism.  The domain walls separate regions whose potential energy differs by 
\begin{equation}
\Delta V \sim  g e_3 \phi \sim f g^2\phi/k \,.
\end{equation}
The bounce action computed by~\cite{Ibanez:2015fcv} is not accurately described by the thin-wall approximation, but involves important gravitational backreaction~\cite{Coleman:1980aw}. The result is a bubble nucleation probability
\begin{equation}
P \sim \exp( - B) \,,~~~~B \approx w(b) \frac{2 \pi^2 T_3}{H^3}\,,
\end{equation}
where $H$ is the Hubble scale during inflation. The parameter $b$ is defined as
\begin{equation}
b = \frac{ \Delta V }{ H T_3 },
\end{equation}
and it turns out that in the parameter range of interest, $w(b) \sim O(1)$ and $b \lesssim 1$. Using $b$, we can rewrite the bounce action estimate as:
\begin{equation}
B \sim 2\pi^2 \frac{T_3^4 b^3}{(\Delta V)^3}.
\end{equation}
The WGC provides a constraint, following~\cite{Ibanez:2015fcv}, because we require $B \gg 1$ for an exponentially suppressed tunneling probability, but the WGC implies that $B < B_\mathrm{max}$ where $B_\mathrm{max}$ is obtained when $T_3$ saturates~\eqref{WGCdomain}. These can only be consistent when $B_\mathrm{max} \gg 1$. Together with the estimates $\phi \sim M^2/g$ and $g M^2 \sim \Lambda^4/f$ required for consistent relaxion phenomenology, this inequality translates into a bound on the EFT scale:
\begin{equation}
M \lesssim \left( 4 \pi^2 b^3 \right)^{1/8} \sqrt{  \Lambda   M_{\text{Pl}} }  \,.
\end{equation}
For $\Lambda = \Lambda_{\text{QCD}}$, this bound becomes
\begin{equation}
M \lesssim b^{3/8} \times 2.5 \times 10^9 \text{ GeV},
\end{equation}
which for $b \sim 1$ rivals the bound for consistency of the relaxion model discussed previously.

Thus, the three-form WGC provides an interesting constraint on cosmological relaxation implemented via axion monodromy. On the other hand a similar constraint may also be derived independently of the WGC, and in the original paper it was shown that this constraint could be satisfied without spoiling the model.  Therefore neither the axion WGC nor the three-form WGC seem to pose a fatal challenge to the axion monodromy version of the cosmological relaxation model.  It is worth mentioning, however, that embedding the model into string theory nonetheless seems to be very challenging, if not impossible. In particular, \cite{McAllister:2016vzi} argued that within a string compactification, the huge winding number of the relaxion corresponds to a huge charge carried by branes or fluxes (this is already apparent in the model \eqref{eq:monodromy} since $\star F_4$ is rolling). This charge backreacts on the compactification geometry and eventually spoils the relaxation mechanism.  The relaxion scenario may lie in the Swampland, but if so then the most stringent top-down constraints do not come from the WGC. Even if axion monodromy does not give a viable realization of the relaxion model in string theory however, it is a quite plausible candidate for realizing inflation: we return to this in section \ref{ssec:axioninflation}.

\subsubsection{Mass of the photon or dark photons}

Conventionally, we assume that the photon is a massless gauge field. However, theories of massive, abelian spin-1 particles are perfectly consistent, either with a simple mass term and no gauge invariance at all \cite{Proca:1900nv} or with a real scalar field added to provide the longitudinal mode, together with a gauge invariance to eliminate the redundant degree of freedom~\cite{Stueckelberg:1938zz}. If the photon has a very small mass, the longitudinal mode is extremely weakly coupled, so it is difficult to experimentally distinguish from a massless photon despite the change from two to three independent propagating polarization states~\cite{bass1955must}. Of course, the photon in our universe must be extremely light. There is a large literature on experimental constraints on the mass, to which a few interesting entry points are \cite{Adelberger:2003qx, Goldhaber:2008xy, Wu:2016brq}. 

In four dimensions a massive photon in the Stueckelberg regime can be described by BF theory: we have a 1-form gauge field $A$ with field strength $F = \rmd A$ and a 2-form gauge field $B$ with field strength $H = \rmd B$, interacting via the Lagrangian
\begin{equation}
S = \int \left(- \frac{1}{2f^2} H \wedge \star H - \frac{1}{2e^2} F \wedge \star F + \frac{k}{2\pi} B \wedge F\right),
\end{equation}
where $k \in \mathbb{Z}$ just as in the axion monodromy discussion of the previous subsection. The gauge coupling $f$ of the 2-form field has mass dimension 1. This theory describes a massive gauge field with mass
\begin{equation}
m = \frac{k}{2\pi} e f.
\end{equation}
For $k \neq 0$, taking the gauge field mass to zero requires either $e \to 0$ or $f \to 0$. In either case, we are taking a gauge coupling to zero, and so the WGC imposes some constraint. In particular, if we send $e \to 0$, the magnetic WGC tells us that there is a UV cutoff on the theory at the scale $e M_\mathrm{Pl}$, and the tower/sublattice WGCs suggest that there effective field theory breaks down irrevocably at some scale, possibly a higher one like $e^{1/3} M_\mathrm{Pl}$ (as in Kaluza-Klein theory). If we send $f \to 0$, the mild WGC for the 2-form gauge field $B$ implies that strings charged under $B$ should exist with tension $T \lesssim f M_\mathrm{Pl}$.

In the case where the photon mass arises from the Higgs mechanism, there is no fundamental obstruction to sending $f \to 0$: this corresponds to turning off the Higgs vev, which can be accomplished just by giving the Higgs a positive mass term around the origin. In this case, the $B$-field may be thought of as an emergent gauge field in the IR below the scale of the Higgs vev, and the charged strings predicted by the 2-form WGC are simply ANO strings~\cite{Abrikosov:1956sx, Nielsen:1973cs}. In the core of an ANO string, the Higgs vev is zero; in the limit that the Higgs vev is taken to zero, an ANO string simply becomes more and more diffuse and fades away. The WGC, then, is compatible with small masses arising from the Higgs mechanism.

By contrast, there are massive gauge theories which are fundamentally of Stueckelberg type. In this case, the strings charged under the $B$ field are fundamental (e.g., the F-strings or D-strings of string theory). The core of the string is not well-described by effective field theory, and there is no finite-distance point in field space at which the gauge boson mass can be sent to zero. In this case, the limit $f \to 0$ corresponds to a theory of fundamental, tensionless strings, signaling a complete breakdown of local effective field theory. In such a case, the fundamental quantum gravity cutoff energy is bounded, $\Lambda_\mathrm{QG} \lesssim \sqrt{2\pi f M_\mathrm{Pl}}$.

The WGC, then, imposes an ultraviolet cutoff on theories of a massive gauge boson with mass arising from a fundamental Stueckelberg term~\cite{Reece:2018zvv}. This can also be understood as a consequence of the Swampland Distance Conjecture: for mass terms of fundamental Stueckelberg type, the $m \to 0$ limit is an infinite distance limit, and so an infinite tower of light states appears when one approaches this limit.

This constraint on massive, abelian gauge bosons has potentially important implications for the Standard Model photon and for potential dark photons~\cite{Reece:2018zvv}. First, consider the Standard Model photon. A conservative bound, obtained from the arrival time of different frequencies from Fast Radio Bursts, is that $m_\gamma \lesssim 10^{-14}~\mathrm{eV}$~\cite{Wu:2016brq}. Stronger bounds exist, but involve more assumptions, so we will work with this very simple kinematic bound; our conclusions can be readily adapted to other constraints. If we assume that electromagnetic charge is quantized in the usual way, the only way to obtain a small Standard Model photon mass of fundamental Stueckelberg type is by taking $f$ to be very small: $e f/(2\pi) \lesssim 10^{-14}~\mathrm{eV}$ requires $f \lesssim 10^{-22}~{\rm GeV}$. But then the Weak Gravity Conjecture would require fundamental strings with tension $T \lesssim f M_\mathrm{Pl} \lesssim (20~\mathrm{MeV})^2$. However, we know that gravity does not become strongly coupled near the MeV scale, so we cannot have a fundamental string with its associated tower of high-spin modes at such a low scale. This strongly suggests that the only way for the Standard Model photon to be massive is if it is Higgsed.

Could the Standard Model photon be Higgsed?  The short answer is ``probably not,'' and it becomes ``no'' provided that we assume that the ratios among charges of light particles in the theory are at most $O(1)$ (such an assumption is common in discussions of phenomenological implications of the WGC). With this assumption, if the Standard Model photon were Higgsed then we would already have discovered the corresponding Higgs boson. The only way to avoid this is for the Standard Model photon to obtain a mass from a Higgs field with a charge that is a tiny fraction of the electron's charge, in which case the associated Higgs boson could remain hidden from experiments. For example, suppose that the fundamental unit of electric charge is not $e$ but some $e_0 = e/N$ where $N$ is a very large integer. Then our calculation becomes quite different: we have $m_\gamma = e_0 f/(2\pi)$ and, rather than small $f$, we are free to take very small $e_0$. Furthermore, if $f$ is a Higgs vev, rather than a Stueckelberg scale, then there are no associated fundamental strings providing a UV cutoff. As an example, the choice $f \sim \mathrm{eV}$ and $e_0 \sim 10^{-14}$ could be consistent with experimental bounds on millicharged particles. It implies a WGC tower of states with tiny electric charge beginning at the scale $e_0 M_\mathrm{Pl} \sim 10~\mathrm{TeV}$, which is allowed by data. Apart from a new hierarchy puzzle associated with the small mass of the new Higgs field, the cost of considering such a theory is the introduction of the enormous integer $N \sim 10^{14}$. The Standard Model fermions would carry electric charge on this order, in units of the fundamental charge. There are no known consistent theories of quantum gravity that can produce such large ratios of charges among light particles. On the other hand, there are phenomenological models in which such a large integer could be obtained as a product of smaller integers, as in the clockwork scenario~\cite{Choi:2014rja,Choi:2015fiu, Kaplan:2015fuy}, adapted to this context by~\cite{Craig:2018yld}. It remains to be seen if such scenarios can be found in the Landscape.

The WGC can constrain not only the possibility that the Standard Model photon is massive, but also the possibility that a dark photon $A'$ has a fundamental Stueckelberg mass. One application is to dark photon dark matter, which is now the target of many dedicated experiments. Dark photon dark matter can arise from the primordial fluctuations of a massive vector field $A'$ during inflation, which can account for the observed dark matter relic abundance if $m_{A'} \gtrsim 10^{-5}\,\mathrm{eV}$~\cite{Graham:2015rva}. However, for such a light dark photon, if the mass is of fundamental Stueckelberg type then the WGC implies a cutoff lower than the inflationary Hubble scale assumed in the calculation of the dark matter relic abundance. This constraint excludes a substantial part of the parameter space of such models~\cite{Reece:2018zvv}.

\subsubsection{Axion inflation}\label{ssec:axioninflation}

The spectra of temperature and polarization fluctuations in the Cosmic Microwave Background radiation (CMB) strongly suggest that the universe experienced an early period of accelerated expansion known as inflation \cite{WMAP:2003syu, Planck:2018jri}. This idea \cite{Guth:1980zm,Starobinsky:1980te} is most easily realized by a scalar field rolling slowly down a potential \cite{Linde:1981mu}.  The action describing this scenario is
\begin{equation}
S= \int \textrm{d}^4 x \sqrt{-g} \left[ \frac{M_\mathrm{Pl}^2}{2}R - \frac{1}{2} g^{\mu\nu} \partial_\mu \phi \partial_\nu \phi - V(\phi) \right] \,,
\end{equation}
where the scalar $\phi$ is called the inflaton.  At leading order the metric $g_{\mu\nu}$ is taken to be the FRW metric,
\begin{equation}
\textrm{d}s^2 = -\textrm{d}t^2 + a^2(t) \textrm{d}\vec{x}^2\,,
\end{equation}
and the scalar field $\phi$ is taken to be homogenous in space, $\phi(t, \vec{x}) = \phi(t)$. The equations of motion are 
\begin{align}
\ddot \phi + 3 H \dot \phi + V'(\phi) = 0,~~~~~
\frac{1}{3M_\mathrm{Pl}^2} \left( \frac{1}{2} \dot \phi^2 + V(\phi) \right)=  \left( \frac{\dot a}{a} \right)^2,
\end{align}
and inflation happens when $H\equiv \frac{\dot{a}}{a}$ is approximately constant.  Requiring $|\dot{H}|\ll H^2$ implies that $\dot \phi^2 \ll V$, and it is usually also assumed that the acceleration of $\phi$ is small, $|\ddot{\phi}|\ll \dot{\phi}H$ (see, e.g., \cite{weinberg2008cosmology}).  Together these requirements are equivalent to the ``slow roll conditions''
\begin{align}
\frac{|V'|}{V}\Mp&\ll  1 &
\frac{|V''|}{V}\Mp^2&\ll 1.
\end{align}
 Inflation ends when these conditions are violated, after which the field is usually expected to oscillate about its current minimum and in some manner (called reheating) decay into the dense gas of hot particles we usually call the big bang.

CMB observables give us data about the inflaton potential $V$. Especially noteworthy for our purposes are the primordial scalar and tensor power spectra
\begin{align}
k^3 P_s(k)&= \frac{H^4}{2\dot{\phi}^2}\approx \frac{V^3}{6\Mp^6V'^2}\,, &
k^3P_t(k)&=\frac{4H^2}{\Mp^2}\approx \frac{4V}{3\Mp^4}\,,\label{CMBP}
\end{align}
where quantities on the right are evaluated at the value of $\phi$ corresponding to the time when modes of wave number $k$ were exiting the inflationary horizon (see, e.g., \cite{Maldacena:2002vr}).  The scalar power spectrum has been well-measured by the temperature anisotropy of the CMB, so it is the tensor spectrum, which causes anisotropy in the B-mode polarization of the CMB, which is of most interest in learning more about the physics of inflation.  In particular from \eqref{CMBP} we see that a measurement of tensor modes would give us direct information about the overall scale of the inflationary potential.   The tensor amplitude is usually expressed via the tensor-to-scalar ratio
\be\label{eq:r}
r\equiv \frac{P_t(k_*)}{P_s(k_*)}=\frac{8\dot{\phi}^2}{H^2 \Mp^2}\approx \frac{8 \Mp^2 V'^2}{V^2},
\ee
where $k_*$ is a typical wave number of large scale structure, say $.05\, \mathrm{Mpc}^{-1}$. For high-scale inflation such as might lead to observable tensor modes, this corresponds to the value of the inflaton about $60$ e-foldings before the end of inflation (i.e., at the time $t_*$ when $\log \left(a(t_\text{end})/a(t_{*})\right) \approx 60$).  

From \eqref{eq:r}, we see that the tensor-to-scalar ratio measures how quickly the inflaton is rolling. We also know how long inflation lasts (60 e-folds for high-scale inflation). Putting these together, and integrating $\dot \phi$ over time, we obtain a rough estimate on the distance traveled by the field $\phi$ during the course of inflation, which is known as the Lyth bound \cite{Lyth:1996im}:
\begin{equation}
\Delta \phi \approx \mathcal{O}(1) \left( \frac{r}{0.01} \right)^{1/2}M_\mathrm{Pl}\,.
\label{Lyth}
\end{equation}
The Lyth bound tells us that if tensor modes can be detected in the CMB in the near future (which would require $r \gtrsim 0.001-0.01$), then the inflaton must traverse a distance of order 1 in Planck units.

This ``super-Planckian'' field range is important because such large field excursions can run into trouble with effective field theory: quantum gravity sets an EFT cutoff no larger than the Planck scale, so a perturbative expansion of $V(\phi)$ in powers of $\phi$ suppressed by $\Mp$ will run into trouble when $\Delta \phi$ is $O(M_{\rm Pl})$ or larger. If tensor modes are observed in the CMB, therefore, we will need a model of large-field inflation that is not destroyed by large corrections arising from quantum gravity. 

Historically, axions have been considered the most promising route for circumventing this issue. An axion has a discrete gauge symmetry, $\phi \rightarrow \phi + 2 \pi$. This shift symmetry protects the axion potential from Planck-suppressed operators $\phi^n / M_{\textrm{Pl}}^{n-4}$. The dynamics are instead controlled by a periodic potential, which is often assumed to be dominated by instanton effects, leading to an action of the form
\begin{equation}
    S = \int \rmd^4x \left[ - \frac{1}{2}f^2 (\partial_\mu \phi)^2 - V(\phi) \right] \,,
\end{equation}
where
\begin{equation}
V(\phi) = \Lambda_{\textrm{UV}}^4 \e^{-S_{\textrm{inst}}} \left( 1 - \cos \left( \phi \right) \right) +O(\e^{-2 S_{\textrm{inst}}})\,.
\label{Vnat}
\end{equation}
Here $f$ is the ``axion decay constant,'' $S_{\textrm{inst}}$ is the instanton action, and higher harmonics of the potential are suppressed by additional powers of $\e^{-S_{\textrm{inst}}}$. The resulting model of inflation is called ``natural inflation'' \cite{Freese:1990rb}, and it yields phenomenologically viable models of large-field inflation with detectable tensor modes ($r > 0.01$) for $f \gtrsim 10 M_{\textrm{Pl}}$.  

Although natural inflation is an appealing model, it has proven difficult to implement in string theory when $f\gtrsim \Mp$ and $S_{\mathrm{inst}}\gg 1$, with the basic problem being that such large excursions in scalar field space often lead to additional light degrees of freedom appearing which spoil inflation \cite{Banks:2003sx}.  Indeed this difficulty is part of what led AMNV to propose the WGC in 2007: the axion WGC \eqref{eq:axionWGC} gives an upper bound on $f$, which in $D=4$ and assuming the instantons satisfying the WGC have small instanton number tells us that
\be\label{4daxWGC}
f S_{\mathrm{inst}}\lesssim \Mp.
\ee
Thus any natural inflation model with observable tensor modes and a computable potential (meaning $f\gtrsim 10 \Mp$ and $S_{\mathrm{inst}}\gg 1$) is in strong tension with the axion WGC.

\begin{figure}
\centering
\includegraphics[width=55mm]{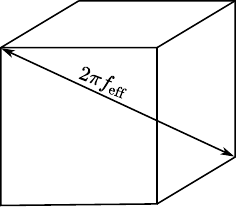}\
\caption{In N-flation, a large number $N$ of axions with individually sub-Planckian decay constants give rise to an effective decay constant $f_{\textrm{eff}}$ that can be arbitrarily large, realized by traveling along the space diagonal of the $N$-dimensional hypercube.}
\label{Nflationfig}
\end{figure}
\begin{figure}
\centering
\includegraphics[width=65mm]{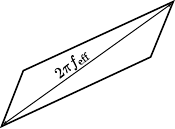}\
\caption{In decay constant alignment, two axions with individually sub-Planckian decay constants are aligned so that their diagonal can be arbitrarily large.}
\label{alignmentfig}
\end{figure}

Already by the time the WGC was introduced, various works had considered possible ways to get around the above difficulties and realize a model of large-field natural inflation consistent with quantum gravity. One such proposal is N-flation \cite{Liddle:1998jc, Dimopoulos:2005ac}. As its name suggested, N-flation invokes not just 1, but $N$ axion fields. If each field has a decay constant $f$, then by traveling along the diagonal in field space, one sees an effective decay constant of $f_{\textrm{eff}} = \sqrt{N} f$ (using the simple fact that an $N$-dimensional hypercube of side length $f$ has a diagonal of length $\sqrt{N} f$; see figure \ref{Nflationfig}). A related idea is decay constant alignment \cite{Kim:2004rp}: here, only two axions are needed, but their decay constants are ``aligned'' so that the fundamental axion domain is not a square, but rather an elongated parallelogram, as shown in figure \ref{alignmentfig}. Even though each individual axion may have a sub-Planckian decay constant, the diagonal direction in field space may be much larger than $M_{\rm Pl}$, thereby generating a model of natural inflation with a super-Planckian effective decay constant $f_{\textrm{eff}} \gg M_{\rm Pl}$.

\begin{figure}
\centering
\includegraphics[width=0.97\textwidth]{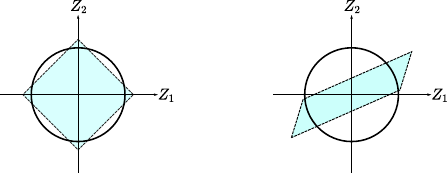}\
\caption{In their simplest incarnations, both N-flation and decay constant alignment violate the axion WGC: the charge-to-action vectors $\vec z_k = \vec{Q}_k/S_k$ of the instantons are superextremal, but their convex hull does not contain the unit ball.}
\label{CHCaxion}
\end{figure}

However, in their simplest incarnations, neither N-flation nor decay constant alignment evade the axion WGC. The individual instantons involved may be superextremal, but together they do not satisfy the convex hull condition, as shown in figure \ref{CHCaxion}. Said differently, there are no superextremal instantons associated with the diagonal directions of field space.

We can make a more general argument. Suppose our theory features $n$ instantons, with action
\begin{equation}
    S = \int \rmd^4x \left[ - \frac{1}{2 } \partial_\mu \vec{\phi} \cdot \mathbf{K} \cdot \partial^\mu \vec{\phi}  - V(\vec{\phi}) \right] \,,
\end{equation}
where $\mathbf{K}$ is the kinetic matrix for the axions. We further
suppose that instantons generate a leading-order potential of the form
\begin{equation}
V( \vec{\phi} ) = \sum_k  \Lambda_{\textrm{UV}}^4 \e^{-S_k} \left( 1 - \cos \left( \vec{Q}_k \cdot \vec{\phi}  \right) \right)\,,
\end{equation}
Next, suppose we want to inflate in the $\vec{e}$ direction of field space, so that the inflaton starts at the point $\vec \phi = \phi_0 \hat{e}$ and rolls to the minimum at the origin in approximately a straight line in field space. We assume that the largest value of $\phi_0$ allowed satisfies
\begin{equation}
\phi_0 \left( \vec{Q}_k \cdot  \hat{e}  \right) \leq \pi ~~ \text{for all $k$}\,,
\label{maxconst}
\end{equation}
since otherwise the inflaton sits in a cosine well of the $k$th potential term which does not contain the origin, which will presumably lead the inflaton to roll into a neighboring vacuum rather than the vacuum at $\vec{\phi} = 0$.

Next, we assume the axion WGC, which implies that for the given direction $\hat{e}$, there exists a superextremal instanton satisfying
\begin{equation}
\frac{ \vec{Q}_k \cdot {\hat e} }{f S_k } \gtrsim \frac{1}{M_{\rm Pl}} \,,
\end{equation}
where $f := \sqrt{\hat{e} \cdot \mathbf{K} \cdot \hat{e}}$ is the axion decay constant for the direction $\hat{e}$.
Finally, we assume $S_k > 1$ for perturbative control of the instanton expansion. Together with \eqref{maxconst}, this gives a bound on the physical displacement of the field,
\begin{equation}
||\vec{\phi}|| := \sqrt{\vec{\phi} \cdot \mathbf{K} \cdot \vec{\phi}} = f \phi_0    \lesssim \pi M_{\rm Pl}\,.
\end{equation}
Thus, the axion WGC constrains the maximum axion field range to be $O( \pi M_{\rm Pl})$: too small to generate a successful model of natural inflation with observable tensor modes \cite{Rudelius:2014wla,Rudelius:2015xta, Montero:2015ofa, Brown:2015iha, Brown:2015lia}.  The generality of this observation  re-invigorated hopes that the consistency of quantum gravity might lead to testable predictions for cosmology, and generated renewed interest in the WGC and the Swampland program more generally.  

There are several caveats to the above argument, however, which need to be discussed.  First of all, it assumes the axion WGC, which as we have seen is on somewhat shakier footing than the higher-form versions of the WGC. In particular, it is not immediately related to black hole evaporation.  Relatedly the $O(1)$ coefficient in the axion WGC is so far not decisively fixed, hence our argument produced only a ``squiggly'' $\lesssim$ statement rather a sharper $\leq$ statement. Various possibilities for the precise $O(1)$ coefficient in the axion WGC bound have been suggested in \cite{Heidenreich:2015nta, Andriolo:2020lul}.  Moreover, the bound \eqref{4daxWGC} relies on assuming that the instantons obeying the axion WGC have instanton number which is $O(1)$: this is natural from the point of view of the tower/sublattice WGCs, and also from the point of view of the idea that there should be objects obeying the WGC which are not black holes (or in the axion case which do not have large gravitational backreaction), but it only follows from tower/sublattice WGCs if we assume the relevant tower/sublattice is not too sparse.  

Another related issue is that the bound \eqref{maxconst} assumes that every instanton whose charge-to-action vector $\vec{Q}_k / S_k$ contributes to the convex hull also contributes significantly to the axion potential. However, it is conceivable that the dominant contributions to the axion potential could come from instantons which violate the axion WGC, whereas the instantons that satisfy the WGC give only subleading, unimportant contributions to the potential. In this case, the inflationary dynamics are unconstrained by the axion WGC.  This ``extra instanton loophole'' has driven a lot of interest in strong forms of the WGC \cite{Hebecker:2015rya, Bachlechner:2015qja}. However, even the lattice WGC is not quite sufficient to close this loophole \cite{Heidenreich:2019bjd}. On the other hand, threading the extra instanton loophole seems to require a fair bit of tuning \cite{Heidenreich:2019bjd}, and so far super-Planckian axion decay constants have yet to be realized in string theory \cite{long:2016jvd}.  

It is also possible to try to relax the assumption that $S_k\gg 1$, which we suggested is required for perturbative control of the instanton expansion.  In string compactifications, $S_k$ is typically the size of some cycle of the Calabi-Yau in string units, so the $\alpha'$ expansion of string theory is not valid unless $S_k \gg 1$. However, in ``extranatural inflation'' \cite{Arkani-Hamed:2003xts}, instantons in 4d come from the particles in 5d wrapping the compactification $S^1$. The instanton action for a particle of mass $m$ wrapping a circle of radius $R$ is given by $S_{\textrm{inst}} = 2 \pi m R$, and the contribution to the axion potential is given by: 
\begin{align}
V(\phi) &= \frac{3 (-)^S}{4 \pi^2} \frac{1}{(2 \pi R)^4}  \sum_{n \in \mathbb{Z}} c_n \e^{- 2 \pi n R m_5} \e^{in \phi}\,, &
c_n &= \frac{( 2 \pi R m_5)^2}{3 n^3} + \frac{2 \pi R m_5}{n^4} + \frac{1}{n^5}\,.
\end{align}
In this context, there is no problem with taking $S_{\textrm{inst}} \ll 1$: this simply corresponds to a light particle with $m \ll 1/R$. Likewise, there is no problem with perturbative control of the potential: the $1/n^5$ term in $c_n$ suppresses higher harmonics even for $S_{\textrm{inst}} \ll 1$ \cite{delaFuente:2014aca}. This significantly weakens the WGC bound \eqref{4daxWGC} on the axion decay constant. By imposing the convex hull condition on both the $U(1)$ of the parent 5d theory and the Kaluza-Klein $U(1)$, one can strengthen the bound to $f \lesssim M_{\textrm{Pl}}/S_{\textrm{inst}}^{1/2}$, but this is still insufficient to close this ``small action loophole'' in the context of extranatural inflation \cite{Heidenreich:2015wga}.

Finally, and perhaps most importantly, the above arguments only apply to theories where the only fields relevant during inflation are axions and the metric.  Including non-periodic scalars would only re-introduce the UV-sensitivity we avoided with axions, but we saw above in our discussion of the axion monodromy model \eqref{eq:monodromy} that the inclusion of a three-form gauge field $A_3$ coupled to an axion by a Chern-Simons term offers a simple mechanism whereby an axion with $f\lesssim \Mp$ can nonetheless lead to a scalar $\star F_4$ which rolls down a potential for many Planck distances.  Moreover we saw that the three-form WGC applied to $A_3$  and stringy corrections of the form $V(\star F_4)$ lead to only weak constraints on the range of this rolling.  We presented this model in the context of cosmological relaxation, but its most compelling application is really inflation, which indeed is what it was originally proposed for \cite{Silverstein:2008sg, McAllister:2008hb}.  Moreover proposals have been given for embedding this model into a consistent string compactification, although there are still certainly details remaining to be worked out and so far no detailed model has been given where a large field range is realized \cite{Baumann:2014nda, McAllister:2016vzi, Kim:2018vgz}.  

It is sometimes suggested that string theory does not allow for models with observable tensor modes, and the axion WGC was proposed in part to give an explanation for this claim, but axion monodromy casts serious doubt on it.  Natural inflation with observable tensor modes may well be in the Swampland, but the right lesson from this may just be that we should see what kind of predictions follow from the more general axion models which do seem to work.  In particular, given the ever-improving observational upper bounds on $r$, it is natural to ask whether axion monodromy models exist which aren't excluded but nonetheless predict observable tensor modes.  The simplest potential, a quadratic one (see equation \eqref{monodromyeom}), as well as simple extensions with other powers~\cite{McAllister:2014mpa}, are now already excluded by the Planck satellite and ground-based experiments including BICEP/Keck~\cite{BICEP:2021xfz, Kallosh:2021mnu}, but variations on the model are possible (e.g.,~\cite{DAmico:2021vka, DAmico:2021fhz}).  Perhaps the detailed issues remaining to be resolved in realizing axion monodromy in a genuine string compactification may yet lead to distinctive predictions.  If so, then the various forms of the WGC will likely be important tools in guiding us towards models that work.  Either way, it is remarkable that ongoing observations are teaching us concrete things about physics near the Planck scale.

\subsection{Implications for mathematics}

The WGC is a statement about the charges and masses of particles in effective field theory. In string/M-theory, supersymmetric effective field theories arise from compactifying on Calabi-Yau manifolds. Charged particles arise from $p$-branes wrapping $p$-cycles of the Calabi-Yau manifold. The charge of such a particle is determined by the homology class $\Sigma$ wrapped by the brane, and the mass of the particle is determined by the volume of the wrapped cycle. Thus, the WGC translates into geometric statements about the volumes of representatives of various cycles in a Calabi-Yau manifold.

For concreteness, let us consider the case of M-theory on a Calabi-Yau threefold $X$. This produces a 5d supergravity theory, and charged particles arise from M2-branes wrapping 2-cycles of $X$. The charge lattice of the theory is identified with the homology lattice $H_2(X, \mathbb{Z})$.

The resulting supergravity theory has a BPS bound: the mass of a particle of charge $q_I$ is constrained to satisfy
\begin{equation}
m \geq  \left( \frac{(2 \pi)^{2}}{2 \kappa_5} \right)^{1/3} |\zeta_q|\,,
\end{equation}
where $\zeta_q$ is the ``central charge,'' a quantity that depends linearly on $q_I$. It is sometimes remarked that the BPS bound is a sort of converse to the WGC bound, and there is a precise sense in which this is true: if there exist BPS black holes in a given direction $\hat{q}$ in the charge lattice, then the BPS bounds and extremality bounds coincide in this direction in the lattice. The only way a particle of charge $q_I \propto \hat{q}$ can satisfy both the WGC bound and the BPS bound is if it saturates both bounds. A particle that saturates the BPS bound is called a BPS particle: therefore, the tower/sublattice WGCs require an infinite tower of BPS particles of increasing mass/charge in every direction in the charge lattice for which the BPS bound coincides with the extremality bound.

Geometrically, BPS particles arise from M2-branes wrapping ``holomorphic'' curves of $X$. Here, a curve $\Sigma$ is ``holomorphic'' if its volume is given by integrating the K\"ahler form $J$ over it, $V_\Sigma = \int_\Sigma J$. Equivalently, we say that the curve is ``calibrated'' by the K\"ahler form.

The upshot of this is that the tower/sublattice WGCs imply the existence of an infinite tower of holomorphic curves in any direction $\hat{q}$ of the homology lattice $H_2(X, \mathbb{Z})$ for which the BPS bound coincides with the extremality bound. In fact, the condition that the BPS bound and extremality bounds coincide in the direction $\hat{q}$ can also be given a geometric interpretation: these bounds necessarily coincide for any $\hat{q}$ that resides in the so-called ``cone of moving curves'' $\mathcal{K}^\vee \subset H_2(X, \mathbb{R})$, which is equal to the cone dual of the ``cone of effective divisors'' \cite{Alim:2021vhs}.\footnote{If the theory in question allows certain $U(1)$ gauge fields to be enhanced to a larger nonabelian group, the effective cone of divisors may change under a phase transformation representing a Weyl reflection of the nonabelian gauge group. In this case, the BPS and extremality bounds coincide in an even larger cone in the charge lattice, which is geometrically the cone dual to the intersection of the effective cone of divisors over all phases of the theory. \cite{CornellGV}} Thus, the tower/sublattice WGCs translate to the nontrivial geometric statement that there must exist an infinite tower of holomorphic curves for any rational direction $\hat{q}$ within the cone of moving curves $\mathcal{K}^\vee$.

This statement is powerful in that (a) it is a purely geometric statement, with no reference to physics, and (b) it can actually be verified in examples. Gopakumar-Vafa (GV) invariants \cite{Gopakumar:1998ii,Gopakumar:1998jq,Gopakumar:1998ki} count the number of BPS particles (i.e., holomorphic curves) of a given charge $q_I$ in the lattice,\footnote{More accurately, GV invariants compute an index of BPS particles of a given charge, meaning that a nonzero GV invariant implies a nonzero number of BPS particles, whereas a vanishing GV invariant could result from an equal number of BPS hypermultiplets and BPS vector multiplets of a given charge. } and these can be computed for many Calabi-Yau hypersurfaces \cite{Hosono:1993qy} and complete intersection Calabi-Yau manifolds (CICYs) \cite{Hosono:1994ax}. At the same time, the cone of moving curves $\mathcal{K}^\vee$ can often be computed (with some input from the 5d supergravity theory) using the methods of \cite{Alim:2021vhs}. Together with the automated computation of GV invariants introduced in \cite{CYTools}, this has enabled thorough checks of this geometric version of the tower/sublattice WGCs in over 1400 Calabi-Yau manifolds \cite{CornellGV}.

So far, we have focused our attention on BPS particles, which are required by the tower/sublattice WGCs in certain directions in the charge lattice, where the BPS bound and extremality bound coincide. In a theory with 8 supercharges, however, there will generically be some directions for which the BPS bound and the extremality bound do not coincide. The tower/sublattice WGCs still require the existence of superextremal particles, which means they still impose constraints on the volumes of cycles of the Calabi-Yau manifold. To be more precise, the charge-to-mass vector of a particle associated with a $p$-cycle $\Sigma$ of a Calabi-Yau manifold is given by \cite{Hebecker:2015zss},
\begin{equation}
\vec{z} = \frac{V_X^{1/2} \vec{q}_\Sigma }{V_\Sigma}\,,
\end{equation}
where $V_X$ is the volume of the Calabi-Yau $X$, $V_\Sigma$ is the volume of $\Sigma$, and $\vec{q}_\Sigma$ labels the charge vector associated with the homology class. The norm $|| \vec q_\Sigma ||$ is the norm of the harmonic form related to $\Sigma$ using the metric on $X$.

The particle is superextremal when $|| \vec{z} || \geq \gamma_4^{1/2}$. In general, $\gamma_4$ depends on the direction $\hat{q}$ as well as the massless scalar fields in the theory, but it necessarily satisfies $\gamma_4 \geq \frac{1}{2}$, which means that a superextremal particle has 
 \begin{equation}
 \frac{V_X^{1/2} || \vec{q}_\Sigma || }{V_\Sigma} \geq \frac{1}{\sqrt{2}} \,.
 \label{geomeq}
\end{equation}
The tower WGC therefore implies that for a given cycle $[\Sigma]$, there exists an integer $n$ and a representative of $n [\Sigma]$ satisfying \eqref{geomeq}. The sublattice WGC further implies that there exists a universal $n$, which is independent of $[\Sigma]$.

In some cases, this bound leads to surprising, nontrivial mathematical results. In particular, any 4-cycle $[\Sigma]$ in a Calabi-Yau threefold $X$ with $h^{2,0}(X)=0$, can be represented as a union of holomorphic and antiholomorphic representatives; upon wrapping D4-branes on these representatives, these correspond to BPS and anti-BPS particles, respectively. The minimal volume representative of $[\Sigma]$ therefore has a volume no larger than the sum of the volumes of these holomorphic and antiholomorphic representatives, which we denote $V(\Sigma_\cup)$. But satisfying \eqref{geomeq} may require this union of holomorphic and antiholomorphic representatives to recombine into a new representative $\Sigma_{\text{min}}$ whose volume is significantly smaller than $V(\Sigma_\cup)$. More precisely, the ``recombination fraction''
\begin{equation}
\tau_\Sigma := \frac{V(\Sigma_\cup)-V(\Sigma_{\text{min}})}{V(\Sigma_{\text{min}})}
\end{equation}
may be much larger than 1 \cite{Demirtas:2019lfi}. Physically, this recombination corresponds to D4-branes wrapping these representatives recombining and fusing, and particles in 4d binding to form bound states of significantly smaller energy. Mathematically, the existence of representatives $\Sigma_{\text{min}}$ with large recombination fraction $\tau_\Sigma \gg 1$ is a nontrivial consequence of the WGC, which has been verified in some examples \cite{Long:2021lon}.

Finally, let us remark that within the context of 5d M-theory compactifications, there are interesting connections between the tower WGC, the WGC for strings, and the Swampland Distance Conjecture and various mathematical conjectures about Calabi-Yau manifolds known as ``cone conjectures'' \cite{Morrison93, Morrison94}. More on these connections can be found in \cite{BPSstrings}.

\subsection{Implications for general relativity}
\label{subsec:GRimplications}

The (weak) cosmic censorship hypothesis holds that for generic initial data, the maximal Cauchy development possesses a complete future null infinity~\cite{Penrose:1969pc}. Colloquially, there can be no naked singularities visible at future null infinity: any such singularity must be hidden behind a horizon.

There is strong numerical evidence that cosmic censorship is violated in more than four spacetime dimensions \cite{Lehner:2011wc}, as black strings may pinch off and develop singularities due to the Gregory-Laflamme instability \cite{Gregory:1993vy}. In four dimensions, however, such instabilities do not exist, and violation of cosmic censorship is much less certain.\footnote{Even in higher dimensions, known violations of cosmic censorship have zero mass and occur in Planck-sized regions where quantum gravitational effects become important. It has been argued that such quantum effects restore some notion of cosmic censorship, so that these counterexamples to cosmic censorship are relatively benign \cite{Emparan:2020vyf}.}

A promising class of counterexamples to cosmic censorship in four dimensions were proposed in \cite{Horowitz:2016ezu}, and strong numerical evidence for these counterexamples was subsequently provided in \cite{Crisford:2017zpi}. These examples involve a $U(1)$ gauge field coupled to gravity in asymptotically AdS space, with action (as usual we set the AdS radius to one)
\begin{equation}
S = \frac{1}{2 \kappa^2} \int \rmd^4x \sqrt{-g} \left( R + 6 - F^{\mu\nu} F_{\mu \nu}  \right)\,,
\label{AdSaction}
\end{equation}
with $F = \rmd A$. The boundary metric is chosen to be flat,
\begin{equation}
\rmd s_\partial^2 = -\rmd t^2 + \rmd r^2 + r^2 \rmd \varphi^2\,,
\end{equation}
and the only nonzero component of the potential at the boundary is the time component:
\begin{equation}
A_\partial =+ \frac{a(t) \rmd t}{(1+r^2)^{n/2}}\,
\end{equation} 
where $n$ is an integer controlling the fall-off of the field at large $r$.

Apparent violations of cosmic censorship arise when $a(t)$ is chosen to vanish at $t=0$ but increases to a constant value larger than some critical value $a_{\text{max}}$. In this case, there is no smooth static endpoint of the evolution, so one expects the curvature $F^2$ will grow indefinitely at late times. Numerical simulations confirm this expectation for $n = 1$ \cite{Crisford:2017zpi}. Note that the curvature does not diverge in finite time, so this example does not quite violate the letter of cosmic censorship, though it does violate the spirit of it.

This class of counterexamples disappears, however, in the presence of a superextremal scalar field \cite{Horowitz:2016ezu,Crisford:2017gsb}. In particular, suppose we add a charged scalar $\Phi$ to the action:
\begin{equation}
S_{\Phi} = - \frac{1}{4 \pi G} \int \rmd^4x \sqrt{-g} \left[ (D_\mu \Phi)(D^\mu \Phi)^\dagger + m^2 \Phi \Phi^\dagger \right]\,,
\label{eq:scalarfieldaction}
\end{equation}
with $D_\mu = \nabla_\mu - i \wt{q} A_\mu$ (here $\wt{q}=\frac{\sqrt{2}e}{\kappa}q$, where $q$ is the integral charge we've been using throughout).\footnote{We observed below equation \eqref{gamma0} that the WGC bound \eqref{gammadef} does not involve any powers of $\hbar$, and thus potentially has classical consequences, but those remarks applied for the case of a classical particle.  A similar statement applies for a charged classical field such as $\Phi$, but we need to be a bit more careful since if we restore $c$ and $\hbar$ then (in Heaviside-Lorentz units) the ``mass'' $m$ and ``charge'' $\wt{q}$ appearing in  \eqref{eq:scalarfieldaction} both have units of inverse length.  The true charge and mass of a particle appearing after this field is quantized are related to these by powers of $\hbar$ and $c$, but in writing the WGC inequality the powers of $\hbar$ and $c$ drop out since $m$ and $\wt{q}$ have the same units and $\kappa$ has already been absorbed into $\wt{q}$.}  The proposed WGC bound in AdS \eqref{eq:WGCAdS} then becomes  
\begin{equation}
\wt{q} \geq \Delta = \frac{3}{2} + \sqrt{\frac{9}{4} + m^2}\,.
\label{Deltaeq}
\end{equation}
When this bound is satisfied, for all choices of $n$ perturbations of $\Phi$ become unstable before $a$ grows to the critical value $a_{\text{max}}$, and cosmic censorship is restored.
When this bound is violated, the solution with the scalar field is still singular, and it is once again likely that cosmic censorship is violated \cite{Horowitz:2016ezu}. Thus, there is evidently a one-to-one correspondence between satisfying the WGC and obeying cosmic censorship in this setup.

Already, this result is quite suggestive. But the connection between the WGC and cosmic censorship becomes even more impressive when including dilatonic couplings and multiple scalar fields, as was done in \cite{Horowitz:2019eum}. In the dilatonic case, one begins with the action
\begin{equation}
S = \frac{1}{2 \kappa^2} \int \rmd^4x \sqrt{-g} \left( R + 6 - \e^{-2 \alpha \phi} F^{\mu\nu} F_{\mu \nu}  - 2 \nabla_\mu \phi \nabla^\mu \phi \right) \,,
\end{equation}
in place of \eqref{AdSaction}. Here, $\phi$ is a massless, uncharged scalar field which we will refer to as the dilaton, not to be confused with the massive, charged  scalar field $\Phi$. In the presence of this dilatonic coupling, the WGC bound for $\Phi$ is modified to
\begin{equation}
\wt{q} \geq \wt{q}^W \equiv \Delta (1+ \alpha^2)^{1/2}\,,
\label{WGCAdSdil}
\end{equation}
where $\Delta$ is given by \eqref{Deltaeq}. Notably, the minimal charge-to-mass ratio $\wt{q}/\Delta$ varies continuously with the parameter $\alpha$.

\begin{figure}
\centering
\includegraphics[width=80mm]{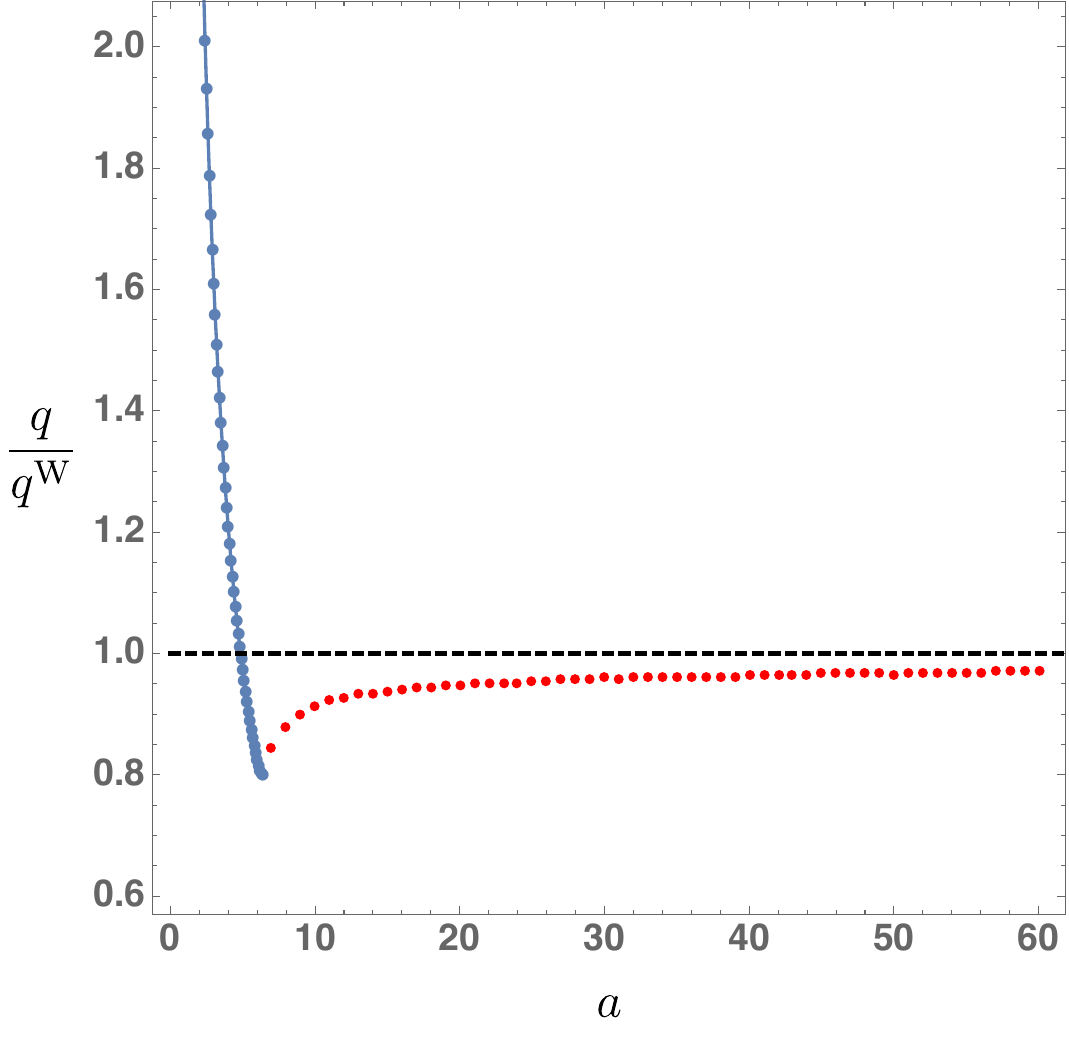}\
\caption{For fixed $n=4$, $\Delta=2$, and dilatonic coupling $\alpha = 0.9$, the condition to preserve cosmic censorship is precisely the WGC bound \eqref{WGCAdSdil}. Here, blue dots indicate the onset of solutions with $\Phi \neq 0$, and red dots indicate the approximate location of singular solutions. Figure from  \cite{Horowitz:2019eum} under a \href{https://creativecommons.org/licenses/by/4.0/}{Creative Commons License}.}
\label{fig:HS5}
\end{figure}

\begin{figure}
\centering
\includegraphics[width=80mm]{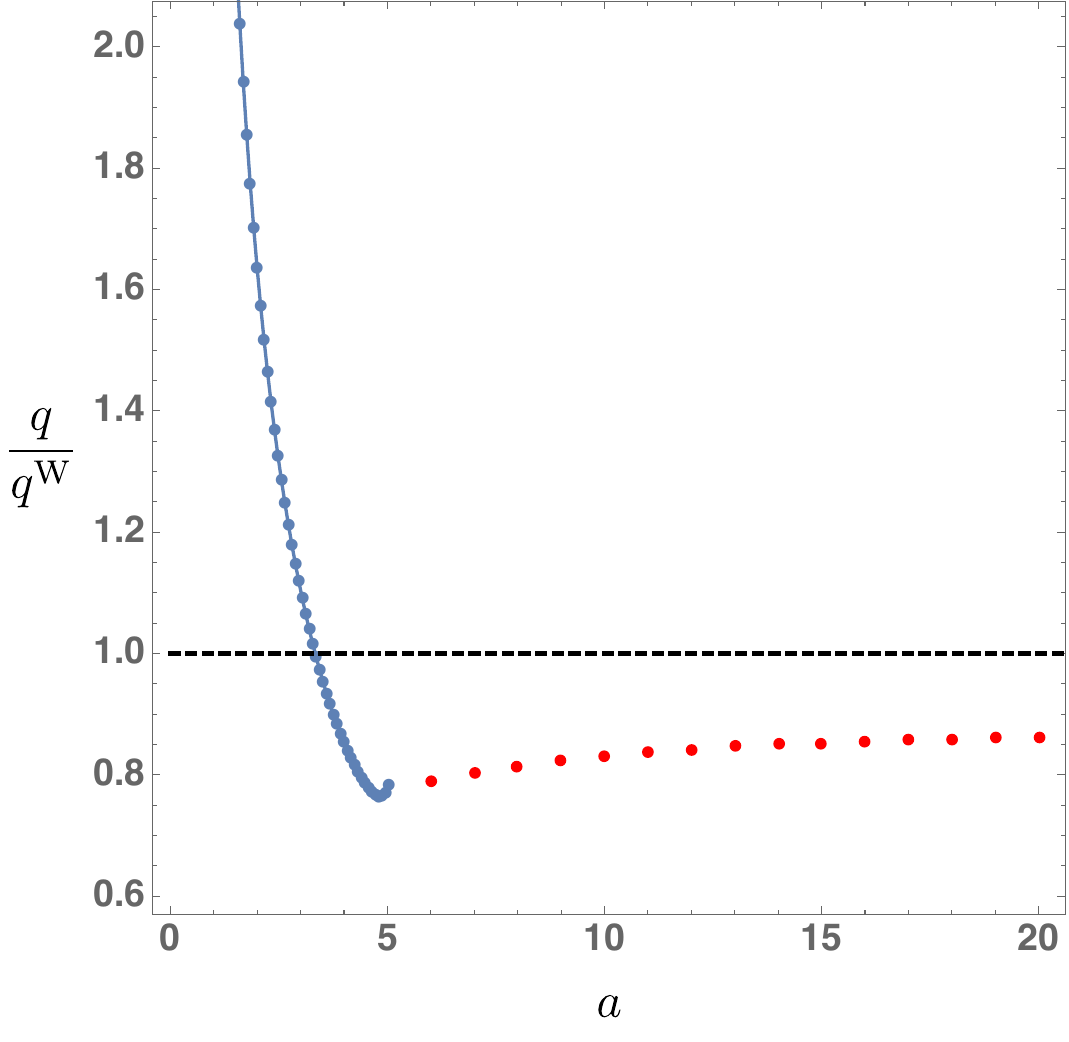}\
\caption{For fixed $n=4$, $\Delta=2$, and dilatonic coupling $\alpha = \sqrt{3}$, tcondition to preserve cosmic censorship does not quite seem to match with the WGC bound \eqref{WGCAdSdil}, though it is possible that modifications at large $a$ could restore the correspondence. Here, blue dots indicate the onset of solutions with $\Phi \neq 0$, and red dots indicate the approximate location of singular solutions. Figure from  \cite{Horowitz:2019eum} under a \href{https://creativecommons.org/licenses/by/4.0/}{Creative Commons License}.}
\label{fig:HS6}
\end{figure}

 \cite{Horowitz:2019eum} constructed numerical solutions to the equations of motion in the presence of the dilaton, with a boundary vector potential given by
\begin{equation}
A_\partial = \frac{a \rmd t}{(1+r^2)^n}\,,
\end{equation}
focusing in particular on the case $n=4$.
They found that for $\alpha < 1$, cosmic censorship is again preserved precisely when the WGC bound \eqref{WGCAdSdil} is satisfied, as shown in figure \ref{fig:HS5}. This is remarkable, in that it establishes a WGC-cosmic censorship connection over a one-parameter family of theories, indexed by $\alpha$. For $\alpha > 1$, numerical solutions suggest that it may be possible to preserve cosmic censorship even when $\wt{q} /\wt{q}^W$ (or equivalently $q/q^W$) is slightly smaller than 1, as shown in figure \ref{fig:HS6}. It is possible that this conclusion could be modified at large values of $a$, and the one-to-one correspondence between the WGC and cosmic censorship could be restored.

\begin{figure}
\centering
\includegraphics[width=80mm]{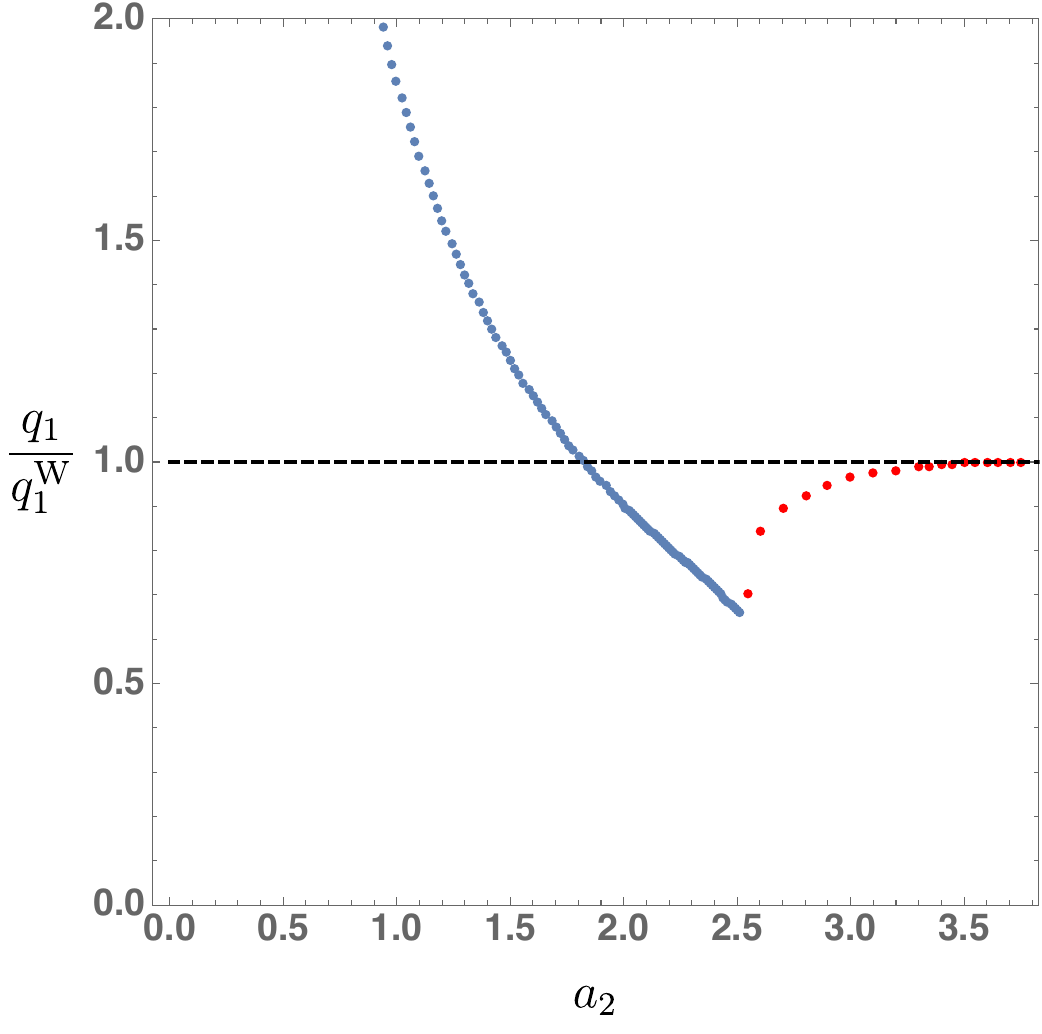}\
\caption{For fixed $n=4$, $a_1 =3 a_2$, $q_2=4$, $\Delta_1=\Delta_2=2$, the condition to preserve cosmic censorship is precisely the convex hull condition \eqref{CHCeq}. Here, blue dots indicate the onset of solutions with $\Phi_1 \neq 0$, and red dots indicate the approximate location of singular solutions. Figure from  \cite{Horowitz:2019eum} under a \href{https://creativecommons.org/licenses/by/4.0/}{Creative Commons License}.}
\label{fig:HS11}
\end{figure}

Finally, \cite{Horowitz:2019eum} also considered the relationship between cosmic censorship and the WGC in theories with two gauge fields. In their analysis, the asymptotic profile of the gauge fields is taken to be
\begin{equation}
A_{I\,\partial} = \frac{a_I}{(1+r^2)^n} \rmd t\,,
\end{equation}
focusing again on the case of $n=4$, with $a_1 = \lambda a_2$. There are now two massive scalar fields, $\Phi_1$ and $\Phi_2$; the former has charge $\wt{q}_1$ under the first gauge field and is uncharged under the second gauge field, whereas the latter carries charge $\wt{q}_2$ under the second gauge field and is uncharged under the first gauge field. Since there are multiple gauge fields, the WGC bound is equivalent to the convex hull condition (see Section \ref{ssec:CHC}), which is given by
\begin{equation}
\frac{1}{z_1^2} + \frac{1}{z_2^2} \leq 1\,,
\label{CHCeq}
\end{equation}
with $z_I = \wt{q}_I/\Delta_I$. By constructing numerical solutions to the equations of motion for various choices of $\lambda$, $\wt{q}_I$, and $\Delta_I$, \cite{Horowitz:2019eum} provided strong evidence that cosmic censorship is preserved precisely when the convex hull condition is satisfied. Figure \ref{fig:HS11} depicts this correspondence for one particular choice of $\lambda$, $\wt{q}_2$, $\Delta_1$ $\Delta_2$, varying $\wt{q}_1$.

The work of \cite{Horowitz:2016ezu, Crisford:2017zpi, Horowitz:2019eum} thus reveals a remarkable and surprising correspondence between the WGC and the weak cosmic censorship conjecture. A couple of aspects of this correspondence are worth further attention. First of all, note that the mildest form of the WGC, which simply requires the existence of a superextremal state, is not sufficient to preserve cosmic censorship. The mild WGC could be satisfied in principle by a finite-sized black hole state due to subleading corrections that slightly increase $Q/M$ relative to $Q/M$ for an infinitely large, extremal black hole. In the present scenario, however, preserving cosmic censorship requires a superextremal, field $\Phi$ (and thus quantum mechanically a superextremal particle) and not merely a superextremal black hole.

Second, we observe that in all examples so far the superextremal field which saves cosmic censorship is a scalar.  The usual formulations of the WGC do not put any restrictions on the spins of the superextremal particles which are required, but it is interesting to consider if there might be some such restriction.  One natural question for future study is whether or not superextremal bosonic fields of nonzero spin similarly prevent violations of cosmic censorship.  More generally one can also consider the question of whether or not fermions can do the same once quantum effects are included.  So far all string compactifications we are aware of in fact do have a superextremal scalar, but this may be an accident of supersymmetry so there is no strong evidence so far for an interplay between the WGC and spin.

Finally, let us mention an interesting connection between the WGC and a different kind of gravitational censorship, namely of super-Planckian spatial field variations. Simple dimensional analysis suggests such a field configuration could collapse into a black hole. Indeed, this is often true~\cite{Nicolis:2008wh}. However, one case in which a classically stable field configuration with an arbitrarily large scalar field variation, not screened by a horizon, can be constructed is a charged Kaluza-Klein bubble stabilized by flux~\cite{Horowitz:2005vp}. The scalar field is the radion, which traverses an infinite distance in field space to the bubble wall where $R \to 0$. Thus, arbitrarily large field values are not classically censored. In the quantum theory, the solution becomes unstable if charged matter satisfying the WGC exists, due to Schwinger pair production, which dynamically censors the large field excursion~\cite{Draper:2019zbb}. 

The general picture painted by these examples is that the WGC can play an important role in ensuring the validity of effective field theory. It prevents a low-energy observer from accessing arbitrarily high energy scales (in the case of cosmic censorship) or field values (in the case of super-Planckian censorship).

\section{Outlook}\label{OUTLOOK}

%!TEX root = WGC_Review_arxiv_v2.tex

In this review, we have seen that the Weak Gravity Conjecture potentially offers a deep organizing principle for unlocking the puzzle of quantum gravity.  In particular the Landscape of string vacua is very large, yet as far as we can tell, the WGC is obeyed in all of them. Suitably strong forms of the WGC place meaningful constraints on particle physics and cosmology, and have further consequences for black holes, pure mathematics, conformal field theories, and more. Thus, whereas most Swampland conjectures fall either into the ``rigorous but uninteresting'' category or the ``interesting but not rigorous'' category, the WGC has a claim at both rigor and importance.

Nonetheless, despite all that we have learned about it, the WGC remains shrouded in mystery. We saw in Section \ref{QUALITATIVE} of this review that several arguments point qualitatively to the validity of the WGC: it is quite plausible that the WGC is satisfied up to $O(1)$ coefficients. However, examples in string theory suggest something stronger: in all such examples, the WGC is satisfied with the precise $O(1)$ coefficient determined by the black hole extremality bound. This suggests that the WGC may be required for consistency of black hole physics, but it is not yet clear what goes wrong if the WGC is violated.

The evidence for the WGC coming from string theory is strong, but there is still a chance that the WGC suffers from a lamppost effect: the known string examples necessarily involve either (a) weak gauge coupling or (b) BPS particles---the only particles we currently know how to track into a strongly coupled regime. An example in which the WGC is satisfied by non-supersymmetric states in a regime of strong coupling is highly desirable, though perhaps unfeasible. Without this, a more compelling black hole argument is likely required to rule out the possibility of a lamppost effect.

Another interesting direction for future research involves the classification of weak coupling limits, as initiated in e.g. \cite{Grimm:2018ohb, Corvilain:2018lgw, Lee:2019wij, Klaewer:2020lfg, Perlmutter:2020buo, Lanza:2020qmt, Lanza:2021udy}. The Emergent String Conjecture of \cite{Lee:2019wij}, in particular, suggests that any infinite distance/weak coupling limit must be either a decompactification limit or an emergent string limit, in which a fundamental string becomes tensionless. In the former case, a tower of light, superextremal Kaluza-Klein modes emerge.\footnote{In the supersymmetric case, these KK modes are BPS and saturate the WGC bound. In the non-supersymmetric case, these KK modes satisfy the WGC bound with room to spare after the radion is stabilized. See Section \ref{REDUCTION} above for more details.} In the latter case, the modular invariance argument of Section \ref{MODULAR} ensures that the particles are superextremal. Thus, at weak coupling, the tower/sublattice WGCs follows from the Emergent String Conjecture (which could have stronger phenomenological implications). 

In Section \ref{EVIDENCE}, we encountered examples in which the sublattice WGC is satisfied by a sublattice of superextremal particles of coarseness $n > 1$ (and index greater than 1). This raises the question of how large the coarseness may become, or equivalently, how sparse the sublattice is allowed to be. This question is important, because the consequences of the sublattice WGC for low-energy physics can be arbitrarily weak if the coarseness is allowed to be arbitrarily large. Fortunately, the maximum coarseness we encountered was $n=3$: the sublattice of superextremal particles is not very sparse, and it is plausible that the sublattice WGC will always be satisfied with a coarseness of $O(1)$.

Before we may attempt to place any sort of universal upper bound on the coarseness/index of the sublattice, however, we must first understand how this sublattice shows up in the low-energy data of the theory. All of the examples with coarseness $n>1$ constructed so far are orbifold models, and $n$ divides the order of the orbifold group. However, we do not yet have a clear understanding of how this UV relationship manifests in the IR. One possibility is that the coarseness is related to the global structure of the low-energy gauge group, but more work is needed to clarify this picture.

As a final direction for future research, let us remark that the statement of the WGC in the presence of Chern-Simons terms is, at present, not well understood. The recent work \cite{Heidenreich:2021yda} presented an example of WGC mixing, in which the WGC for different $p$-form gauge fields are mixed up in the presence of Chern-Simons terms (see also \cite{Montero:2017yja, Heidenreich:2020pkc, Brennan:2020ehu, mixing}). These Chern-Simons terms imply that the gauge symmetry acquires a higher-group structure \cite{Sharpe:2015mja, Cordova:2018cvg}, and it seems likely that the full statement of the WGC is modified in the presence of such higher-group symmetries, reminiscent of how the WGC is modified to the Convex Hull Condition in theories with multiple $U(1)$s.  Understanding this better might have interesting implications for axion monodromy.

The Weak Gravity Conjecture has produced no shortage of surprises over the course of its fifteen-year existence, providing us with new insights into quantum gravity and unexpected connections between disparate areas of theoretical physics. Yet many of the most important questions remain open: is the Weak Gravity Conjecture true? If so, why? And which version(s) of the conjecture are the right ones? The answers to these questions may well lead us to even greater surprises than the ones we have already met.

\section{Acknowledgments}

We are grateful to many physicists for conversations that have shaped our understanding of the Weak Gravity Conjecture over the years. We would particularly like to thank Murad Alim, Nima Arkani-Hamed, Tom Banks, Clifford Cheung, Clay C\'ordova, Patrick Draper, Thomas Dumitrescu, Netta Engelhardt, Muldrow Etheredge, Gary Horowitz, Isabel Garcia Garcia, Manki Kim, Cody Long, Matteo Lotito, Javier Magan, Juan Maldacena, Liam McAllister, Jacob McNamara, Miguel Montero, Hirosi Ooguri, Kantaro Ohmori, Grant Remmen, Steve Shenker, Gary Shiu, John Stout, Leonard Susskind, Cumrun Vafa, and Irene Valenzuela for collaboration or discussions, but this is by no means an exhaustive list of those to whom we owe thanks.

DH is supported by the Simons Foundation as a member of the ``It from Qubit'' collaboration, the Sloan Foundation as a Sloan Fellow, the Packard Foundation as a Packard Fellow, and the Air Force Office of Scientific Research under the award number FA9550-19-1-0360, and the US Department of Energy under grants DE-SC0012567 and DE-SC0020360.  The work of BH is supported by NSF grant PHY-1914934. The work of MR is supported in part by the NASA Grant 80NSSC20K0506 and the DOE Grant DE-SC0013607. The work of TR is supported by NSF grant  PHY-1820912, the Simons Foundation, and the Berkeley Center for Theoretical Physics.

\appendix

\section{The black hole extremality bound}
\label{app:extremalitygeneral}

%!TEX root = WGC_Review_arxiv_v2.tex

In a gravitational theory with no naked singularities, a black hole of non-zero charge $Q$ cannot be arbitrarily light:
\begin{equation}
  M_{\text{BH}} \geqslant M_{\text{ext}} (Q)>0 \,, \label{eqn:AbstractExtBound}
\end{equation}
where the \emph{extremal mass} $M_{\text{ext}} (Q)$ is defined as the
infinimum of the set of possible masses for black holes of charge $Q$.  This bound arises because the energy stored in the electromagnetic field is a positive-energy source for the gravitational field, and in a theory with matter obeying reasonable energy conditions this gravitational flux cannot be cancelled without introducing a naked singularity as a negative-energy source.  A black
hole saturating this bound is \emph{extremal}, whereas all others are \emph{subextremal}.

To determine the extremal mass $M_{\text{ext}}(Q)$ and thereby the extremality bound \eqref{eqn:AbstractExtBound} it would obviously suffice to find an extremal black hole solution of charge $Q$ and read off its mass. However, an extremal black hole solution of a given charge does not always exist because taking the extremal limit sometimes generates a singularity at the event horizon. Furthermore, identifying whether a given solution is extremal is not straightforward. Experience with Reissner-Nordstr\"om black holes suggests that vanishing surface gravity is closely connected to extremality, but as we'll see in \S\ref{ssec:ExtBoundExamples}, not every black hole solution with vanishing surface gravity is extremal!

To solve the first problem, we expand our field of interest to include charged solutions that are merely \emph{limits} of black holes, not necessarily black holes themselves. In a convenient abuse of terminology, we call such limiting cases ``singular black holes.'' Familiar examples of singular black holes include, e.g., the background generated by $N \gg 1$ D0 branes in type IIA string theory.  Understanding what happens near a singular black hole usually requires UV information which goes beyond effective field theory, for example near a stack of D0 branes the string coupling gets large and there is a dual M-theory description.  

To solve the second problem, note that Hawking radiation must shut off in the extremal limit to satisfy cosmic censorship, so either the surface gravity $g_h$ (Hawking temperature) or horizon area $A_h$ (Bekenstein-Hawking entropy) must go to zero in this limit.
%%in an abuse of terminology we will still call such singular limits ``black holes.''
%\fixme{*****}
%To satisfy cosmic censorship, extremal black holes cannot emit Hawking radiation, and so should either have
%vanishing surface gravity (zero Hawking temperature) or a horizon of zero size
%(vanishing Bekenstein-Hawking entropy). 
We refer to black holes with either of
these two properties as \emph{quasiextremal}, where those with $A_h \to 0$ are singular. Cosmic censorship
requires extremal black holes to be quasiextremal, but as noted above
quasiextremal black holes are not always extremal; see~\S\ref{ssec:ExtBoundExamples} for examples. We call black holes that are neither extremal
nor quasiextremal \emph{nonextremal}.

In this appendix, we review general techniques for determining the extremality
bound in the large $Q$ limit (where derivative corrections can be ignored) for theories with multiple $U(1)$ gauge fields $A^A$, massless scalars $\phi^i$, and vanishing cosmological constant. For simplicity, we assume that
the lightest black holes of a given charge $Q$ are spherically
symmetric.\footnote{Without this assumption, the problem is unsolved in
general, except in special cases where a BPS-like bound can be derived using
spinor methods as in, e.g.,~\cite{Gibbons:1982jg}.} Temporarily ignoring the possibility of magnetic charge, the relevant terms in the
low-energy Einstein-frame effective action at  two-derivative order are
\begin{align}
  S_0 = \int \dd^d x \sqrt{- g}  \biggl[ \frac{1}{2 \kappa^2} R - \frac{1}{2} t_{A
  B} (\phi) F^A_2 \cdot F^B_2  - \frac{1}{2} G_{i j} (\phi) \nabla \phi^i \cdot
  \nabla \phi^j \biggr],
\end{align}
where $F_2^A = \dd A_1^A$. The types of two-derivative terms omitted above do not affect spherically symmetric black holes with purely
electric charge~\cite{Heidenreich:2020upe}.

\subsection{Black hole solutions}

In a convenient gauge, a spherically symmetric metric ansatz takes the
form:
\begin{equation}
  \dd s^2 = - \e^{2 \psi (r)} f (r) \dd t^2 + \e^{- \frac{2 \psi (r)}{d - 3}} 
  \left[ \frac{\dd r^2}{f (r)} + r^2 \dd \Omega^2_{d - 2} \right]
\end{equation}
for functions $\psi (r), f (r)$ to be determined, where $\dd \Omega^2_{d - 2}$ is the round metric of unit radius on $S^{d-2}$. The electric charge of the
solution is
\begin{equation}
  Q_A = \oint_{S^{d - 2}} t_{A B} (\phi) \star F^B_2,
\end{equation}
where the integral is taken over a sphere enclosing the horizon.
Spherical symmetry fixes the electric field to be
\begin{equation}
  F^A_2 = - \frac{t^{A B} (\phi) Q_B}{V_{d - 2}}  \frac{\e^{2 \psi} \dd t \wedge
  \dd r}{r^{d - 2}},
\end{equation}
where $t^{A B} (\phi)$ is the inverse of $t_{A B} (\phi)$ and $V_{d - 2} = 2
\pi^{\frac{d - 1}{2}} / \Gamma \left( \frac{d - 1}{2} \right)$ is the volume
of $S^{d - 2}$.

One component of Einstein's equations is now
\begin{equation}
f''(r) + \frac{3d-8}{r} f'(r) + 2 \frac{(d-3)^2}{r^2} (f(r)-1) = 0 \,,
\end{equation}
with the solution $f(r) = 1 + A/r^{d-3} + B/r^{2(d-3)}$. To interpret $A$ and $B$, we switch to ingoing Eddington–Finkelstein coordinates:
\begin{equation}
  \dd s^2 = - \frac{F (\rho) \, \dd v^2}{R^{2 (d - 3)} (\rho)} + \frac{2\, \dd
  v \,\dd \rho}{(d - 3) R^{d - 4} (\rho)} + R^2 (\rho) \dd \Omega_{d -
  2}^2,
\end{equation}
where $\rho = r^{d - 3}$, $R (\rho) = r \e^{- \frac{\psi}{d - 3}}$ and $F(\rho) = r^{2 (d - 3)} f (r)$.
A smooth event horizon occurs when $F \rightarrow 0$ with $R$ finite.
If one exists, $F (\rho) = \rho^2 + A \rho +
B$ can be factored
\begin{equation}
  F (\rho) = (\rho - \rho_+) (\rho - \rho_-), \quad \rho_+ \ge \rho_- ,
\end{equation}
leading to an outer (inner)
horizon at $\rho = \rho_+$ ($\rho = \rho_-$). We set $\rho_- = 0$ using the residual gauge symmetry $\rho \rightarrow \rho + \text{constant}$ with $F(\rho)$ and $R(\rho)$ held fixed,  so that
\begin{equation}
  F (\rho) = \rho (\rho - \rho_h) \qquad \Leftrightarrow \qquad f (r) = 1 -
  \frac{r_h^{d - 3}}{r^{d - 3}}   
\end{equation}
with $\rho_h = r_h^{d - 3}$.

In terms of $z \df \frac{1}{(d - 3) V_{d - 2} r^{d - 3}}$,\footnote{In comparison with~\cite{Heidenreich:2020upe},
$z^{(\text{here})} = z^{(\text{there})} / V_{d - 2}$.} $f(z) = 1 - z/z_h$ and the remaining equations of motion are:
\begin{align}
  \frac{\dd}{\dd z}  [f \dot{\phi}^j] + f \Gamma^i_{\; j k}  \dot{\phi}^j 
  \dot{\phi}^k & = \frac{1}{2} G^{i j} Q^2_{, j} (\phi) \e^{2 \psi}, 
  \label{eqn:phieqn}\\
  k_N^{- 1}  \frac{\dd}{\dd z} [f \dot{\psi}] & = \e^{2 \psi} Q^2 (\phi),
  \label{eqn:psieqn}\\
  k_N^{- 1}  \dot{\psi}  (f \dot{\psi} + \dot{f}) + f G_{i j} (\phi) 
  \dot{\phi}^i  \dot{\phi}^j & = \e^{2 \psi} Q^2 (\phi),  \label{eqn:consEqn}
\end{align}
where $\dot~{} = \frac{\dd}{\dd z}$, $G^{i j}(\phi)$ is the inverse of $G_{i j}(\phi)$,
$\Gamma^i_{\; j k} = \frac{1}{2} G^{i l}  (G_{l j, k} + G_{l k, j} - G_{j k,
l})$ are the associated Christoffel symbols, 
\be
Q^2 (\phi) \df t^{A B}
(\phi) Q_A Q_B,
\ee 
and 
\be
k_N \df \frac{d - 3}{d - 2} \kappa^2
\ee
is the
rationalized Newton force constant (such that $F_{\text{grav}} = -
\frac{k_N}{V_{d - 2}}  \frac{m m'}{r^{d - 2}}$).

Note that (\ref{eqn:phieqn}--\ref{eqn:consEqn}) are $z$-translation
invariant,\footnote{To preserve the boundary condition $\psi_{\infty} = \psi
(z = 0) = 0$ we then shift $\psi$ and rescale $z$ to compensate.} so for any
 solution passing through a point $\phi^i_0 = \phi^i (z_0)$ there is
corresponding solution ${\phi^i}' (z) = \phi^i (z + z_0)$ with
${\phi^i_{\infty}}' = \phi^i_0$. Moreover, (\ref{eqn:consEqn}) is a consistent constraint in that the
derivative of $f$ times (\ref{eqn:consEqn}) is a linear combination of
(\ref{eqn:phieqn}, \ref{eqn:psieqn}).

When $r_h > 0$ ($z_h$ is finite), a smooth horizon requires $\dot{\psi}(z_h) = -k_N z_h \mathrm{e}^{2 \psi_h} Q^2(\phi_h) \le 0$ by evaluating \eqref{eqn:psieqn} at $z = z_h$ where $\psi_h = \psi(z_h)$. Likewise, when $r_h = 0$ ($z_h = \infty$), a smooth horizon requires $z \mathrm{e}^{\psi} \propto R^{-(d-3)}$ to approach a non-zero constant as $z \to \infty$, hence $\psi \to - \log z + \text{constant}$, implying that $\dot{\psi} \to -\frac{1}{z} < 0$.
Combining (\ref{eqn:psieqn}, \ref{eqn:consEqn}),
\begin{equation}
  \ddot{\psi} = \dot{\psi}^2 + k_N G_{i j} (\phi)  \dot{\phi}^i  \dot{\phi}^j,
  \label{eqn:noQeqn}
\end{equation}
hence $\ddot{\psi} \ge 0$ and we conclude that %$\dot{\psi} \le 0$ for all $z \le z_h$.  Since $\ddot{\psi} \ge 0$, we conclude that
\begin{equation}
\dot{\psi} \le 0 \quad \text{for all} \quad z \le z_h \label{eqn:BHcondition}
\end{equation}
is required for a smooth horizon. Conversely, \eqref{eqn:BHcondition} together with $\ddot{\psi} \ge 0$ and $\psi_{\infty} = \psi
(z = 0) = 0$ gives $z \dot{\psi}_\infty \le \psi(z) \le 0$ for $0 \le z \le z_h$, so when $r_h > 0$ ($z_h$ is finite), $\psi_h$ is finite and the horizon is smooth.

Because a condition of the form \eqref{eqn:BHcondition} is preserved under limits, singular black holes must also satisfy \eqref{eqn:BHcondition}. Likewise, because any $r_h > 0$ solution satisfying \eqref{eqn:BHcondition} is smooth, $r_h = 0$ solutions satisfying \eqref{eqn:BHcondition} are limits of smooth solutions, and so solutions to \eqref{eqn:phieqn}--\eqref{eqn:consEqn} are (possibly singular) black holes if and only if~\eqref{eqn:BHcondition} holds.

% \eqref{eqn:BHcondition}.

%\fixme{*****}

Such black hole solutions have ADM mass
\begin{equation}
M=k_N^{- 1}  \bigl[ - \dot{\psi}_{\infty} + \frac{1}{2 z_h} \bigr]
\end{equation} 
(positive by \eqref{eqn:BHcondition}) and surface gravity and horizon area 
\begin{align}
g_h=\frac{d - 3}{2 r_h} \e^{\frac{d - 2}{d - 3} \psi_h}, \;\;\; A_h=V_{d - 2} r_h^{d - 2} \e^{- \frac{d - 2}{d - 3} \psi_h} 
\;\;\;\Longrightarrow\;\;\; g_h A_h = (d - 3) V_{d - 2} r_h^{d - 3} = \frac{1}{2 z_h}\,.
\end{align}
Therefore $r_h = 0$ ($z_h = \infty$) is the quasiextremal case, with coincident (possibly singular) inner and outer horizons, and $r_h > 0$ ($z_h$ finite) is the nonextremal (invariably smooth) case.

\subsubsection{Magnetic charge}

In 4d, spherical symmetry allows black holes to carry both electric and
magnetic charge. The theta term---which had no effect on purely electrically charged black holes---then becomes important
\begin{equation}
  S = S_0 - \frac{1}{8 \pi^2}  \int \theta_{A B} (\phi) F_2^A \wedge F_2^B .
\end{equation}
 The electric and magnetic charges are defined by
\begin{align}
  Q_A = \oint \left[ t_{A B} \star F^B + \frac{\theta_{A B}}{4 \pi^2}
   F^B \right], ~~~~~~ \tilde{Q}^A &= \frac{1}{2 \pi} \oint F^A .
\end{align}
The black hole
equations (\ref{eqn:phieqn}--\ref{eqn:consEqn}) take the same form as before (see, e.g., \cite{Heidenreich:2020upe})
but with
\begin{align}
  Q^2 (\phi) = t^{A B} (\phi)  \biggl[ Q_A - \frac{\theta_{A C} (\phi)}{2
  \pi} \tilde{Q}^C \biggr]  \biggl[ Q_B - \frac{\theta_{B D} (\phi)}{2 \pi}
  \tilde{Q}^D \biggr] + 4 \pi^2 t_{A B} (\phi) \tilde{Q}^A \tilde{Q}^B .
\end{align}

\subsubsection{Black branes}

Generalizing to homogenous, isotropic, and
spherically symmetric black $(p - 1)$-branes, the relevant effective action is
\begin{align}
  S = \int \rmd^d x \sqrt{- g}  \biggl[ \frac{1}{2 \kappa^2} R - \frac{1}{2} t_{A
  B} (\phi) F^A_{p + 1} \cdot F^B_{p + 1}  - \frac{1}{2} G_{i j} (\phi) \nabla
  \phi^i \cdot \nabla \phi^j \biggr] .
\end{align}
With the appropriate ansatz (see \cite{Heidenreich:2020upe} with
$z^{(\text{here})} = z^{(\text{there})} / V_{d - p - 1}$) the black hole
equations again take the form (\ref{eqn:phieqn}--\ref{eqn:consEqn}) with $k_N$ replaced by
the rationalized gravitational force constant
for $(p - 1)$-branes $k_{(p)} = \frac{p (d - p - 2)}{d - 2}
\kappa^2$. Magnetic charge can be added consistent with spherical
symmetry when $d = 2 p + 2$; see
\cite{Heidenreich:2020upe} for details.

\subsection{Quasiextremal black holes}

The quasiextremal case, in which $f (r) = 1$, has several interesting
properties that play an important role in determining the extremality bound.

\subsubsection{Vanishing self-force}

Evaluating (\ref{eqn:consEqn}) at $z = 0$ ($r = \infty$), we obtain
\begin{equation}
k_N M^2 + G^{i j}_{\infty} \mu_i \mu_j = t^{A
B}_{\infty} Q_A Q_B \label{eqn:noforce}
\end{equation}
where $\mu_i$ is the scalar charge appearing in, e.g.,
$\phi^i (z) = \phi^i_{\infty} - G^{i j}_{\infty} \mu_j z + O (z^2)$. Thus,
the long-range self-force between an identical pair of quasiextremal black
holes vanishes \cite{Heidenreich:2020upe}. As a corollary, $M = \sqrt{\frac{1}{k_N}  (t^{A B}_{\infty} Q_A Q_B
- G^{i j}_{\infty} \mu_i \mu_j)} \leqslant \sqrt{\frac{1}{k_N} t^{A
B}_{\infty} Q_A Q_B}$, hence quasiextremal black holes
coupled to moduli are no heavier than an extremal Reissner-Nordstr{\"o}m black
hole of the same charge that would result if the moduli were artificially frozen in place at their asymptotic values.

At this point it is interesting to note that even the \emph{short-range} forces between identical quasiextremal black holes vanishes if we restrict our attention to the classical, two-derivative effective action. This can be shown by explicitly constructing the static, multicenter solutions corresponding to several such black holes at rest near each other. While numerous examples of such solutions date back many years \cite{Majumdar:1947,Papaetrou:1947ib,Breitenlohner:1987dg}, a nice summary was recently given in~\cite{VanRiet:2020csu}. Of course, generic quantum and derivative corrections not only alter the short-range forces, but also change the extremal black hole solutions and thereby the long-range forces, see~\S\ref{subsec:BHcorrection}.

\subsubsection{The attractor mechanism}

A smooth horizon requires $R (\rho)$ to remain finite as $\rho \rightarrow
0$. Defining $\chi \df \psi + \log z$, $\e^{- \chi} = (d
- 3) V_{d - 2} R^{d - 3}$, so $\chi (z)$ must remain finite a $z \rightarrow
\infty$. Written in terms of $\chi$ and $\tau = - \log z$, the equations
of motion become:
\begin{align}
  \frac{\dd^2 \phi^i}{\dd \tau^2} + \Gamma^i_{\; j k}  
  \frac{\dd \phi^j}{\dd \tau}  \frac{\dd \phi^k}{\dd \tau} &= -
  \frac{\dd \phi^i}{\dd \tau} + \frac{1}{2} G^{i j} Q^2_{, j}
   \e^{2 \chi},  \label{eqn:chiphi}\\
  \frac{\dd^2 \chi}{\dd \tau^2} &= - \frac{\dd \chi}{\dd \tau} +
  k_N \e^{2 \chi} Q^2 - 1,  \label{eqn:chichi}\\
  \biggl[ \frac{\dd \chi}{\dd \tau} \biggr]^2\!\!\! + k_N G_{i j} 
  \frac{\dd \phi^i}{\dd \tau}  \frac{\dd \phi^i}{\dd \tau} &= - 2
  \frac{\dd \chi}{\dd \tau} + k_N \e^{2 \chi} Q^2 - 1 . 
  \label{eqn:chicons}
\end{align}
(\ref{eqn:chiphi}) and (\ref{eqn:chichi}) describe (generalized) Newtonian
motion in the potential $V (\chi, \phi^i) = k_N^{- 1} \chi - \frac{1}{2} \e^{2
\chi} Q^2 (\phi^i)$ with metric $G_{I J} = \diag (k_N^{- 1}, G_{i j})$
and a linear drag force. A smooth horizon requires $\chi (\tau)$ and $\phi^i
(\tau)$ to approach finite values $\chi_h$ and $\phi_h^i$ respectively as
$\tau \rightarrow - \infty$, which can only occur at a critical point of $V (\chi, \phi^i)$, i.e.,
\begin{equation}
  k_N \e^{2 \chi_h} Q^2 (\phi_h) = 1, \qquad Q^2_{, i} (\phi_h) = 0,
  \label{eqn:attractorconds}
\end{equation}
from which the constraint (\ref{eqn:chicons}) automatically follows. Thus,
$\phi_h$ is a critical point of $Q^2 (\phi)$ and $Q^2(\phi_h)$ determines the horizon area:
\begin{equation}
  A_h = V_{d -
  2} R_h^{d - 2} = V_{d - 2}  \left[ \frac{\sqrt{k_N} Q (\phi_h)}{(d - 3) V_{d
  - 2}} \right]^{\frac{d - 2}{d - 3}} .
\end{equation}
This is the attractor mechanism \cite{Ferrara:1995ih,Cvetic:1995bj,Strominger:1996kf,Ferrara:1996dd,Ferrara:1996um}. The trivial solution
$\phi^i (z) = \phi^i_h$ is Reissner-Nordstr{\"o}m, with mass $M_0 = k_N^{- 1 /
2} Q (\phi_h)$. All other solutions are strictly heavier since
$\mathcal{W} (z) = k_N^{- 1}  \frac{\dd}{\dd z}  (\e^{- \psi})$ 
evaluates to $M_{\text{BH}} = k_N^{- 1}  (- \dot{\psi}_{\infty})$ and
$M_0 = k_N^{- 1} \e^{- \chi_h}$ at $z=0$ and $z=\infty$ respectively, and 
$\dot{\mathcal{W}} = - \e^{- \psi} G_{i j} (\phi)  \dot{\phi}^i  \dot{\phi}^j
\leqslant 0$ per~\eqref{eqn:noQeqn}.\footnote{Thus, $k_N^{-1/2} Q (\phi_h) \le M \le k_N^{- 1/2} Q (\phi_\infty)$.}

If $\phi_h^i$ is a local minimum of $Q^2 (\phi)$ (the attractor point is
``stable'') then $(\chi_h, \phi_h^i)$ is a local maximum of $V (\chi,
\phi^i)$ and we can roll off the hill in any direction, hence there
are attractor solutions for any nearby choice of $\phi_{\infty}^i$.
This is not necessarily the case farther from the attractor point, where the
family of solutions to 
(\ref{eqn:chiphi}--\ref{eqn:chichi}) beginning at $\phi_h$ may encounter
turning points and/or caustics.

Unstable critical points of $Q^2 (\phi)$ also admit attractor solutions but by
the same reasoning these do not exist for generic values of $\phi_{\infty}^i$,
and so play little role in the determing the extremality bound.

\subsubsection{Fake superpotentials and a Bogomol'nyi bound}

Combining the preceeding observations, we see that there are families of
quasiextremal solutions corresponding to each stable attractor point
$\phi_h^i$, with mass $M$ determined by the choice of vacuum
$\phi_{\infty}^i$. The resulting mass function $M = W (\phi_{\infty})$ is also
known as the ``fake superpotential'' \cite{Ceresole:2007wx,Andrianopoli:2007gt,Andrianopoli:2009je,Andrianopoli:2010bj,Trigiante:2012eb} associated to the
attractor point in question. Because each member of the family has the same
charge and horizon-area (entropy), we can identify $\frac{\partial W}{\partial
\phi^i}$ as the scalar charge $\mu_i$ via the first law $\delta M = \mu_i
\delta \phi_{\infty}^i + \Phi_h^A \delta Q_A + \frac{1}{\kappa^2} g_h \delta
A_h$ \cite{Gibbons:1996af}. Therefore, due to the no-force condition~\eqref{eqn:noforce}, the
fake superpotential satisfies
\begin{equation}
  k_N W^2 (\phi) + G^{i j} (\phi) \partial_i W (\phi) \partial_j W (\phi) = Q^2 (\phi), \label{eqn:Mcond}
\end{equation} 
where $W (\phi)$ has a global minimum at $\phi_h^i$. Solutions to the
non-linear first-order differential equation (\ref{eqn:Mcond}) are in general
highly non-unique. However, the condition that $W (\phi)$ has a minimum at
$\phi = \phi_h$ is enough to fix this ambiguity, at least locally. 

This can be shown using a Bogomol'nyi bound, as follows. Consider any black
hole solution $\psi (r), \phi^i (r), f (r)$ (not necessarily quasiextremal).
The functional
\begin{align}
  I [\psi, \phi, f] \df \int_0^{z_h} \biggl[ \frac{1}{2 k_N}  (f \dot{\psi}
  + \dot{f})^2 + \frac{1}{2} f^2 G_{i j} (\phi)  \dot{\phi}^i  \dot{\phi}^j +
  \frac{1}{2} \e^{2 \psi} Q^2 (\phi) \biggr] \rmd z,
\end{align}
evaluates to the black hole mass upon imposing the equations of
motion:
\begin{align}
  I [\psi, \phi, f] = k_N^{- 1} \biggl[ \int_0^{z_h}  \frac{\dd}{\dd z}
  \left( \frac{1 + f}{2} f \dot{\psi} \right) \rmd z + \frac{1}{2 z_h} \biggr]  =
  k_N^{- 1} \biggl[ - \dot{\psi}_{\infty} + \frac{1}{2 z_h} \biggr] =
  M_{\text{BH}} ,
\end{align}
where the horizon boundary term vanishes in the quasiextremal $(f = 1)$ case because $\ddot{\psi} \ge \dot{\psi}^2$ and $\dot{\psi} \le 0$ from \eqref{eqn:noQeqn}, \eqref{eqn:BHcondition} imply $\dot{\psi} \to 0$ as $z \to \infty$. 

Given any function $W(\phi)$ satisfying $k_N W (\phi)^2 + G^{i j} W_{, i} W_{, j} \leqslant Q^2 (\phi)$
along the entire trajectory $\phi^i (r)$, $I$ can be factored as follows:
\begin{align}
  I [\psi, \phi, f] = & \int_0^{z_h} \biggl( \frac{k_N}{2}  \Bigl[ \frac{f
  \dot{\psi} + \dot{f}}{k_N} + \e^{\psi} W (\phi) \Bigr]^2 + \frac{1}{2} G_{i
  j}  [f \dot{\phi}^i + \e^{\psi} G^{i k} W_{, k}] [f \dot{\phi}^j + \e^{\psi}
  G^{j l} W_{, l}] \biggr) \dd z \nonumber\\
  & + \frac{1}{2} \int_0^{z_h} \e^{2 \psi} [Q^2 (\phi) - k_N W (\phi)^2 - G^{i
  j} W_{, i} W_{, j}] \dd z + W (\phi_{\infty}), \label{eqn:BogolmolnyiEqn}
\end{align}
where we use $\int_0^{z_h} \frac{\dd}{\dd
z} [f \e^{\psi} W (\phi)] \dd z = - W (\phi_{\infty})$ with the horizon boundary term vanishing in the quasiextremal $(f = 1)$ case because $e^{\psi} W(\phi) \le k_N^{-1/2} e^\psi |Q(\phi)| = k_N^{-1} \sqrt{\ddot{\psi}}$ and $\ddot{\psi} \ge \dot{\psi}^2$ and $\dot{\psi} \le 0$ likewise imply $\ddot{\psi} \to 0$ as $z \to \infty$.
Every term in \eqref{eqn:BogolmolnyiEqn} but the last is positive-definite, so we conclude that
\begin{equation}
  M_{\text{BH}} \geqslant W (\phi_{\infty}) . \label{eqn:Wbound}
\end{equation}
Saturating the bound requires (\ref{eqn:Mcond}) along the trajectory together
with $f = 1$ (quasiextremality)\footnote{The Bogomol'nyi equations $f \dot{\psi} + \dot{f} = - \e^{\psi} k_N W$, $f
\dot{\phi}^i = - \e^{\psi} G^{i j} W_{, j}$ and (\ref{eqn:Mcond}) imply
$\frac{\dd}{\dd z} [f \e^{\psi} W] = - \e^{2 \psi} Q^2 (\phi)$, whereas $\frac{\dd}{\dd
z} [\e^{\psi} W] = - k_N^{- 1}  \frac{\dd}{\dd z} [f \dot{\psi} +
\dot{f}] = - \e^{2 \psi} Q^2$ using (\ref{eqn:psieqn}). Thus
$\frac{\dd}{\dd z} [(1 - f) \e^{\psi} W] = 0$, implying $(1 - f)
\e^{\psi} W = 0$ by integrating from $z = 0$. Since $\e^{\psi} W>0$, $f = 1$ follows.} and the Bogomol'nyi equations
\begin{align}
  \dot{\psi} = - \e^{\psi} k_N W (\phi), ~~~~~~ \dot{\phi}^i = - \e^{\psi} G^{i
  j} W_{, j} . \label{eqn:Wflow}
\end{align}
These equations, implying (\ref{eqn:phieqn}--\ref{eqn:consEqn}), have a unique
solution for each choice of $\phi_{\infty}^i$ provided that the $W (\phi)$
gradient flow remains entirely within the region $R_0$ where (\ref{eqn:Mcond})
is satisfied, and the solution has a smooth horizon provided that the gradient
flow ends at a critical point of $W (\phi)$ with $W \left( \phi_{\text{crit}}
\right) > 0$ (ensuring $\e^{\psi} \propto \frac{1}{z}$ as $z \rightarrow
\infty$). Thus, given $W (\phi)$ and $\phi_{\infty}^i$, the black hole
solution saturating (\ref{eqn:Wbound}) is unique if it exists.

If both $W_1 (\phi)$ and $W_2 (\phi)$ satisfy (\ref{eqn:Mcond}) in a region
$R_0$ encompassing coincident local minima at $\phi = \phi_h$,\footnote{This
implies that $Q^2 (\phi)$ also has a local minimum at $\phi_h$.} $W_1$
gradient flows ending at $\phi_h$ produce quasiextremal solutions of mass
$M_{\text{BH}} = W_1 (\phi_{\infty})$, which must satisfy $M_{\text{BH}} = W_1
(\phi_{\infty}) \geqslant W_2 (\phi_{\infty})$ per (\ref{eqn:Wbound}). By the same token $W_2
(\phi_{\infty}) \geqslant W_1(\phi_{\infty})$ hence $W_1
(\phi_{\infty}) = W_2 (\phi_{\infty})$ for all $\phi_{\infty}$ in $R_0$
flowing to $\phi_h$.

Thus, the fake superpotential $W (\phi)$ associated to a given stable
attractor point $\phi^i_h$ is uniquely fixed near the attractor point by
(\ref{eqn:Mcond}) and the condition that $W (\phi)$ has a minimum at
$\phi_h^i$, and the corresponding attractor solutions can be obtained from $W
(\phi)$ by solving the gradient flow equations (\ref{eqn:Wflow}). (The family
of attractor solutions beginning at $\phi_h^i$ may have turning points and/or
caustics farther away from the attractor point, so $W (\phi)$ does not
necessarily extend uniquely throughout moduli space.)

\subsubsection{Asymptotic attractors}

A special case of the attractor mechanism occurs when $\phi_h^i$ lies at
infinite distance in the moduli space, usually in a direction where $Q^2
(\phi) \rightarrow 0$.\footnote{In principle, $\phi_h^i$ could lie at an
infinite distance point where $Q^2 (\phi_h) > 0$, but this does not occur
anywhere in the landscape to our knowledge.} Such ``asymptotic attractors''
technically do not lead to smooth black hole solutions (e.g., $Q^2 (\phi_h) =
0$ implies vanishing horizon area), but can be understood as the limit of a
family of smooth nonextremal solutions, and therefore play a role in
determining the extremality bound. 

Asymptotic attractors are typically also characterized by a fake
superpotential. Heuristically, because modifying $Q^2 (\phi)$ very far out in the moduli space can turn the asymptotic attractor into a standard attractor at finite distance with an associated unique fake superpotential, taking a limit where the new attractor point is sent off to infinity while restoring $Q^2 (\phi)$ to its original form should yield a fake superpotential for the original asymptotic attractor. This argument could fail in several ways when the asymptotic behavior of $Q^2(\phi)$ is sufficiently strange, but within the realm of actual quantum gravities, we know of no issue with it.

\subsection{The extremality bound}

Cosmic censorship requires extremal black holes to be quasiextremal, and
therefore for a given choice of $Q_A$ and $\phi_{\infty}^i$ the lightest
quasiextremal black hole should be extremal. Combined with the discussion of
fake superpotentials in the previous section, this suggests that the
extremality bound is given by
\begin{equation}
  M_{\text{BH}} \geqslant W (\phi) \df \min_{\{ a \}} W_a (\phi),
  \label{eqn:extBound}
\end{equation}
where $W_a (\phi)$ are the fake superpotentials associated to the various
stable or asymptotic attractors and the
minimum is taken amongst all the fake superpotentials defined at the point in
question.

To verify (\ref{eqn:extBound}), we assume that the ``global fake
superpotential'' $W (\phi)$ specified in (\ref{eqn:extBound}) is defined and
continuous everywhere in moduli space. Thus, the moduli space is partitioned
into different ``attractor basins'' associated to the various stable or asymptotic attractors, with $W(\phi)$ equal to the corresponding fake superpotential $W_a(\phi)$ within each basin.
Since each constituent fake superpotential $W_a (\phi)$ satisfies
(\ref{eqn:Mcond}), $W (\phi)$ also satisfies (\ref{eqn:Mcond}) except possibly
at the boundaries between attractor basins. However, as these boundaries are
sets of measure zero and the continuity of $W (\phi)$ precludes delta-function
contributions to $\nabla_i W$, the argument leading to the Bogomol'nyi bound
(\ref{eqn:Wbound}) is unaffected, hence $M_{\text{BH}} \geqslant W (\phi)$.
As this bound can be saturated by construction, it is indeed the
extremality bound.

Naviely, to construct $W (\phi)$ we must first classify \emph{all}
quasiextremal solutions. Fortunately, this is not the case: given any
(possibly incomplete) collection of local fake superpotentials $W_a
(\phi)$ such that $\hat{W} (\phi) \df \min_{\{ a \}} W_a (\phi)$ is
everywhere defined and continuous, the same reasoning as above implies that
$M_{\text{BH}} \geqslant \hat{W} (\phi)$ is the extremality bound, hence $\hat{W} (\phi) = W (\phi)$.

\subsection{Examples} \label{ssec:ExtBoundExamples}

A simple example that has played an outsized role in the development of the
WGC is Einstein-Maxwell-Dilaton theory, with the effective action:
\begin{equation}
  S = \int \rmd^d x \sqrt{- g}  \left[ \frac{1}{2 \kappa^2}  \left( R -
  \frac{1}{2} (\nabla \phi)^2 \right) - \frac{1}{2 \hat{e}^2} \e^{- \alpha \phi}  |
  F_2 |^2 \right] .
\end{equation}
Then (\ref{eqn:Mcond}) becomes:
\begin{equation}
  \xi \kappa^2  [W (\phi)]^2 + 2 \kappa^2  [W' (\phi)]^2 = \e^{\alpha \phi} 
  (\hat{e} Q)^2, \qquad \xi \df \frac{d - 3}{d - 2} .
\end{equation}
There is an asymptotic attractor at $\phi_h = - \infty$. Guessing a solution
of the form $W (\phi) = \hat{M} \e^{\alpha \phi / 2}$, we obtain
\begin{align}
  \left[ \xi + \frac{\alpha^2}{2} \right]  (\kappa \hat{M})^2 = (\hat{e} Q)^2  
 \qquad \Rightarrow \qquad \kappa \hat{M} = \gamma^{- \frac{1}{2}} \hat{e}  | Q |, \qquad
  \gamma = \xi + \frac{\alpha^2}{2} .
\end{align}
Since $W (\phi)$ is globally defined, positive, and satisfies
(\ref{eqn:Mcond}) everywhere, it defines a global fake superpotential, and the
extremality bound is
\begin{equation}
  \kappa M_{\text{BH}} \geqslant \kappa W (\phi_{\infty}) = \gamma^{-
  \frac{1}{2}} e | Q |,
\end{equation}
where $e = \hat{e}\, \e^{\alpha \phi_{\infty} / 2}$ is the vacuum gauge coupling. The same result applies to the corresponding $(p -
1)$-brane theory with $\xi \rightarrow \xi_{(p)} = \frac{p (d - p - 2)}{d -
2}$, as in~\eqref{eqn:PBraneBound}, \eqref{gammaPD}.

As a somewhat less trivial example, consider two gauge fields with different
dilaton couplings:
\begin{align}
  S = \int \rmd^d x \sqrt{- g}  \biggl[ \frac{1}{2 \kappa^2}  \left( R -
  \frac{1}{2} (\nabla \phi)^2 \right) - \frac{1}{2 \hat{e}^2_1} \e^{- \alpha_1 \phi} 
  | F_2 |^2  - \frac{1}{2 \hat{e}^2_2} \e^{\alpha_2 \phi}  | H_2 |^2 \biggr] .
\end{align}
Then (\ref{eqn:Mcond}) becomes
\begin{equation}
  \xi \kappa^2  [W (\phi)]^2 + 2 \kappa^2  [W' (\phi)]^2 = \e^{\alpha_1 \phi} 
  (\hat{e}_1 Q_1)^2 + \e^{- \alpha_2 \phi}  (\hat{e}_2 Q_2)^2 . \label{eqn:dilatonTwoQ}
\end{equation}
Provided that $\alpha_1 \alpha_2 > 0$ and $Q_{1, 2} \neq 0$, there is a stable
attractor point at $\phi_h = \frac{1}{\alpha_1 + \alpha_2} \log \left[
\frac{\alpha_2  (\hat{e}_2 Q_2)^2}{\alpha_1  (\hat{e}_1 Q_1)^2} \right]$.

Guessing a solution of the form $W (\phi) = \hat{M}_1 \e^{\alpha_1 \phi / 2} + \hat{M}_2
\e^{- \alpha_2 \phi / 2}$, the left-hand-side of \eqref{eqn:dilatonTwoQ} has cross terms proportional to $\e^{(\alpha_1 - \alpha_2) \phi}$ whose cancellation requires $\alpha_1
\alpha_2 = 2 \xi$. With this condition, we obtain
\begin{align}
  \kappa \hat{M}_{1,2} = \gamma_{1,2}^{- \frac{1}{2}} \hat{e}_{1,2}  | Q_{1,2} |, ~~~~~~ \gamma_{1, 2} = \xi +
  \frac{\alpha_{1, 2}^2}{2},
\end{align}
so the extremality bound in this case is
\begin{equation}
  \kappa M_{\text{BH}} \geqslant \gamma_1^{- \frac{1}{2}} e_1 | Q_1 | +
  \gamma_2^{- \frac{1}{2}} e_2 | Q_2 |,
\end{equation}
where $e_1 = \hat{e}_1 \e^{\alpha_1 \phi_{\infty} / 2}$ and $e_2 = \hat{e}_2 \e^{- \alpha_2
\phi_{\infty} / 2}$ are the gauge couplings in the vacuum in question. When
$\alpha_1 \alpha_2 \neq 2 \xi$, the fake superpotential solving
(\ref{eqn:dilatonTwoQ}) is not known in closed form (apart from some special
cases) but it is easily found by numerical integration.

\begin{figure}
\centering
\includegraphics[width=85mm]{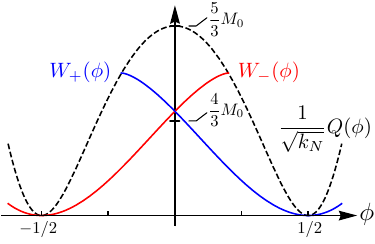}
\caption{The fake superpotentials for the two stable attractor points of \eqref{eqn:multipleattractor} with $\lambda = 4/3$ and $\phi_0 = \sqrt{3}/4$. Because $W_{\pm}(\phi)$ cross over each other at $\phi = 0$, $W_-(\phi)$ gradient flow solutions with $\phi_\infty > 0$ are quasiextremal but not extremal, as are $W_+(\phi)$ gradient flow solutions with $\phi_\infty < 0$.}
\label{fig:FirstOrderPT}
\end{figure}

\begin{figure}
\centering
\includegraphics[width=85mm]{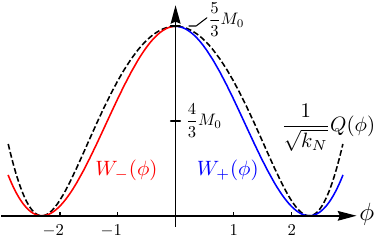}
\caption{The fake superpotentials for the two stable attractor points of \eqref{eqn:multipleattractor} with $\lambda = 4/3$ and $\phi_0 = 2$. In this case, all quasiextremal solutions are extremal.}
\label{fig:SecondOrderPT}
\end{figure}

So far, we have considered examples with a single attractor basin. A simple
(if contrived) example that exhibits multiple attractor basins is
\begin{align}
  Q^2 (\phi) = k_N M_0^2  \Bigl( 1 + \bigl[ (\phi/\phi_0)^2 - \lambda \bigr]^2 \Bigr), \quad G_{\phi \phi} = k_N^{- 1} \quad
\nonumber  \\ \Rightarrow \quad W (\phi)^2 + W' (\phi)^2 = M_0^2  \Bigl( 1 + \bigl[ (\phi/\phi_0)^2 - \lambda \bigr]^2 \Bigr), \label{eqn:multipleattractor}
\end{align}
with attractor points at $\phi = \pm \sqrt{\lambda} \phi_0$. The associated
fake superpotentials can be found by numerical integration, see
figures~\ref{fig:FirstOrderPT} and \ref{fig:SecondOrderPT}. Note that for some values of $\phi_0, \lambda$ there
are quasi-extremal solutions that are not extremal due to finite overlap
between the domains of the local fake superpotentials $W_{\pm} (\phi)$
associated to the attractor points $\phi = \pm \sqrt{\lambda} \phi_0$.

Some examples where numerical integration was used to determine the extremality bound in an actual theory of quantum gravity are discussed in \cite{Alim:2021vhs}.

\subsection{Closing comments}

To derive the extremality bound \eqref{eqn:extBound} we assumed that $W (\phi) \df \min_{\{ a \}} W_a (\phi)$ was everywhere defined and continuous. This is easily proved for a one-dimensional moduli space, since (1) the domain $\mathcal{D}_a$ of the fake superpotential $W_a(\phi)$ associated to a minimum $\phi_{(a)}$ of $Q^2(\phi)$ is at least as large as the interval between the two adjacent maxima of $Q^2(\phi)$ and (2) $W_a(\phi) = |Q(\phi)|/\sqrt{k_N}$ at the boundary of this domain (see, e.g., figure~\ref{fig:FirstOrderPT}), saturating the upper bound $W_b(\phi) \le |Q(\phi)|/\sqrt{k_N}$ on all fake superpotentials, where these two properties ensure (1) the existence and (2) the continuity of $W(\phi)$ at each point. While we do not know a general proof for a higher-dimensional moduli spaces, the existence and continuity of $W(\phi)$ can be checked on a case-by-case basis.

Phase transitions in the moduli space can create additional subtleties. Firstly, while $t_{A B} (\phi)$ and $G_{i j} (\phi)$ need not be analytic at a phase transition, in itself this has little effect on the foregoing analysis. More importantly, different branches of moduli space can meet at a phase transition, opening up the possibility of black hole solutions that cross from one branch to another. This is a rather complicated question that has not been worked out in the literature to our knowledge, but it seems probable that some version of fake superpotentials will still be applicable. Yet more drastically, the moduli space could have finite-distance boundaries where a strongly-coupled CFT appears, as in some examples from \cite{Alim:2021vhs}. Black hole solutions that reach this CFT boundary outside their event horizon lie outside the regime of validity of the weakly-coupled EFT that our analysis is based on, and require a separate analysis.

%\footnotesize
\setstretch{0.85}

\bibliographystyle{utphys}
\bibliography{ref}

\end{document}